\documentclass[aps,showpacs,prb,reprint,longbibliography,superscriptaddress]{revtex4-2}

\usepackage[colorlinks=true,linkcolor=blue,citecolor=blue,urlcolor=blue]{hyperref}
\usepackage{setspace} 
\usepackage{graphicx}
\usepackage{amsmath}
\usepackage{color}
\usepackage{amsmath}
\usepackage{mathtools}
\usepackage{amssymb}
\usepackage{verbatim}
\usepackage{physics}
\usepackage{latexsym}
\usepackage{enumerate} 
\usepackage{bm} 
\usepackage[caption=false]{subfig}
\usepackage{multirow}
\usepackage{booktabs}
\usepackage{siunitx}
\usepackage{comment}

\newcommand*{\Ang}{\, \mbox{\normalfont\AA}}

\usepackage{float}

\usepackage{lipsum}

\begin{document}





\title{\textbf{\Large{Theory of \emph{ab initio} downfolding with arbitrary range electron-phonon coupling}}}


\author{Norm M. Tubman}
\email{norman.m.tubman@nasa.gov}
\affiliation{NASA Ames Research Center, Moffett Field, California 94035, United States}
\author{Christopher J. N. Coveney}
\affiliation{Department of Physics, University of Oxford, Oxford OX1 3PJ, United Kingdom}
\author{Chih-En Hsu}
\affiliation{Department of Physics, Tamkang University, New Taipei City 251301, Taiwan}
\affiliation{Mork Family Department of Chemical Engineering and Materials Science, University of Southern California, Los Angeles, California 90089, USA}
\author{Andres Montoya-Castillo}
\affiliation{Department of Chemistry, University of Colorado Boulder, Boulder, CO 80309, USA}
\author{Marina R. Filip}
\affiliation{Department of Physics, University of Oxford, Oxford OX1 3PJ, United Kingdom}
\author{Jeffrey B. Neaton}
\affiliation{Materials Sciences Division, Lawrence Berkeley National Laboratory, Berkeley, California 94720, United States}
\affiliation{Department of Physics, University of California Berkeley, Berkeley, California 94720, United States}
\affiliation{Kavli Energy NanoScience Institute at Berkeley, Berkeley, California 94720, United States}
\author{Zhenglu Li}
\affiliation{Mork Family Department of Chemical Engineering and Materials Science, University of Southern California, Los Angeles, California 90089, USA}
\author{Vojtech Vlcek}
\affiliation{Department of Chemistry and Biochemistry, University of California, Santa Barbara, CA 93106, USA}
\affiliation{Materials Department, University of California, Santa Barbara, CA 93106-9510, USA}
\author{Antonios M. Alvertis}
\email{antonios.alvertis@oden.utexas.edu}
\affiliation{KBR, Inc., NASA Ames Research Center, Moffett Field, California 94035, United States}
\affiliation{Department of Physics, The University of Texas at Austin, Austin, TX 78712}
\affiliation{Oden Institute for Computational Engineering and Sciences, The University of Texas at Austin, Austin, TX 78712}

\date{\today}

\begin{abstract}
\emph{Ab initio} downfolding
describes the electronic structure of materials within a low-energy subspace, often
around the Fermi level. Typically starting from mean-field calculations, this framework
allows for the calculation of 
one- and two-electron
interactions, and the parametrization of a many-body Hamiltonian representing the active space of interest. The subsequent solution of such Hamiltonians can provide insights into
the physics of strongly-correlated materials. 
While phonons can substantially screen electron-electron interactions, electron-phonon coupling
has been commonly ignored within 
\emph{ab initio}
downfolding, and when considered this is done
only for short-range coupling. Here we propose
a theory of \emph{ab initio} downfolding
that accounts for short- and long-range electron-phonon 
coupling on equal footing. Our practical computational implementation is readily compatible with current downfolding approaches. We apply our approach to 
polar materials MgO and GeTe,
and we reveal the importance of 
both short-range and long-range
electron-phonon coupling in determining the magnitude of electron-electron interactions. Our results show that in the static limit, phonons reduce the on-site repulsion between electrons by $40$\% for MgO, and by $79$\% for GeTe.
Our framework also predicts that
overall attractive nearest-neighbor interactions arise
between electrons in GeTe, consistent with superconductivity in this material. 

\end{abstract}

\maketitle

\section{Introduction}

Understanding the properties of materials with
strong electronic correlations is a central challenge of condensed matter physics. Strong correlations are often primarily present
within a low-energy subspace around the Fermi
level, and can be described using so-called downfolding schemes~\cite{Nakamura2008,Nakamura2010,Arita2015,Zheng2018,Pham2020,Bauman2022,weisburn2024multiscaleembeddingquantumcomputing}. Within these approaches, an accurate many-body representation of the material is derived within the subspace,  typically from first-principles calculations. The high-energy electrons outside the subspace 
are treated at a lower level of theory, such
as mean-field approaches that include density functional theory (DFT).
A general Hamiltonian describing the electronic system within an active space is of the form
\begin{align}
    \label{eq:general_Hamiltonian}
    H=\sum_{ij\sigma}t_{ij}a_i^{\sigma \dagger}a_{j}^{\sigma}+\frac{1}{2}\sum_{ijkl\sigma \rho}U_{ijkl}a_{i}^{\sigma \dagger}a_{j}^{\rho \dagger}a_{l}^{\rho}a_{k}^{\sigma},
\end{align}
where $\sigma,\rho$ are spin indices, and $i,j,k,l$ run over the electronic basis states of the system. The one-body
terms $t_{ij}$ represent hopping
for $i\neq j$, and a chemical potential
term for $i = j$. The two-body terms $U_{ijkl}$ represent
the two-body electronic interactions within the subspace. These interactions are screened
by the electronic states outside the subspace
and their magnitude may be written as $U_{ijkl}=\int d\mathbf{r}\int d\mathbf{r}'\phi_i^*(\mathbf{r})\phi_j(\mathbf{r})W^{el}(\mathbf{r},\mathbf{r}',\omega)\phi_k^*(\mathbf{r}')\phi_l(\mathbf{r}')$, where
$W^{el}$ the frequency-dependent electronically-screened Coulomb
interaction, and $\phi$ denotes the electronic basis states. In the context of downfolding, $W^{el}$ is usually obtained within the constrained random-phase approximation (cRPA)~\cite{PhysRevB.70.195104,PhysRevB.105.235104,Chang2024}.

The ground and excited state properties
of downfolded Hamiltonians have been successfully obtained through a 
variety of methods, including exact diagonalization~\cite{Yoshimi2021,Botzung2024}, quantum Monte Carlo~\cite{Ma2015}, and variational
quantum eigensolvers~\cite{10.1063/5.0213525,alvertis2024downfolding}. 
Downfolding methods
have yielded insights into the ground state
properties of a variety of strongly-correlated materials, including 
high-temperature superconductors~\cite{Arita2015,Hirayama2019,Ohgoe2020,Been2021,Schmid2023},
excitonic insulators~\cite{PhysRevLett.124.197601,alvertis2024downfolding}, charge-ordered systems~\cite{Yoshimi2023,Schobert2024} and beyond, as well as into the excited state
properties in molecular materials~\cite{10.1063/5.0213525,Chang2024}.

Despite the above successes of \emph{ab initio} downfolding approaches, the vast
majority of studies in this context only
consider electronic effects within Hamiltonians of the form of Eq.\,\eqref{eq:general_Hamiltonian}. However,
it is established by this point that
lattice vibrations (phonons) can
renormalize the electronic properties of solids, such as their band gap~\cite{PhysRevLett.105.265501,PhysRevLett.112.215501,PhysRevB.92.140302}, excited state energies~\cite{Marini2008,PhysRevB.102.081122,Huang2021}, electron-electron interactions~\cite{PhysRevB.23.3627,PhysRevLett.99.236802}, and beyond. Indeed some studies have introduced electron-phonon coupling effects within a downfolding context~\cite{Nomura2015,PhysRevB.90.115435,PhysRevB.103.205103,10.21468/SciPostPhys.16.2.046,PhysRevX.13.041009}. Specifically, Ref.~\cite{PhysRevB.90.115435} included the effect of short-range electron-phonon coupling from a single optical phonon on the downfolded Hamiltonian. Ref.~\cite{Nomura2015} developed constrained density functional perturbation 
theory (cDFPT), as a means of appropriately screening the properties of phonons within
the subspace without double-counting interactions. Ref.~\cite{Nomura2015} also derived expressions for
electronic interactions mediated by short-range coupling to phonons, and subsequent works have shown that this coupling can 
cause up to $27$\% reduction of the on-site electron-electron repulsion, in systems that are predicted 
to be metallic by DFT~\cite{PhysRevB.108.L220508}. Ref.~\cite{PhysRevB.103.205103} explored
the implications of using cDFPT schemes for
the phonon properties of molecular systems, while Ref.~\cite{10.21468/SciPostPhys.16.2.046} demonstrated the potential of using downfolded electron-ion Hamiltonians to efficiently drive molecular dynamics simulations. Moreover, Ref.~\cite{PhysRevX.13.041009} provided a detailed investigation of how different approaches to screening affect phonon properties, and provided evidence that a combination of the cRPA and cDFPT schemes is
appropriate for the downfolding of electron-phonon systems on a subspace, without introducing double-counting errors. 

Despite this progress, a theoretical framework
that accounts for arbitrary-range electron-phonon coupling on the downfolded electronic Hamiltonian is still missing. The formulations outlined above only account for
the effect of \emph{short-range} electron-phonon coupling on electronic interactions. 
While this can be appropriate in metallic systems and non-polar semiconductors and insulators, 
polar, long-range electron-phonon coupling is not
only important in the most general case, but can in fact to dominate the electron-phonon interactions in a wide range of systems~\cite{Verdi2015,PhysRevB.99.235139}. Several of the systems within this category exhibit
strong electronic interactions near the Fermi level, and the rigorous application of downfolding frameworks could greatly improve our microscopic understanding of their physics. 

Here we develop a theory of \emph{ab initio} downfolding that includes short-
and long-range coupling between electrons and
phonons on equal footing. We derive expressions which allow us to quantify
the effect of phonons on electron-electron interactions
of arbitrary range. We formulate our theoretical approach in the basis
of maximally localized Wannier functions~\cite{RevModPhys.84.1419}, making
it immediately compatible with current approaches for
downfolding electronic degrees of freedom, 
and allowing us to present an efficient 
computational implementation. We apply our formalism to semiconducting MgO and GeTe, 
and show that apart
from the previously discussed renormalization of the
on-site Coulomb repulsion, there can be a strong influence
of lattice motion on long-range electron-electron interactions as well. 
For MgO, phonons reduce the on-site
electron-electron repulsion by $40$\%,
primarily through the deformation potential mechanism. For GeTe, phonon screening results in attractive long-range electron-electron
interactions due to the interplay
of the Fr\"ohlich, piezoelectric,
and deformation potential mechanisms
of electron-phonon coupling, offering
a potential explanation for superconductivity in this system. 

The structure of this paper is as follows. In Section\,\ref{downfolding} we provide an overview of 
\emph{ab initio} downfolding and the ingredients for its
computational implementation. Section\,\ref{phonon_screening} develops our inclusion of
phonon effects within \emph{ab initio} downfolding,
and gives the main theoretical results of this paper. In
Section\,\ref{lang_firsov} we connect these theoretical
results to the picture obtained from the Lang-Firsov transformation to a coupled electron-phonon system. Section\,\ref{computational_approach} discusses the
computational implementation of our formalism, to efficiently account for phonons within
\emph{ab initio} downfolding. We show the results of
our computational approach for MgO and GeTe
in Section\,\ref{results}, and we conclude the paper
with a discussion in Section\,\ref{discussion}.

\section{Brief overview of \emph{ab initio} downfolding}
\label{downfolding}

\begin{figure}[tb]
    \centering
    \includegraphics[width=0.8\linewidth]{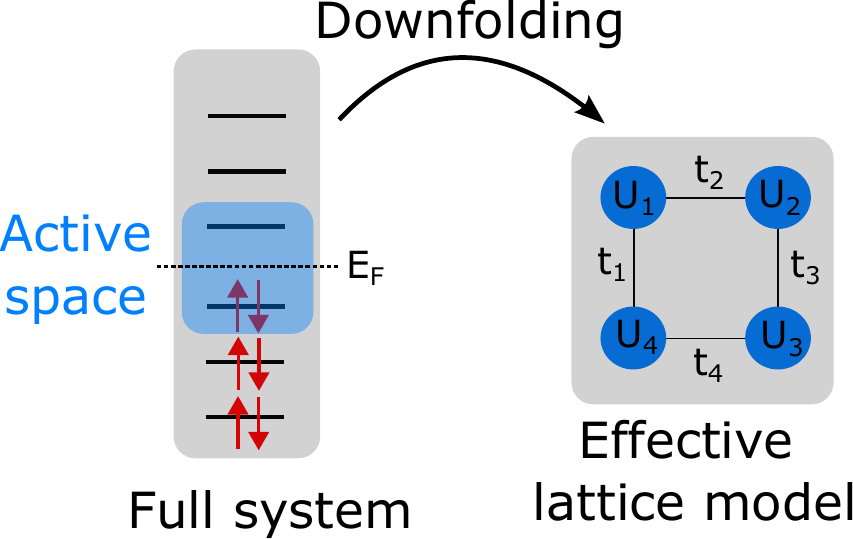}
    \caption{Schematic illustration of \emph{ab initio} downfolding. Starting from a description of the full system, an active space, typically around the Fermi level, is downfolded onto an effective many-body Hamiltonian on a lattice.}
    \label{fig:downfolding}
\end{figure}

Within \emph{ab initio} downfolding, the aim
is to derive a many-body Hamiltonian of the form of Eq.\,\eqref{eq:general_Hamiltonian}, which
represents the physics of a material within
an active space - typically around the Fermi level. It is often convenient that the downfolded many-body Hamiltonian describes
interactions of electrons on a lattice, 
making the choice of maximally localized Wannier functions as the computational basis a natural one. Denoting the $i$th Wannier orbital of our active space centered at lattice vector $\mathbf{R}$ as $\phi_{i\mathbf{R}}$, we can write the Hamiltonian in the case where we ignore Coulomb terms beyond density-density interactions as
\begin{align}
    \label{eq:Hamiltonian}
    H= \sum_{\sigma}\sum_{\mathbf{R}\mathbf{R}'}\sum_{ij}t_{i\mathbf{R}j\mathbf{R}'}a_{i\mathbf{R}}^{\sigma \dagger}a_{j\mathbf{R}'}^{\sigma}\nonumber\\ +\frac{1}{2}\sum_{\sigma \rho}\sum_{\mathbf{R}\mathbf{R}'}\sum_{ij}U_{i\mathbf{R}j\mathbf{R}'}a_{i\mathbf{R}}^{\sigma \dagger}a_{j\mathbf{R}'}^{\rho \dagger}a_{j\mathbf{R}'}^{\rho}a_{i\mathbf{R}}^{\sigma}.
\end{align}
The downfolding procedure is schematically illustrated in Fig.\,\ref{fig:downfolding}.

The one-body and
two-body terms $t_{i\mathbf{R}j\mathbf{R}'},U_{i\mathbf{R}j\mathbf{R}'}$ appearing in the Hamiltonian of Eq.\,\eqref{eq:Hamiltonian} are computed from first principles. The hopping term is
\begin{equation}
    \label{eq:hopping}
    t_{i\mathbf{R}j\mathbf{R}'}=\int_V d\mathbf{r}\phi_{i\mathbf{R}}^*(\mathbf{r})H_{o}\phi_{j\mathbf{R}'}(\mathbf{r}'),
\end{equation}
where $H_{o}$ a single-particle description
of the electronic system, often taken to
be equal to the 
Kohn-Sham Hamiltonian $H_{KS}$. 
The two-body Coulomb interaction is written as~\cite{Yoshimi2021}
\begin{align}
    \label{eq:coulomb}
    U_{i\mathbf{R}j\mathbf{R}'}(\omega)=\nonumber \\ \int_V d\mathbf{r} \int_V d\mathbf{r}' \phi_{i\mathbf{R}}^*(\mathbf{r})\phi_{i\mathbf{R}}(\mathbf{r})W(\mathbf{r},\mathbf{r}',\omega)\phi_{j\mathbf{R}'}^*(\mathbf{r}')\phi_{j\mathbf{R}'}(\mathbf{r}').
\end{align}
The frequency-dependent screened Coulomb interaction $W(\mathbf{r},\mathbf{r}',\omega)$ is commonly taken to be equal to the electronically-screened Coulomb interaction, \emph{i.e.}, $W=W^{el}$, within the cRPA, and in those cases we denote $U=U^{el}$.

It is important to emphasize at this point that the Hamiltonian of Eq.\,\eqref{eq:Hamiltonian} is of the extended Hubbard form and constitutes
a specific choice for the representation of the electronic structure
within the active space of interest. In addition to the
one-body hopping and two-body Coulomb terms, a two-body
exchange term is also included in some cases~\cite{Yoshimi2021}. Moreover, it is possible to incorporate terms beyond
density-density Coulomb interactions~\cite{Romanova2023}, as in the Hamiltonian of Eq.\,\eqref{eq:general_Hamiltonian}. 
Our theoretical framework outlined below may 
be readily extended to incorporate phonon
effects on general $4$-center integrals. Due to the more cumbersome notation that this entails, we will present here our results
primarily for $2$-center integrals, however, in Appendix~\ref{derivation_Uph} we give the
expressions for the more general case. 

An important aspect of the two-body terms of Eq.\,\eqref{eq:coulomb} is the frequency dependence of
the screened Coulomb interaction. This derives from
the frequency dependence of the inverse dielectric matrix $\epsilon^{-1}$, which screens the bare Coulomb repulsion $v$
~\cite{Strinati1988,Rohlfing1998,Rohlfing2000}:
\begin{equation}
\label{eq:screening}
    W^{el}(\mathbf{r},\mathbf{r}',\omega)=\int d\mathbf{r}''\epsilon^{-1}(\mathbf{r},\mathbf{r}'',\omega)v(\mathbf{r}'',\mathbf{r}'),
\end{equation}
where $v$ the bare (unscreened) Coulomb interaction. This expression may in turn be written as~\cite{Deslippe2012}:
\begin{equation}
\label{eq:w_crpa}
    W^{el}_{\mathbf{G}\mathbf{G}'}(\mathbf{q},\omega)=\epsilon^{-1}_{\mathbf{G}\mathbf{G}'}(\mathbf{q},\omega)v(\mathbf{q}+\mathbf{G}'),
\end{equation}
where here we work in reciprocal space and a plane wave basis, denoting $\mathbf{q}$
as a vector within the first Brillouin zone, and $\mathbf{G}$ as a reciprocal lattice vector. Within the
cRPA, the dielectric matrix is written as
\begin{equation}
    \label{eq:eps_crpa}
    \epsilon_{\mathbf{G}\mathbf{G}'}(\mathbf{q},\omega)=\delta_{\mathbf{G}\mathbf{G}'}-v(\mathbf{q}+\mathbf{G})\chi_{\mathbf{G}\mathbf{G}'}(\mathbf{q},\omega),
\end{equation}
where the bare Coulomb interaction is
\begin{equation}
v(\mathbf{q}+\mathbf{G})=4\pi/|\mathbf{q}+\mathbf{G}|^2, 
\end{equation}
and the polarizability is written as
\begin{align}
    \label{eq:chi_crpa}
    \chi_{\mathbf{G}\mathbf{G}'}(\mathbf{q},\omega)=\sum_{\mathbf{k}}\sum_{n}^{occ}\sum_{n'}^{unocc}(1-T_{n'\mathbf{k}+\mathbf{q}}T_{n\mathbf{k}}) \nonumber \\  \times M_{nn'}(\mathbf{k},\mathbf{q},\mathbf{G})M^*_{nn'}(\mathbf{k},\mathbf{q},\mathbf{G}')
    \nonumber \\ \times [\frac{1}{\omega-E_{n'\mathbf{k}+\mathbf{q}}+E_{n\mathbf{k}}+i\delta}-\frac{1}{\omega+E_{n'\mathbf{k}+\mathbf{q}}-E_{n\mathbf{k}}-i\delta}],
\end{align}
where $\delta \rightarrow 0^+$.
Here we sum over $\mathbf{k}$-points, occupied and unoccupied bands, and we use the notation $M_{nn'}(\mathbf{k},\mathbf{q},\mathbf{G})=\bra{n\mathbf{k}+\mathbf{q}}e^{i(\mathbf{q}+\mathbf{G})\cdot \mathbf{r}}\ket{n'\mathbf{k}}$. Moreover, we take the definition 
\begin{equation}
    \label{eq:crpa_operators}
    T_{n\mathbf{k}}=\sum_{i}^{N_{wan}}w^{\mathbf{k}}_{ni}w^{\mathbf{k}*}_{ni},
\end{equation}
where we sum over the number of Wannier states in the subspace and $w^{\mathbf{k}}_{ni}$ are the rotation matrices that take us from the plane-wave to the Wannier basis~\cite{Yoshimi2021}. The term $(1-T_{n'\mathbf{k}+\mathbf{q}}T_{n\mathbf{k}})$ excludes
the screening of the states within the subspace, hence
avoiding double-counting. In regular RPA, rather than cRPA, the polarizability is computed by  setting $T_{n'\mathbf{k}+\mathbf{q}}T_{n\mathbf{k}}=0$.

A common approximation to simplify the frequency dependence
of the screened Coulomb interaction is to take the static limit $\omega=0$ in eqs.\,\eqref{eq:w_crpa},\,\eqref{eq:eps_crpa} and\,\eqref{eq:chi_crpa}. This is a significant
simplification; finding the eigenvalues of a downfolded Hamiltonian which includes frequency-dependence constitutes a non-linear problem~\cite{canestraight2024renormalizationstatesquasiparticlesmanybody}, the accurate solution of
which is an active area of research. 
The static approximation can lead to both under-screening and over-screening of Coulomb
interactions depending on the details of a given system~\cite{Romanova2023,PhysRevLett.132.076401}, nevertheless its simplicity and practicality has
led to its widespread adoption. Our formalism 
for including phonon effects in \emph{ab initio} downfolding, as developed in the following Section\,\ref{phonon_screening}, includes the full frequency dependence of phonon screening, however, we will
in some cases discuss the static limit in order to develop
intuition, and to understand the impact of our approach on
results similar to those reported in the literature. 

\section{Inclusion of phonon effects}
\label{phonon_screening}

The key element in determining the electron-electron interactions $U_{i\mathbf{R}j\mathbf{R}'}$ within a subspace is the screened Coulomb interaction
$W$, as expressed in Eq.\,\eqref{eq:coulomb}. So far we have only considered the case where
the Coulomb interaction is screened electronically, \emph{i.e.}, $W=W^{el}$. However, it is established that phonons can also screen the Coulomb interaction between
electrons. Within many-body perturbation theory and to the lowest order in the electron-phonon interaction, the phonon correction to the screened Coulomb interaction is written as~\cite{Baym1961,Hedin1965,Giustino2017}
\begin{equation}
\label{eq:phonon_screening}
W^{ph}(\mathbf{r},\mathbf{r}',\omega) = \sum_{\mathbf{q},\nu} D_{\mathbf{q},\nu}(\omega)g_{\mathbf{q},\nu}(\mathbf{r})g^*_{\mathbf{q},\nu}(\mathbf{r}').
\end{equation}
Here $D_{\mathbf{q},\nu}(\omega)$ is the propagator of a phonon with branch index $\nu$ and at wavevector $\mathbf{q}$, and $g_{\mathbf{q},\nu}$ is the electron-phonon
vertex. The phonon propagator here already includes inter-phonon mixing~\cite{ruhman2025commentarxiv250117230arxiv250200103phononmediated} as described by a nonanalytical contribution~\cite{PhysRevB.55.10355}. 
Phonon screening can result in attractive electron-electron interactions and phenomena such as superconductivity~\cite{MahanGeraldD2013MP}, and has also recently been incorporated
in the \emph{ab initio} calculation of excited states
within the $GW$-Bethe-Salpeter equation formalism~\cite{Alvertis2024},
where it was shown that it can lead to strongly temperature-
dependent exciton binding energies. Moreover, it was found 
in Refs.~\cite{PhysRevLett.127.067401,Alvertis2024} that Fr\"ohlich-like long-range coupling of electrons to
longitudinal phonons dominates this effect in semiconductors. 

We now include phonon screening in the screened Coulomb interaction appearing in Eq.\,\eqref{eq:coulomb}, \emph{i.e.}, we take the total screened Coulomb interaction
to be
\begin{equation}
    \label{eq:total_screening}
    W=W^{el}+W^{ph},
\end{equation}
where $W^{el}$ is given by Eq.\,\eqref{eq:screening}, and
$W^{ph}$ is given by Eq.\,\eqref{eq:phonon_screening}. We
therefore obtain the \emph{total} Coulomb interaction
\begin{align}
    \label{eq:total_coulomb}
    U^{tot}_{i\mathbf{R}j\mathbf{R}'}(\omega)=\nonumber \\ \int_V d\mathbf{r} \int_V d\mathbf{r}' \phi_{i\mathbf{R}}^*(\mathbf{r})\phi_{i\mathbf{R}}(\mathbf{r})W^{el}(\mathbf{r},\mathbf{r}',\omega)\phi_{j\mathbf{R}'}^*(\mathbf{r}')\phi_{j\mathbf{R}'}(\mathbf{r}')\nonumber \\ + \int_V d\mathbf{r} \int_V d\mathbf{r}' \phi_{i\mathbf{R}}^*(\mathbf{r})\phi_{i\mathbf{R}}(\mathbf{r})W^{ph}(\mathbf{r},\mathbf{r}',\omega)\phi_{j\mathbf{R}'}^*(\mathbf{r}')\phi_{j\mathbf{R}'}(\mathbf{r}').
\end{align}
The term $U^{el}_{i\mathbf{R}j\mathbf{R}'}(\omega)=\bra{\phi_{i\mathbf{R}}\phi_{j\mathbf{R}'}}W^{el}(\omega)\ket{\phi_{i\mathbf{R}}\phi_{j\mathbf{R}'}}$ in the second row is the usual (extended) Hubbard term.

The phonon term of the third row is written as $U^{ph}_{i\mathbf{R}j\mathbf{R}'}(\omega)=\bra{\phi_{i\mathbf{R}}\phi_{j\mathbf{R}'}}W^{ph}(\omega)\ket{\phi_{i\mathbf{R}}\phi_{j\mathbf{R}'}}$, so
that overall we can write the total Coulomb term in the compact form $U^{tot}_{i\mathbf{R}j\mathbf{R}'}(\omega)=U^{el}_{i\mathbf{R}j\mathbf{R}'}(\omega)+U^{ph}_{i\mathbf{R}j\mathbf{R}'}(\omega)$. We now evaluate the phonon-mediated Coulomb term $U^{ph}_{i\mathbf{R}j\mathbf{R}'}(\omega)$.
The phonon propagator appearing in Eq.\,\eqref{eq:phonon_screening} is written as 
\begin{equation}
    \label{eq:propagator}
        D_{\mathbf{q},\nu}(\omega)=\frac{1}{\omega-\omega_{\mathbf{q},\nu}+i\delta}-\frac{1}{\omega+\omega_{\mathbf{q},\nu}-i\delta},
\end{equation}
where $\omega_{\mathbf{q},\nu}$ the phonon frequency. 
Substituting the
definition of the phonon-screened Coulomb interaction of Eq.\,\eqref{eq:phonon_screening} in the integral for $U^{ph}_{i\mathbf{R}j\mathbf{R}'}(\omega)$:
\begin{align}
    \label{eq:Uph}
    U^{ph}_{i\mathbf{R}j\mathbf{R}'}(\omega)=\bra{\phi_{i\mathbf{R}}\phi_{j\mathbf{R}'}}W^{ph}(\omega)\ket{\phi_{i\mathbf{R}}\phi_{j\mathbf{R}'}} =\nonumber\\ \sum_{\mathbf{q},\nu}g_{ii\mathbf{q}\nu}(\mathbf{0})g^*_{jj\mathbf{q}\nu}(\mathbf{R}'-\mathbf{R})\nonumber \\
    \times [\frac{1}{\omega-\omega_{\mathbf{q},\nu}+i\delta}-\frac{1}{\omega+\omega_{\mathbf{q},\nu}-i\delta}].
\end{align}
A detailed derivation of this expression is given in Appendix\,\ref{derivation_Uph}, where it is also generalized to the case of $4$-center integrals. Here we use the mixed Wannier-Bloch representation of the electron-phonon coupling matrix
element:
\begin{align}
        g_{mn\mathbf{q}\nu}(\mathbf{R}) = \bra{\phi_{m\mathbf{R}}}g_{\mathbf{q}\nu}\ket{\phi_{n\mathbf{R}}}\nonumber \\
        = \sum_{n'm'\mathbf{k}}e^{i\mathbf{q}\cdot\mathbf{R}}w^\dag_{mn'}(\mathbf{k+q})g_{n'm'\nu}(\mathbf{k},\mathbf{q})w_{m'n}(\mathbf{k}),
\end{align}
where $w_{mn}$ are Wannier rotation matrices,
and $g_{nm\nu}(\mathbf{k},\mathbf{q})$ the
electron-phonon matrix element in the Bloch
basis. As outlined in Appendix\,\ref{lang_firsov_derivation}, in principle one should also account for non-local electron-phonon coupling, captured through matrix elements of the form $g_{mn\mathbf{q}\nu}(\mathbf{R},\mathbf{R'}) = \bra{\phi_{m\mathbf{R}}}g_{\mathbf{q}\nu}\ket{\phi_{n\mathbf{R'}}}$. These terms will couple electron hopping to
phonons, and correspond to a Su-Schrieffer-Heeger mechanism~\cite{PhysRevLett.42.1698,RevModPhys.60.781}. Given the localized nature of Wannier functions, integrals of the form $\bra{\phi_{m\mathbf{R}}}g_{\mathbf{q}\nu}\ket{\phi_{n\mathbf{R'}}}$ will be small compared to the local couplings $\bra{\phi_{m\mathbf{R}}}g_{\mathbf{q}\nu}\ket{\phi_{n\mathbf{R}}}$, and we will ignore non-local coupling for the
remainder of this manuscript. However, incorporating
these terms from first-principles calculations is an
interesting future avenue to pursue within downfolding aproaches. 

By taking the static limit $\omega=0$ of Eq.\,\eqref{eq:Uph}, and by considering intra-band terms ($i=j$) we obtain the simplified expression
\begin{equation}
    \label{eq:Uph_st}
    U^{ph,st}_{i\mathbf{R}i\mathbf{R}'}=-2\sum_{\mathbf{q}\nu}\frac{g_{ii\mathbf{q\nu}}(\mathbf{0})g^*_{ii\mathbf{q}\nu}(\mathbf{R}'-\mathbf{R})}{\omega_{\mathbf{q},\nu}}.
\end{equation}
The superscript ``$st$" will henceforth denote that
we take the static limit $\omega=0$. If additionally we consider on-site terms $\mathbf{R}=\mathbf{R}'$, we get 
\begin{equation}
    \label{eq:Uph_st_on_site}
    U^{ph,st}_{i\mathbf{R}i\mathbf{R}}=-2\sum_{\mathbf{q}\nu}\frac{|g_{ii\mathbf{q\nu}}(\mathbf{0})|^2}{\omega_{\mathbf{q},\nu}}.
\end{equation}
From Eq.\,\eqref{eq:Uph_st_on_site} it is clear that in the static limit phonons contribute
an attractive term to the on-site Coulomb interaction of electrons.

The effect of phonon screening on on-site Coulomb interactions, similar to Eq.\,\eqref{eq:Uph_st_on_site}, has
been computed previously for metals and non-polar semiconductors, where only short-range electron-phonon coupling is present~\cite{Arita2015,PhysRevB.108.L220508}.
However, such an approach
breaks down for polar materials, where long-range electron-phonon coupling emerges. In these cases, the matrix element is typically written as a
sum of a short-range and a long-range term~\cite{Giustino2007,Verdi2015}:
\begin{equation}
    g_{ij\mathbf{q}\nu} = g^{\mathcal{S}}_{ij\mathbf{q}\nu}+g^{\mathcal{L}}_{ij\mathbf{q}\nu}.
\end{equation}
The reason for separating these two contributions is that the long-range part $g^{\mathcal{L}}$ scales as $1/|\mathbf{q}|$, leading to a divergence at $\mathbf{q}\rightarrow \mathbf{0}$. It is therefore subtracted from
the total electron-phonon coupling computed within density functional perturbation theory (DFPT) on a coarse grid, in order to perform
a Wannier-Fourier interpolation of the well-behaved short-range part onto a fine grid. The long-range contribution is then added post fact, using an analytical expression~\cite{Giustino2007}. 

Here we follow a similar approach, 
where the short-range and long-range electron-phonon coupling are treated
separately.
The long-range coupling $g^{\mathcal{L}}_{ij\mathbf{q}\nu}$ is computed using the \emph{ab initio} expression~\cite{Verdi2015}:
\begin{align}
\label{eq:generalized_Frohlich}
    g^{\mathcal{L}}_{ij\mathbf{q}\nu}(\mathbf{R})=i\frac{4\pi}{V}\sum_{\kappa}(\frac{1}{2NM_{\kappa}\omega_{\mathbf{q},\nu}})^{1/2}\frac{\mathbf{q}\cdot \mathbf{Z}_{\kappa}\cdot \mathbf{e}_{\kappa \nu}(\mathbf{q})}{\mathbf{q}\cdot \mathbf{\epsilon}_{\infty}\cdot \mathbf{q}} \nonumber \\ \times \bra{\phi_{i\mathbf{R}}(\mathbf{r})}e^{i\mathbf{q}\cdot \mathbf{r}}\ket{\phi_{j\mathbf{R}}(\mathbf{r})},
\end{align}
which is the first-principles generalization of the Fr\"ohlich model. Here $\mathbf{q}$ is the phonon momentum, with the wave vector belonging to a regular grid of $N$ points within the first Brillouin zone. For an atom $\kappa$, $M_{\kappa}$ and $\mathbf{Z}_{\kappa}$ are its
mass and Born effective charge tensor respectively, $\mathbf{e}_{\kappa \nu}$ a vibrational eigenvector, and
$\mathbf{\epsilon}_{\infty}$ the dielectric matrix. In writing expression Eq.\,\eqref{eq:generalized_Frohlich}, 
we have made the approximation $\mathbf{q}+\mathbf{G}\rightarrow \mathbf{q}$, where $\mathbf{G}$ is a reciprocal lattice vector, and also $e^{i\mathbf{q}\cdot \mathbf{\tau}_{\kappa}}\rightarrow 1$, 
where $\mathbf{\tau}_{\kappa}$ the atomic positions. These approximations become accurate in the long-wavelength $\mathbf{q}\rightarrow \mathbf{0}$ limit, which dominates 
the Fr\"{o}hlich coupling. Moreover, by not considering the contribution of finite $\mathbf{G}$ vectors, we are ignoring local field effects and considering the macroscopic high-frequency dielectric
response as defined by $\mathbf{\epsilon}_{\infty}$. It is
also worth noting that Eq.\,\eqref{eq:generalized_Frohlich}
only allows for longitudinal phonons to couple to the
electrons. Transverse phonons have a polarization that is
perpendicular to the phonon wavevector $\mathbf{q}$, and in
principle including the contribution of reciprocal lattice
vectors $\mathbf{q}+\mathbf{G}$ could result in a finite
contribution by these lattice motions. Nevertheless here we focus on the effect of longitudinal phonons, which is known to generally dominate
the long-range coupling to electrons by far~\cite{Verdi2015}.

We now insert the electron-phonon vertex of Eq.\,\eqref{eq:generalized_Frohlich} into Eq.\,\eqref{eq:Uph} in order to derive the contribution
of long-range coupling to the phonon-mediated Hubbard term, denoted as  
$U^{ph,\mathcal{L}}_{i\mathbf{R}j\mathbf{R}'}(\omega)$. As detailed in Appendix\,\ref{derivation_long_range}, this gives us the long-range contribution
to the phonon-mediated Coulomb interaction:

\begin{align}
\label{eq:Uph_long_range_full}
U_{i\mathbf{R}j\mathbf{R}'}^{ph,\mathcal{L}}(\omega)
=
\left(\frac{4\pi}{V}\right)^2
\sum_{\mathbf qm}\frac{1}{2}(\frac{1}{\omega-\omega_{\mathbf qm}}-\frac{1}{\omega+\omega_{\mathbf qm}})
\nonumber \\ \times\frac{1}{N\omega_{\mathbf qm}}\cdot
\frac{e^{i\mathbf q\cdot(\mathbf R-\mathbf R')}}{|\mathbf q|^2}
\cdot
\frac{\left|q_\alpha \mathcal Z_{m,\alpha}\right|^2}
{\left(q_\alpha \epsilon^\infty_{\alpha\beta} q_\beta\right)^2}.
\end{align}
where we have used the fact that Wannier functions are highly localized around their centers, the index $m$ runs over LO phonons, and the frequency $\omega^2_{\mathbf{q}m}$ refers to the LO phonon.
Here we have defined the mode effective charge:
\begin{equation}
\mathbf q\cdot \boldsymbol{\mathcal Z}_\nu
=
\sum_{\kappa}\frac{1}{\sqrt{M_\kappa}}\,
\mathbf q\cdot \mathbf Z_\kappa\cdot \mathbf e_{\kappa\nu}(\mathbf{q}\rightarrow\mathbf{0}).
\end{equation}

The expression of Eq.\,\eqref{eq:Uph_long_range_full} is amenable to first-principles computation, and its static limit is consistent with  the ion part of the total dielectric function that has been derived by Dolgov, Kirzhnits and Maksymov ~\cite{RevModPhys.53.81}. 
This phonon-mediated Coulomb interaction due to the Fr\"ohlich mechanism does not overcome electron-electron repulsion in the static limit $\omega=0$~\cite{RevModPhys.53.81,ruhman2025commentarxiv250117230arxiv250200103phononmediated}. Negative values of the total static dielectric function are in principle permissible and consistent with a stable dielectric at $\mathbf{q}\neq \mathbf{0}$~\cite{RevModPhys.53.81}, but the Fr\"ohlich interaction alone does not give rise to such an effect.

The expression of Eq.\,\eqref{eq:Uph_long_range_full} may be simplified significantly for isotropic, or mildly anisotropic systems, as we demonstrate in Appendix~\ref{derivation_long_range}. Here we give the final expressions in the static limit, which can be generalized to finite frequencies in a straightforward manner. In every case, the long-range phonon-mediated electron-electron interaction assumes the general form: 
\begin{equation}
\label{eq:ph_mediated_interaction_general}
U_{\mathbf{R}\mathbf{R}'}^{ph,\mathcal{L}}(\omega=0)
=
-\frac{\gamma^2}{\epsilon|\mathbf{R-R'}|},
\end{equation}
which is always an attractive interaction.
For isotropic systems, the dielectric constant $\epsilon$ is simply equal to $\epsilon_{\infty}$ and 
\begin{equation}
\label{eq:gamma_general_isotropic}
\gamma^2
=
\left(\frac{4\pi}{V}\right)
\sum_{m}
\frac{|\boldsymbol{\mathcal Z}_m|^2}
{\epsilon_\infty\,\omega_{LO,m}^2}.
\end{equation}
For systems with a mild anisotropy we instead show in Appendix\,\ref{derivation_long_range} that $\epsilon=\epsilon^{\rm eff}_{\infty}$, with:
\begin{equation}
\label{eq:eps_eff}
\frac{1}{\epsilon^{\rm eff}_{\infty}}
\equiv
\int\frac{d\Omega}{4\pi}\,
\frac{1}{q_\alpha \epsilon^\infty_{\alpha\beta} q_\beta},
\end{equation}
and
\begin{equation}
\label{eq:gamma_general_anisotropic}
\gamma^2
=
\epsilon^{\rm eff}_{\infty}\left(\frac{4\pi}{V}\right)
\sum_{m}\frac{1}{\omega_{LO,m}^2}
\int\frac{d\Omega}{4\pi}\,
\frac{\left|q_\alpha \mathcal Z_{m,\alpha}\right|^2}
{\left(q_\alpha \epsilon^\infty_{\alpha\beta} q_\beta\right)^2},
\end{equation}
where here we take an orientational average over the unit sphere. 

Another interesting limit for the long-range phonon-mediated interaction is obtained for systems where the phonon eigenvectors at $\mathbf{q\rightarrow 0}$ are equal to those at $\mathbf{q=0}$, despite a jump in the frequency value, which is guaranteed by symmetry in some cases. In those cases, the following condition holds~\cite{PhysRevB.55.10355}:
\begin{align}
\label{eq:Born_LO_TO}
\left(\frac{4\pi}{V}\right)
\frac{|q_\alpha \mathcal Z_{\nu,\alpha}|^2}
{q_\alpha \epsilon^{\infty}_{\alpha\beta} q_\beta}
=
\omega^{2}_{LO,\nu}-\omega^{2}_{TO,\nu},
\end{align}
which for the isotropic case leads to the following simple expression for the value of $\gamma^2$:
\begin{equation}
\label{eq:gamma_simple}
\gamma^2
=
\sum_m
\frac{\omega^{2}_{LO,m}-\omega^{2}_{TO,m}}{\omega^{2}_{LO,m}},
\end{equation}
and to the long-range phonon-mediated interaction:
\begin{equation}
\label{eq:ph_mediated_interaction_simple}
U_{i\mathbf{R}j\mathbf{R}'}^{ph,\mathcal{L}}(\omega=0)
=
-\sum_{m}
\left(
\frac{\omega^{2}_{LO,m}-\omega^{2}_{TO,m}}{\omega^{2}_{LO,m}}
\right)
\frac{1}{\epsilon_{\infty}|\mathbf{R-R'}|}.
\end{equation}
We emphasize here that using this simpler expression of Eq.\,\eqref{eq:gamma_simple} requires one to check the validity of the condition $\mathbf{e}(\mathbf{q}\rightarrow \mathbf{0})=\mathbf{e}(\mathbf{q}= \mathbf{0})$ on a material-by-material basis. 
While we will not use Eq.\,\eqref{eq:gamma_simple} to obtain our computational results for the long-range interaction in the main part of our article, we discuss its validity in Appendix~\ref{gamma_LO_TO} for MgO and GeTe, where it is demonstrated that it holds to good accuracy for these systems.

Moreover, let us consider the long-range phonon-mediated interaction under the assumptions
in which the \emph{ab initio} long-range vertex of Eq.\,\eqref{eq:generalized_Frohlich} reduces to the standard Fr\"ohlich electron-phonon matrix element~\cite{doi:10.1080/14786445008521794}:
\begin{align}
    \label{eq:Frohlich}
    g^{Fr}_{ij\mathbf{q}}(\mathbf{R})=\frac{i}{|\mathbf{q}|}\sqrt{\frac{4\pi}{NV}\frac{\omega_{LO}}{2}(\frac{1}{\epsilon_{\infty}}-\frac{1}{\epsilon_o})}\nonumber \\ \times \bra{\phi_{i\mathbf{R}}(\mathbf{r})}  e^{i\mathbf{q}\cdot \mathbf{r}}\ket{\phi_{j\mathbf{R}}(\mathbf{r})},
\end{align}
where $\epsilon_o$ the static dielectric constant.
The assumptions that reduce Eq.\,\eqref{eq:generalized_Frohlich} to Eq.\,\eqref{eq:Frohlich} are that of~\cite{PhysRevB.105.115414} taking the $\mathbf{q}\rightarrow 0$ limit for all smoothly-varying quantities, having a band structure identical to that of the electron gas model, the permittivity being isotropic, and having coupling between electrons and single Einstein LO phonon (hence we have dropped the phonon index $\nu$ in Eq.\,\eqref{eq:Frohlich}).

Using the Fr\"ohlich vertex of Eq.\,\eqref{eq:Frohlich} in Eq.\,\eqref{eq:Uph} yields
\begin{align}
    \label{eq:Uph_long_range_Frohlich}
    U^{ph,Fr,\mathcal{L}}_{i\mathbf{R}j\mathbf{R}'}(\omega)=\frac{\omega_{LO}}{2}\cdot (\frac{1}{\omega-\omega_{LO}}-\frac{1}{\omega+\omega_{LO}})\nonumber \\ \times(\frac{1}{\epsilon_{\infty}}-\frac{1}{\epsilon_o})\cdot \frac{1}{|\mathbf{R}-\mathbf{R}'|},
\end{align}
which in the static limit gives
the simple form
\begin{align}
    \label{eq:Uph_long_range_Frohlich_static}
    U^{ph,st,Fr,\mathcal{L}}_{i\mathbf{R}j\mathbf{R}'}=- (\frac{1}{\epsilon_{\infty}}-\frac{1}{\epsilon_o})\cdot \frac{1}{|\mathbf{R}-\mathbf{R}'|},
\end{align}
which is an attractive screened electron-electron interaction, with both the high- and the low-frequency dielectric constants determining the amount of screening. 
For $\epsilon_o \rightarrow \infty$, the phonon-mediated attraction approaches the value $-1/(\epsilon_{\infty}|\mathbf{R}-\mathbf{R}'|)$, perfectly canceling the screened electron-electron repulsion, but unable to result in net attraction, consistent with the single-phonon limit of a dielectric discussed in Ref.~\cite{RevModPhys.53.81}. 

\section{Connection to a Lang-Firsov transformation picture}
\label{lang_firsov}

We now connect our results of
the previous Section\,\ref{phonon_screening}
to those obtained within a popular approach
within polaron physics, which is taking a Lang-Firsov transformation of a coupled electron-phonon Hamiltonian~\cite{MahanGeraldD2013MP}. This will allow us to develop
better intuition of the underlying assumptions
to arrive at the result of Eq.\,\eqref{eq:Uph}. Here we will
present the key steps of the derivation, which is given in greater detail in Appendix\,\ref{lang_firsov_derivation}.

Consider the coupled electron-phonon
Hamiltonian
\begin{align}
\label{eq:Holstein_Hamiltonian}
        H_{pol} = \sum_{\sigma}\sum_{i\mathbf{R},j\mathbf{R'}}t_{i\mathbf{R},j\mathbf{R'}}a^{\sigma\dag}_{i\mathbf{R}}a^{\sigma}_{j\mathbf{R'}}  \nonumber \\+\frac{1}{2}\sum_{\sigma \rho}\sum_{i\mathbf{R},j\mathbf{R'}} U_{i\mathbf{R},j\mathbf{R'}} a^{\sigma\dag}_{i\mathbf{R}}a^{\rho\dag}_{j\mathbf{R'}}a^{\rho}_{j\mathbf{R'}}a^{\sigma}_{i\mathbf{R}}\nonumber \\
        +\sum_{\mathbf{q}\nu}\omega_{\mathbf{q}\nu}b^\dag_{\mathbf{q}\nu}b_{\mathbf{q}\nu} + \sum_{\sigma}\sum_{ij\mathbf{R}\mathbf{q}\nu} g_{ij\mathbf{q}\nu}(\mathbf{R})(b_{\mathbf{q}\nu}+b^\dag_{-\mathbf{q}\nu})a^{\sigma\dag}_{i\mathbf{R}}a^{\sigma}_{j\mathbf{R'}}
\end{align}
where the operators $a_{i\mathbf{R}}^{\sigma \dagger}$ ($a_{i\mathbf{R}}^{\sigma}$) create (destroy) an electron of spin $\sigma$ in band $i$
and on lattice site $\mathbf{R}$, and $b^\dag_{\mathbf{q}\nu}$ ($b_{\mathbf{q}\nu}$) creates (destroys) a single phonon at branch $\nu$ with momentum $\mathbf{q}$. Let us now perform a Lang-Firsov transformation of
the Hamiltonian of Eq.\,\eqref{eq:Holstein_Hamiltonian}. We define: 
\begin{equation}
    \label{eq:transform}
        S = \sum_{\sigma}\sum_{i\mathbf{R}\mathbf{q}\nu}a^{\sigma\dag}_{i\mathbf{R}}a^{\sigma}_{i\mathbf{R}}\frac{g_{ii\mathbf{q}\nu}(\mathbf{R})}{\omega_{\mathbf{q}\nu}}(b^\dag_{\mathbf{q}\nu}-b_{-\mathbf{q}\nu}) \ 
\end{equation}
and perform the polaron transformation $\tilde{H}_{pol}=e^SH_{pol}e^{-S}$ to eliminate the electron-phonon coupling term. Without relying on any approximation, this leads to the renormalized two-body interaction
\begin{equation}
            \tilde{U}_{i\mathbf{R},j\mathbf{R'}} = U_{i\mathbf{R},j\mathbf{R'}}-2\sum_{\mathbf{q}\nu}\frac{g_{ii\mathbf{q}\nu}(\mathbf{R})g^{*}_{jj\mathbf{q}\nu}(\mathbf{R'})}{\omega_{\mathbf{q}\nu}}.
\end{equation}
Considering the Bloch periodicity $\tilde{U}_{i\mathbf{R},j\mathbf{R'}}=\tilde{U}_{i\mathbf{0},j\mathbf{R'}-\mathbf{R}}$, this suggests that 
\begin{equation}
            \tilde{U}_{i\mathbf{R},j\mathbf{R'}}(\omega) = U_{i\mathbf{R},j\mathbf{R'}}(\omega)+U^{ph,st}_{i\mathbf{R},j\mathbf{R'}},
\end{equation}
with $U^{ph,st}_{i\mathbf{R},j\mathbf{R'}}$ the static limit of
the phonon-mediated electron-electron interaction, derived in Eq.\,\eqref{eq:Uph_st}.

Additionally, if we assume the phonon degrees of freedom to be in thermal
equilibrium, and trace them out
within a weak-coupling approximation, 
we can replace the phonon operators with their expectation values, in a mean-field fashion.
Within its regime of validity, this approximation yields an effective electronic Hamiltonian in the presence of phonons at temperature $T$:
\begin{align}
\label{eq:eff_el}
        H^{el}_{eff} = \sum_{\sigma}\sum_{i\mathbf{R},j\mathbf{R'}}\tilde{t}_{i\mathbf{R},j\mathbf{R'}}a^{\sigma\dag}_{i\mathbf{R}}a^{\sigma}_{j\mathbf{R'}}\nonumber \\+ \frac{1}{2}\sum_{\sigma \rho}\sum_{i\mathbf{R},j\mathbf{R'}} \tilde{U}_{i\mathbf{R},j\mathbf{R'}}a^{\sigma\dag}_{i\mathbf{R}}a^{\rho \dag}_{j\mathbf{R'}}a^{\rho}_{j\mathbf{R'}}a^{\sigma}_{i\mathbf{R}},
\end{align}
where we have ignored constant shifts to the electronic energy. 
Here the modulated off-site interaction is
\begin{align}
\label{eq:hopping_renormalization}
\tilde{t}_{i\mathbf{R},j\mathbf{R'}}(T) = t_{i\mathbf{R},j\mathbf{R'}}\nonumber \\ \times \exp[-\sum_{\mathbf{q}\nu}\frac{|g_{jj\mathbf{q}\nu}(\mathbf{R'})-g_{ii\mathbf{q}\nu}(\mathbf{R})|^2}{\omega^2_{\mathbf{q}\nu}}(N_{\mathbf{q}\nu}(T)+\frac{1}{2})],
\end{align}
where $N_{\mathbf{q}\nu}(T)$ the Bose-Einstein distribution.
The on-site energies are renormalized as 
\begin{align}
        \tilde{t}_{i\mathbf{R},i\mathbf{R}} = t_{i\mathbf{R},i\mathbf{R}}-\sum_{\mathbf{q}\nu}\frac{|g_{ii\mathbf{q}\nu}(\mathbf{0})|^2}{\omega_{\mathbf{q}\nu}}.
\end{align}
It should again be emphasized that the above result for the temperature-dependent modulation of the hopping term relies on the weak-coupling approximation, which is violated in systems with strong electron-phonon interactions, such as MgO.

Let us pause to consider what has been achieved at this point. Eq.\,\eqref{eq:eff_el} is a generalized Hubbard model, including only electronic terms, with the hopping and Coulomb integrals renormalized by the magnitude
of the electron-phonon interactions. One can solve this Hamiltonian in exactly the
same manner as the original Hamiltonian Eq.\,\eqref{eq:Hamiltonian}, with the only
difference that the parameters entering it
are renormalized. Moreover, the correction to the two-body
Coulomb term due to phonons is found to be identical to the
static phonon-mediated correction of Eq.\,\eqref{eq:Uph_st}. Therefore, starting from the coupled electron-phonon Hamiltonian of Eq.\,\eqref{eq:Holstein_Hamiltonian},
performing a Lang-Firsov transformation and tracing out phonon degrees of freedom, 
yields
the static limit of including phonon screening in the
Coulomb integral of Eq.\,\eqref{eq:coulomb}. 

It is interesting to note at this point some characteristics of the temperature dependence of the effective electronic Hamiltonian of Eq.\,\eqref{eq:eff_el}.
The phonon-induced renormalization of the hopping terms, as described by Eq.\,\eqref{eq:hopping_renormalization}, is
temperature-dependent, with the hopping integrals reduced with increasing temperature, indicative of a flattening of
the electronic bands, and real-space localization of electronic states, in agreement with previous derivations~\cite{PhysRevB.69.075211,PhysRevResearch.1.013017}.
However, the correction to the Coulomb term describes virtual phonon
processes and hence is temperature-independent. For the
remainder of this paper we will focus on the renormalization of the Coulomb interactions. 

\section{Computational implementation}
\label{computational_approach}

\begin{table*}[tb]
\centering
  \setlength{\tabcolsep}{6pt} 
\begin{tabular}{cccccccc}
\hline
Material & Crystal System & Structure & $a$ ($\Ang$) & $\alpha$ & $V$ ($\Ang^3$)  &Space Group & Identifier \\
\hline
MgO & Cubic & Halite, Rock Salt & $2.97$ & $60.0$\textdegree & $26.198$ & $\text{F}\overline{m}3\text{m}$ & mp-1265 \\
GeTe & Trigonal & Halite, Rock Salt & $4.19$ & $59.08$\textdegree & $50.93$ & $\text{R}\overline{3}\text{m}$  & mp-938 \\
\hline
\end{tabular}
\caption{Studied materials, their structure, primitive lattice constants, angles, and volumes, space group, and identifier in the Materials Project database~\cite{Jain2013}. For GeTe, we performed geometry optimization for the atomic positions and lattice parameters, using DFT within the LDA, which has been 
discussed to yield accurate structural properties for
this system~\cite{PhysRevB.78.205203}.}
\label{table:structures}
\end{table*}

We now return to Eq.\,\eqref{eq:Uph} for the general
frequency-dependent phonon correction to the Coulomb
interactions, and its computational implementation. Moreover, below we discuss the practical calculation of  the long-range part
of the electron-phonon interaction, which is treated separately 
in order to allow for the Wannier-Fourier interpolation
of the short-range coupling $g^{\mathcal{S}}$.

A notable characteristic of Eq.\,\eqref{eq:Uph} is
that the electron-phonon matrix elements appear in a mixed Wannier-Bloch basis, 
for the electronic and phonon degrees of freedom respectively. 
Within our formalism, the lattice vectors $\mathbf{R}$
indicate the centers of Wannier functions, with $\mathbf{R}=\mathbf{R}'$ representing an on-site term
and $\mathbf{R}\neq \mathbf{R}'$ off-site terms, the
range of which is determined by the distance of the Wannier centers. The effect of phonons on each of the
lattice terms $U^{ph}_{i\mathbf{R}j\mathbf{R}'}(\omega)$ is
determined by summing over the phonon momentum grid $\mathbf{q}$, and the sum needs to be converged by using
a fine mesh. However, unlike most studies that evaluate
electron-phonon self-energies from first principles~\cite{Alvertis2024,doi:10.1021/acs.nanolett.4c01508,PhysRevB.101.165102}, here we do not need to also evaluate the electronic
degrees of freedom on a dense grid. There is therefore
only a need to perform Wannier-Fourier interpolation
for $\mathbf{q}$-vectors (and not $\mathbf{k}$-vectors).
As discussed previously, we only perform Wannier-Fourier
interpolation for the short-range electron-phonon matrix
element $g^{\mathcal{S}}$, in order to avoid the singular
behavior of the long-range term, which we add analytically, directly on a fine $\mathbf{q}$-grid. 

Similarly to using constrained RPA
rather than regular RPA to compute the polarizability of Eq.\,\eqref{eq:chi_crpa}, as a means of excluding the
screening of the states of the active space within which
we derive the Coulomb interactions, here we employ constrained density functional perturbation theory (cDFPT)~\cite{Nomura2015} rather than usual DFPT, to obtain the short-range electron-phonon matrix elements $g^{\mathcal{S}}$ entering
the phonon-renormalized Coulomb terms of Eq.\,\eqref{eq:Uph}. 
We use the modified version of Quantum Espresso~\cite{QE} of Ref.~\cite{PhysRevX.13.041009} to perform cDFPT calculations on a coarse $\mathbf{q}$-grid, and we determine the Wannier representation of our systems using
Wannier90~\cite{Pizzi2020}. Within the cDFPT calculations, we obtain phonon frequencies and electron-phonon matrix elements of states
within an active space which is defined on a case-by-case basis for studied materials. We perform Wannier-Fourier interpolation
of short-range electron-phonon coupling matrix elements
on a fine $\mathbf{q}$-grid within the EPW code~\cite{Lee2023}, and
we extract $g^{\mathcal{S}}_{ij\mathbf{q}\nu}(\mathbf{R})$ using a modified version of the code. Therefore, we obtain converged values for the short-range contribution to the phonon renormalization of the Hubbard
term $U^{ph,\mathcal{S}}_{i\mathbf{R}j\mathbf{R}'}(\omega)=\sum_{\mathbf{q},\nu}g^{\mathcal{S}*}_{ij\mathbf{q}\nu}(\mathbf{0})g^{\mathcal{S}}_{ij\mathbf{q}\nu}(\mathbf{R}'-\mathbf{R})
    [\frac{1}{\omega-\omega_{\mathbf{q},\nu}+i\delta}-\frac{1}{\omega+\omega_{\mathbf{q},\nu}-i\delta}]$.
We obtain the electronic Coulomb interactions within the same active space using cRPA, and specifically Eq.\,\eqref{eq:coulomb} and Eq.\,\eqref{eq:chi_crpa} as implemented in RESPACK~\cite{Nakamura2021} in each case. 
We use Eq.\,\eqref{eq:ph_mediated_interaction_general} to compute
the long-range contribution $U^{ph,\mathcal{L}}_{i\mathbf{R}j\mathbf{R}'}(\omega)$, in its isotropic or anisotropic manifestation depending on the studied material, and yielding the overall result 
$U^{ph}_{i\mathbf{R}j\mathbf{R}'}(\omega)=U^{ph,\mathcal{S}}_{i\mathbf{R}j\mathbf{R}'}(\omega)+U^{ph,\mathcal{L}}_{i\mathbf{R}j\mathbf{R}'}(\omega)$. 

\section{Studied structures and computational details}
\label{computational_details}

\begin{figure}[tb]
    \centering
    \includegraphics[width=\linewidth]{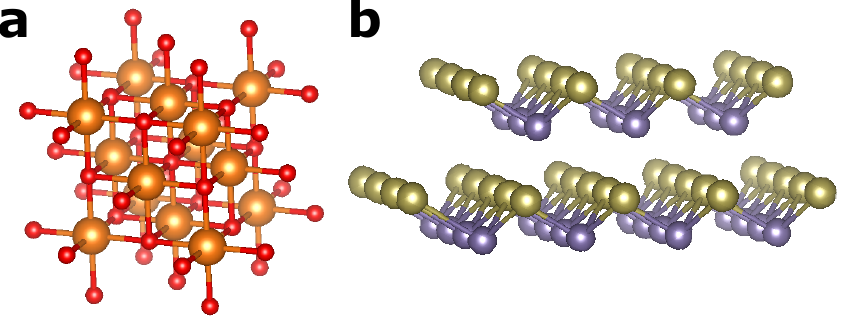}
    \caption{Structures of MgO (panel \textbf{a}) and GeTe (panel \textbf{b}). Mg atoms are shown in orange, O in red, Ge in purple, and Te in gold. The structures are visualized in VESTA~\cite{Momma:db5098}. }
    \label{fig:structures}
\end{figure}

Below we outline the details of our calculations for MgO
and GeTe. We have chosen MgO since it is a prototypical bulk insulator with strong long-range electron-phonon coupling, hence providing
an excellent material to demonstrate the impact
of our formalism and of polar contributions to its electron-electron interactions. Moreover, short-range electron-phonon interactions are known to also modify the electronic structure
by a non-negligible amount~\cite{PhysRevB.93.100301}, making the
interplay of these effects at different length scales an interesting one to explore. Semiconducting GeTe on the other hand is known
to exhibit superconductivity and strong electron-phonon interactions~\cite{PhysRev.177.704,PhysRevLett.124.047002}, we were therefore interested in understanding whether our framework could provide
an explanation for the formation of Cooper pairs in this material. 
The structural details of MgO and GeTe are given in Table\,\ref{table:structures},
and the studied structures
are also visualized in Fig.\,\ref{fig:structures}. The structures were taken
from the Materials Project database~\cite{Jain2013}. 
MgO forms a cubic rocksalt structure
GeTe is studied in its low-temperature trigonal phase, which consists of GeTe bilayers along the $c$ axis. 

We employ Quantum Espresso~\cite{QE} for all DFT and DFPT calculations, Wannier90~\cite{Pizzi2020} in order to generate the Wannier
representation of the active space of interest, as discussed in Section\,\ref{results}, and EPW~\cite{Lee2023} in
order to perform Wannier-Fourier interpolation of electron-phonon matrix elements. 
We employ scalar-relativistic optimized norm-conserving Vanderbilt pseudopotentials (ONCV)~\cite{Hamann2013} with standard accuracy, taken from Pseudo Dojo~\cite{VanSetten2018}. Specifically, these include three valence electrons for Mg, two valence electrons for O, three valence electrons for Ge, and three valence electrons for Te.

For MgO we work within the generalized gradient approximation (GGA) of DFT as formulated by Perdew, Burke and Ernzerhof (PBE)~\cite{pbe}. We use a plane
wave cutoff of $80$\,Ry. As a starting point for 
downfolding calculations of the electronic structure, we
compute the electronic density on a $6\times 6\times 6$ $\Gamma$-centered $\mathbf{k}$-grid, and we include $300$
bands in the cRPA calculations, using a cutoff of $7$\,Ry, and having excluded the three
highest-lying valence bands as discussed in Section\,\ref{results}. For cDFPT calculations we use an electronic density computed on a $6\times 6\times 6$ half-shifted $\mathbf{k}$-grid, and we obtain the
phonons on a $6\times 6\times 6$ grid of $\mathbf{q}$-points. We interpolate the phonon frequencies and electron-phonon matrix elements on a $20\times 20\times 20$ $\mathbf{q}$-grid, which we find is sufficient to converge
the values of the short-range phonon-mediated contribution
to the Coulomb interactions within the valence bands, \emph{i.e.}, $U^{ph,\mathcal{S}}_{i\mathbf{R}j\mathbf{R}'}(\omega)$. When it comes to computing the long-range phonon-mediated Coulomb term $U^{ph,\mathcal{L}}_{i\mathbf{R}j\mathbf{R}'}(\omega)$, we use the isotropic limit $-\gamma^2/(\epsilon_{\infty}r)$, with $\gamma$ given by eq.\,\eqref{eq:gamma_general_isotropic}, given the isotropy of the crystal and its dielectric properties.

For
GeTe we work within the local density approximation (LDA), which
is known to lead to accurate phonon and dielectric properties for this system~\cite{PhysRevB.78.205203}.
We optimize the atomic positions and lattice parameter
of the system prior to any other calculations, using a plane
wave cutoff of $100$\,Ry, on a $12\times 12\times 12$
half-shifted $\mathbf{k}$-grid. The resulting structural parameters of Table\,\ref{table:structures} are in close
agreement to previously reported theoretical values~\cite{PhysRevB.78.205203}. Beyond the geometry optimization, we use a plane
wave cutoff of $100$\,Ry across all DFT calculations on this system. As a starting point for downfolding calculations of the electronic structure, we
compute the electronic density on a $6\times 6\times 6$ $\Gamma$-centered $\mathbf{k}$-grid, and we include $700$
bands in the cRPA calculations, using a cutoff of $7$\,Ry, and having excluded the three
highest-lying valence bands as discussed in Section\,\ref{results}. For cDFPT calculations we use an electronic density computed on a $12\times 12\times 12$ half-shifted $\mathbf{k}$-grid, and we obtain the
phonons on a $12\times 12\times 12$ grid of $\mathbf{q}$-points. We compute the short-range term $U^{ph,\mathcal{S}}_{i\mathbf{R}j\mathbf{R}'}(\omega)$ on the same $12\times 12\times 12$ grid, we therefore do not need to perform any interpolation of phonon frequencies and 
electron-phonon matrix elements in this case. Given the layered character of this material, we utilize the anisotropic definition of Eq.\,\eqref{eq:gamma_general_anisotropic} for $\gamma$ and obtain  $U^{ph,\mathcal{L}}_{i\mathbf{R}j\mathbf{R}'}=-\gamma^2/(\epsilon_{\rm eff}r)$.

\section{Computational results}
\label{results}
Below we outline our computational results for \emph{ab initio} downfolding of the electronic structure
of MgO and GeTe, and we include the effects
of phonon screening. 

\subsection{MgO}

\begin{figure}[tb]
    \centering
    \includegraphics[width=\linewidth]{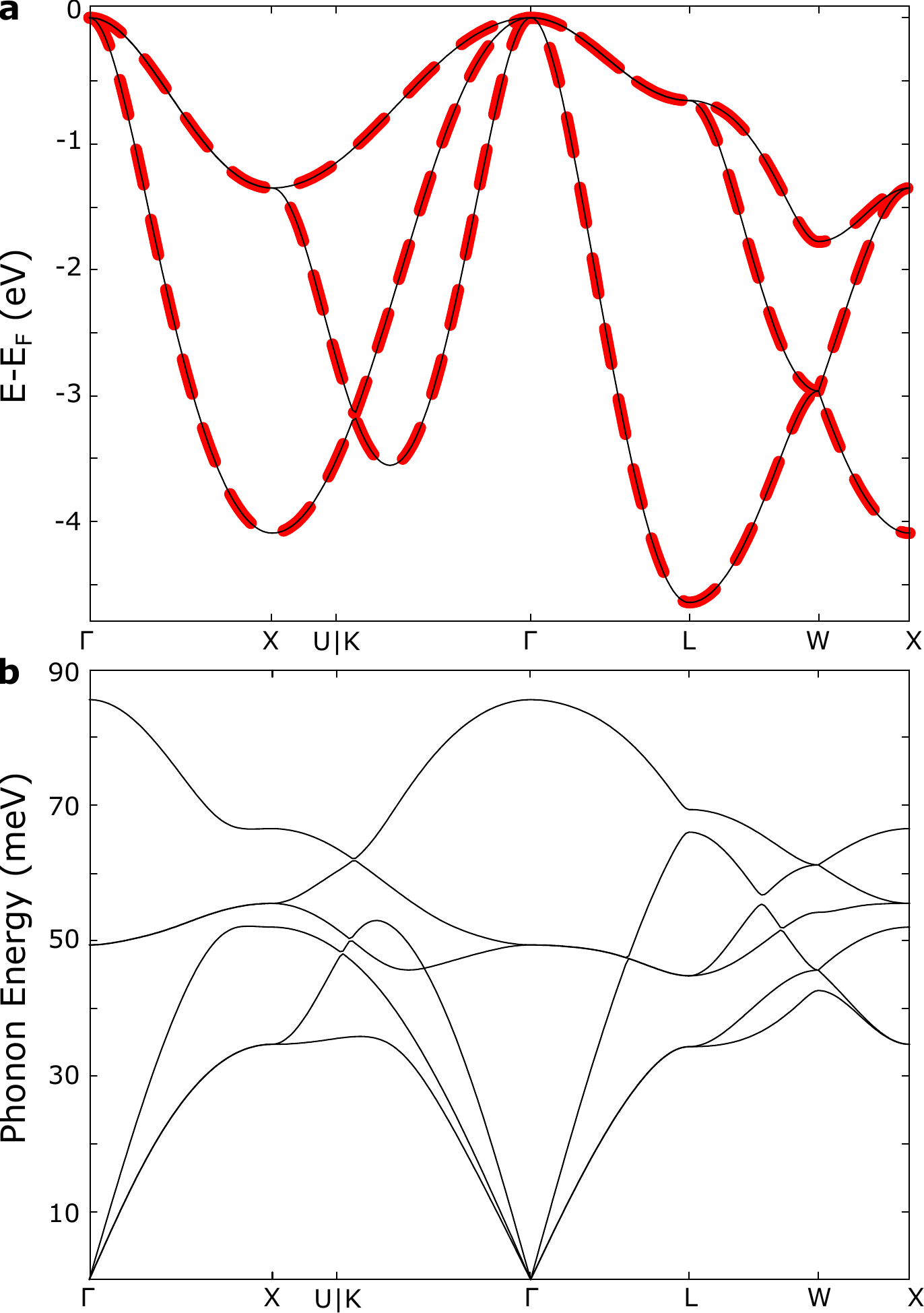}
    \caption{Electronic band structure (panel \textbf{a}) and phonon dispersion (panel \textbf{b}) of MgO computed
    at the DFT-GGA level of theory as discussed in the main text. The red dashed lines in panel \textbf{a} indicate our Wannier-interpolated bands for the three highest valence bands, which constitute our active space for \emph{ab initio} downfolding.}
    \label{fig:MgO_bands}
\end{figure}

We begin our analysis of MgO with a
discussion of its electronic band structure, visualized in Fig.\,\ref{fig:MgO_bands}a. 
We focus on computing the electronic interactions of carriers within the subspace of the three highest-lying
valence bands, for which our Wannier-interpolated band
structure is visualized in red in Fig.\,\ref{fig:MgO_bands}a. These
bands are dominated by oxygen $p$ orbitals, and the Wannier functions
are centered on $O$ atoms, with a distance between neighboring sites that is equal to the primitive lattice constant of  $2.97$\,\AA. 
Below we focus on computing
the phonon-induced renormalization of the Coulomb interactions within this subspace 
using \emph{ab initio} downfolding and our extension of
it to account for phonon screening.
The phonon dispersion of MgO is visualized in Fig.\,\ref{fig:MgO_bands}b. The unit cell of this system
consists of two atoms, and there are therefore six phonons, with a single longitudinal optical (LO) mode
with a $\Gamma$ frequency of $86$\,meV, and two transverse optical (TO) phonons with  a $\Gamma$ frequency of $49$\,meV. We also compute the Born effective charges associated with the Mg and O atoms, reported in Table\,\ref{table:MgO_Born}, which are 
very close to the oxidation numbers of these
elements, indicating the strong ionic character of the bonding. 

\begin{table*}[tb]
\centering
  \setlength{\tabcolsep}{6pt} 
\begin{tabular}{cc}
\hline
atom & $Z^*$  \\
\hline
Mg & $1.989$ \\
O & $-1.985$ \\
\hline
\end{tabular}
\caption{Born effective charges of MgO.}
\label{table:MgO_Born}
\end{table*}

\begin{table*}[tb]
\centering
  \setlength{\tabcolsep}{6pt} 
\begin{tabular}{cccccccc}
\hline
$\epsilon_{\infty}$ & $\epsilon_o$ & $\gamma^2$ & $U^{el}$ (eV) & $V^{el}$ (eV) & $U^{ph,st}$ (eV)  & $V^{ph,st,\mathcal{S}}$ (eV) & $V^{ph,st,\mathcal{L}}$ (eV) \\
\hline
$3.22$ & $9.67$ & $0.660$ & $6.826$ & $1.393$ & $-2.720$ & $-0.011$ & $-1.005$ \\
\hline
\end{tabular}
\caption{Numerical results for MgO: High-frequency and static dielectric constants, $\gamma^2$ following Eq.\,\eqref{eq:gamma_general_isotropic}, static on-site and nearest neighbor Coulomb repulsion, static on-site and nearest-neighbor phonon-induced attraction due to short-range electron-phonon coupling, and static nearest-neighbor phonon-induced attraction due to long-range electron-phonon coupling.}
\label{table:MgO_results}
\end{table*}

In Table\,\ref{table:MgO_results} we give the values for the static intra-band (band-averaged) on-site
Hubbard term $U^{el}=\frac{1}{3}\sum_{i}U^{el}_{i\mathbf{R}i\mathbf{R}}(\omega=0)$ of the MgO valence bands within
the active space, where we have used the expression of Eq.\,\eqref{eq:coulomb}. We also report 
the nearest-neighbor (band-averaged) Coulomb repulsion $V^{el}=\frac{1}{3}\sum_{i}U^{el}_{i\mathbf{R}i\mathbf{R}'}(\omega=0)$, with the lattice vectors $\mathbf{R},\mathbf{R}'$ indicating nearest-neighbor positions in this case. 
Here ``nearest-neighbor'' refers to the
distance between Wannier function centers, which for MgO coincide with the positions of O atoms.
We find a $U^{el}=6.826$\,eV on-site repulsion of the electrons in the valence bands, and a $V^{el}=1.393$\,eV nearest-neighbor repulsion. While here we have focused on the intra-band electron-electron repulsion, the inter-band terms are also found to be of comparable magnitude.
Moreover, we report in Table\,\ref{table:MgO_results} the high-frequency and static
dielectric constants as obtained from DFPT. 

We now consider the phonon-induced
renormalization of the intra-band terms $U^{el},V^{el}$, in the static limit. 
The short-range electron-phonon coupling contribution 
$U^{ph,st},V^{ph,st,\mathcal{S}}$ to the on-site and
nearest-neighbor electron-electron interaction respectively is computed using Eq.\,\eqref{eq:Uph_st}.
As we report in Table\,\ref{table:MgO_results}, we find a renormalization of the on-site term of $U^{ph,st}=-2.720$\,eV, bringing the overall on-site
interaction to $U^{tot}(\omega=0)=U+U^{ph,st}=4.106$\,eV,
\emph{i.e.}, phonons reduce the repulsive interaction by
$39.8$\% compared to the case where their effect is ignored. Short-range electron-phonon coupling also leads to a weak attraction $V^{ph,st,\mathcal{S}}=-11$\,meV between nearest neighbors. Moreover, we find for both the on-site and nearest-neighbor terms, that the correction to the inter-band electron-electron interaction due to short-range
electron-phonon coupling is of comparable magnitude to the intra-band values of these quantities.   

Importantly, we find that long-range electron-phonon coupling induces a large
attraction between electrons in MgO, particularly when these are on neighboring sites (the on-site correction $\mathbf{R}=\mathbf{R}'$ is ill-defined for the long-range
contribution).
Given the isotropic character of MgO, we compute $\gamma^2$ using Eq.\,\eqref{eq:gamma_general_isotropic}, and obtain the phonon-induced attraction in the static limit as $-\gamma^2/(\epsilon_{\infty}r)$. We find an attraction
$V^{ph,st,\mathcal{L}}=-1.005$\,eV between
electrons on neighboring sites in MgO, as  summarized in Table\,\ref{table:MgO_results} together with our computed value for $\gamma^2$. This strong effect is \emph{in addition} to the weak effect of short-range
electron-phonon coupling on nearest neighbor interactions, as incorporated within
the established cDFPT approach~\cite{Arita2015}. As we saw, for MgO this short-range contribution leads to
an attraction of $V^{ph,st,\mathcal{S}}=-11$\,meV between neighboring electrons. It therefore becomes
clear that within downfolded representations of polar materials, long-range electron-phonon coupling can dominate, and including it is imperative
for the derived model Hamiltonians to faithfully represent the low-energy physics.

\begin{figure}[tb]
    \centering
    \includegraphics[width=\linewidth]{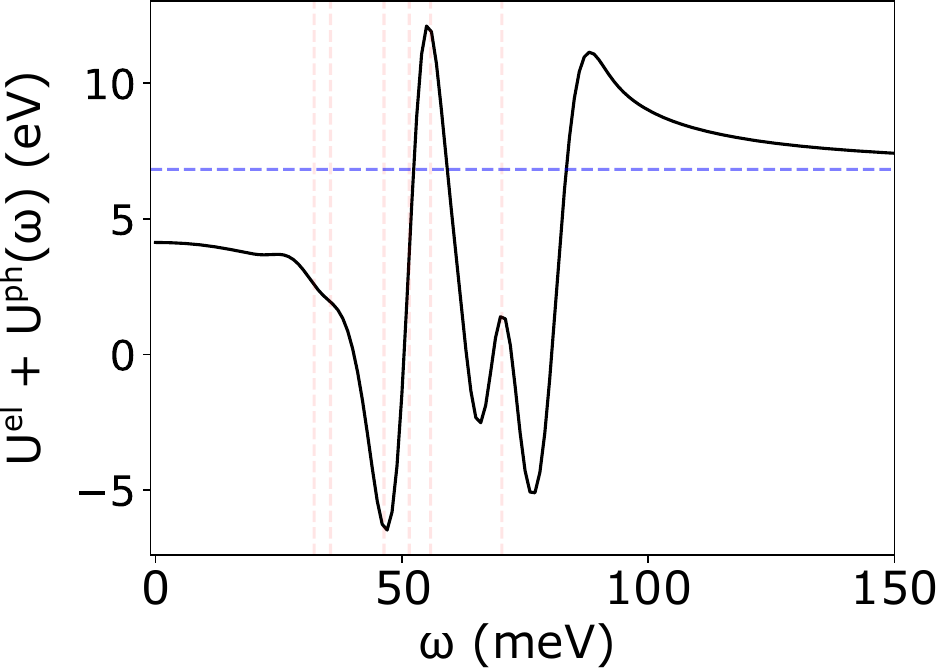}
    \caption{Frequency-dependence of the total on-site Hubbard term for the highest-lying valence band of MgO. The red vertical lines indicate the average phonon frequencies on the $20\times20\times20$ $\mathbf{q}$-grid where $U^{ph}$ is computed, while
    the blue horizontal line marks the value of the electronic Hubbard term of $6.826$\,eV for this band, in the $\omega=0$ limit.}
    \label{fig:MgO_frequency}
\end{figure}

We now turn to the frequency dependence of the correction $U^{ph}(\omega)$ to the on-site Hubbard term, following Eq.\,\eqref{eq:Uph}. In Fig.\,\ref{fig:MgO_frequency} we visualize
$U^{tot}(\omega)=U^{el}(\omega)+U^{ph}(\omega)$ over a range of frequencies that 
covers the entire phonon spectrum of the material. Within
this range, the frequency dependence of the electronic term
is negligible, and oscillations are almost entirely due to the phonon term. To improve visibility, which can be obscured by strong oscillations in the vicinity of poles, we have  applied a Gaussian smoothing
filter to reduce noise and emphasize the overall trend. 
It is evident from Fig.\,\ref{fig:MgO_frequency} that
the Hubbard term exhibits a strong frequency dependence,
particularly around phonon energies (red dashed lines), where it may even
assume negative values depending on how finely we sample 
on the frequency grid. As expected, at frequencies above
the highest phonon energy of the system, the Coulomb term approaches the purely electronic value $U^{el}$ (blue dashed line). The importance of the frequency
dependence for obtaining quantitatively accurate eigenstates of the downfolded Hamiltonian
has been discussed previously in the context
of excited state calculations~\cite{Romanova2023}. 

\begin{figure}[tb]
    \centering
    \includegraphics[width=\linewidth]{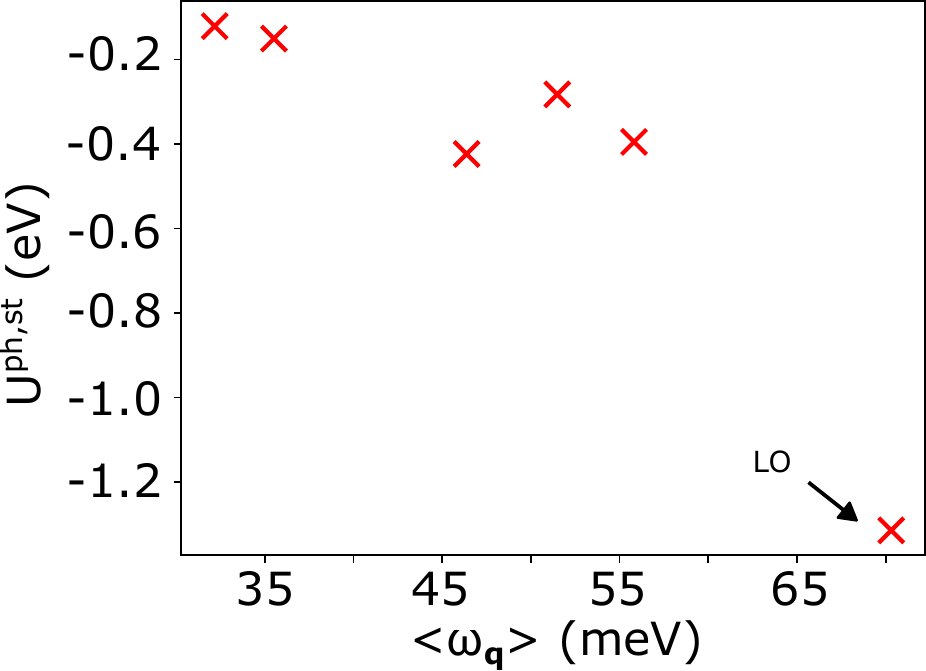}
    \caption{Contributions of individual phonon modes to
    the renormalization of the on-site Hubbard term of MgO.}
    \label{fig:MgO_modes}
\end{figure}

We now analyze how different phonons contribute
to the overall renormalization of the electronic interactions. We visualize in Fig.\,\ref{fig:MgO_modes} 
the phonon-resolved values of the on-site correction $U^{ph,st}$. We see that the most important contribution
arises from the LO phonon, but there are also significant contributions from the TO and
acoustic phonons of MgO. Unlike the long-range
electron-phonon interaction, which is dominated by the
Fr\"ohlich mechanism where exclusively LO phonons 
contribute, short-range electron-phonon coupling may be
significant also for transverse optical and acoustic
phonons, as we observe in Fig.\,\ref{fig:MgO_modes}. There have been previous reports in the literature for the magnitude of the short-range electron-phonon interaction in materials such as $\text{TiO}_2$~\cite{Verdi2015}, where it was found to be
significant, although substantially weaker than the long-range coupling. Indeed, for off-site terms $V$, the long-range
phonon-mediated electronic interaction dominates over the short-range one as seen in Table\,\ref{table:MgO_results}. Nevertheless, short-range electron-phonon coupling results in a significant renormalization
of the on-site electronic repulsion. 

\begin{figure}[tb]
    \centering
    \includegraphics[width=\linewidth]{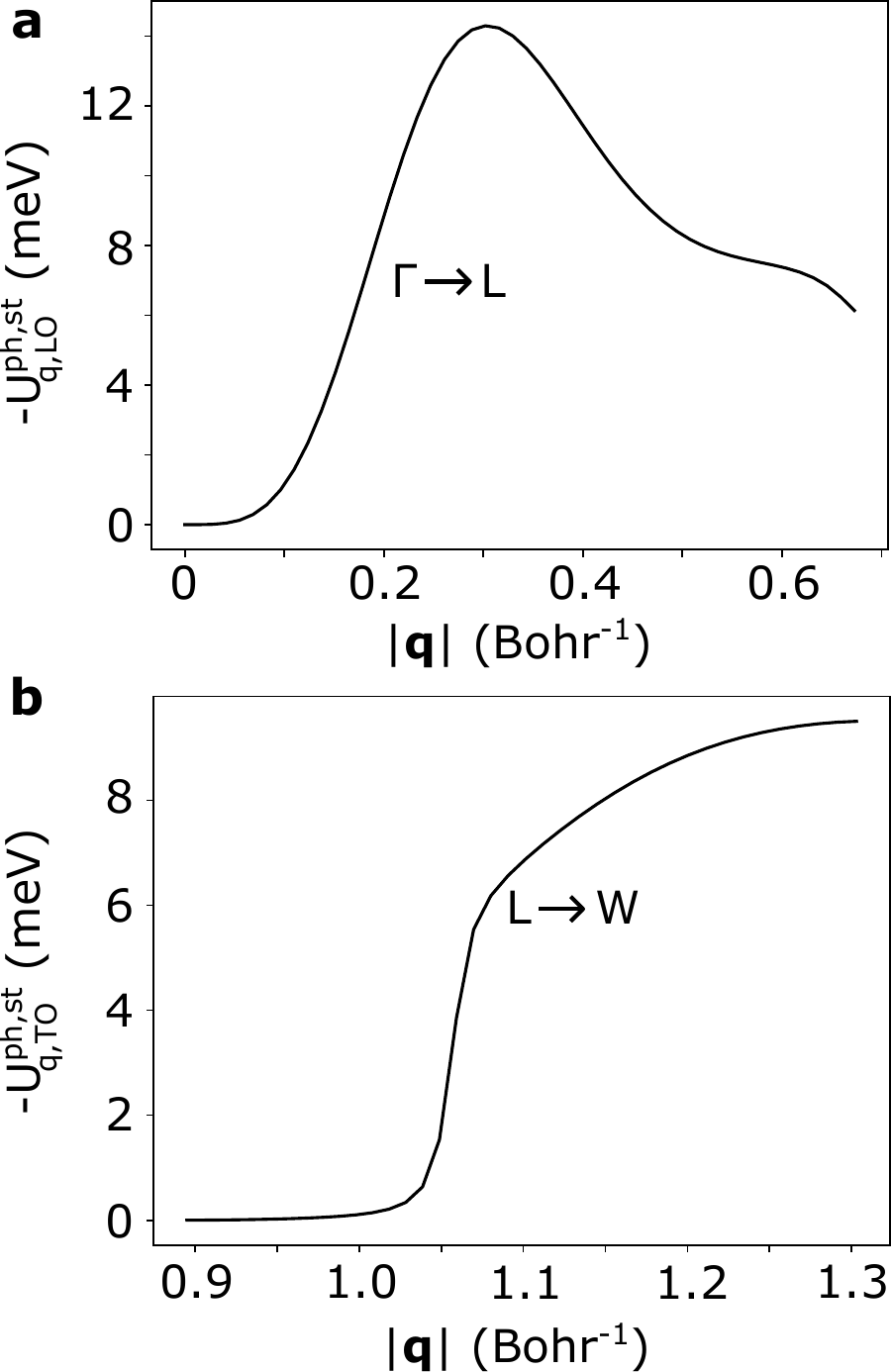}
    \caption{Dependence of the short-range phonon-induced
    correction of the on-site Coulomb interaction on the phonon momentum, in the static limit. This is visualized for the LO phonon of MgO, along the $\Gamma \rightarrow L$ path in reciprocal space (panel $\textbf{a}$), and for a TO phonon along the $L \rightarrow W$ path (panel $\textbf{b}$).}
    \label{fig:MgO_q}
\end{figure}

At this point, it is worth emphasizing some of the physical
characteristics of the contribution of the short-range electron-phonon interaction to the renormalization of the
Coulomb interactions within the valence bands. In Fig.\,\ref{fig:MgO_q} we visualize the on-site phonon-mediated Coulomb interaction $U^{ph,st}$ in the static
limit, due to the LO phonon (panel $\textbf{a}$) and due to
one of the TO phonons (panel $\textbf{b}$), as a function of 
the phonon momentum $|\mathbf{q}|$. Here we have chosen two paths along the Brillouin zone where the effect is significant. We see that short-range electron-phonon coupling
can cause an attraction with complex dependence on $|\mathbf{q}|$, unlike the $1/|\mathbf{q}|^2$ dependence
due to long-range Fr\"ohlich coupling. This short-range interaction is due to the so-called
deformation potential mechanism. As discussed in analytic
theories of electron-phonon coupling~\cite{PhysRevB.13.694}, the matrix elements
describing the interaction within this mechanism generally consist of a constant part, as well as octopole and higher-order terms that depend
on $\mathbf{q}$, which vanish as $\mathbf{q} \rightarrow 0$.
The $\mathbf{q}$-independent part of the deformation potential is highly band-dependent and can vanish due to symmetry in several cases, for example for the LO phonons
of direct gap semiconductors, and for coupling within conduction bands~\cite{PhysRevB.85.035201}. Indeed we see that for MgO, the
short-range contribution to the phonon-mediated Coulomb 
term vanishes for $\mathbf{q} \rightarrow 0$ (Fig.\,\ref{fig:MgO_q}a). 

\subsection{GeTe}

\begin{figure}[tb]
    \centering
    \includegraphics[width=\linewidth]{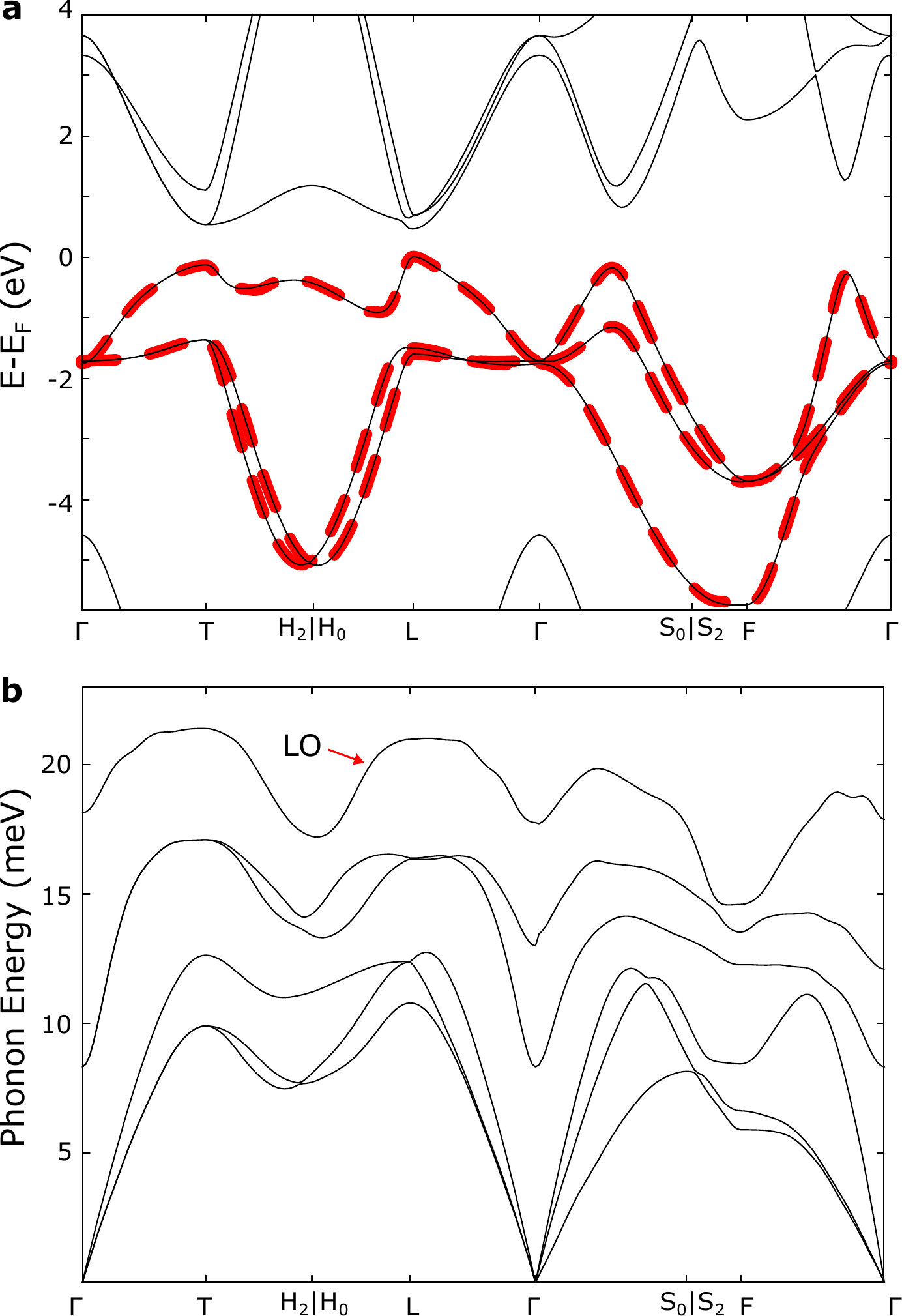}
    \caption{Electronic band structure (panel \textbf{a}) and phonon dispersion (panel \textbf{b}) of GeTe computed
    at the DFT-LDA level of theory as discussed in the main text. The red dashed lines in panel \textbf{a} indicate our Wannier-interpolated bands for the highest three valence bands, which constitute our active space for \emph{ab initio} downfolding.}
    \label{fig:GeTe_bands}
\end{figure}

Next we consider the effect of phonon screening on the electron-electron interactions of semiconducting GeTe. The electronic band structure and phonon dispersion of this system are visualized in Fig.\,\ref{fig:GeTe_bands}. We see that there
exists a manifold of three fairly isolated valence bands near the Fermi level, and these form the active space within
which we compute the electron-electron interactions. Approximately $68$\% of the wave function amplitude for these
bands corresponds to contributions from
Te $p$-orbitals, with the remaining
$32$\% coming from Ge $p$-orbitals. As
a result, the resulting Wannier functions
have strong $p$-orbital character, and
are not perfectly centered on either
atom within the unit cell, having a nearest neighbor distance of $4.19$\,\AA, \emph{i.e.}, equal to the primitive lattice constant. Given the proximity of the conduction bands to the valence band manifold within the DFT-LDA level of theory, we repeated the analysis presented below, this time including the three lowest conduction states in the active space. At least for the phonon-mediated electron-electron interactions within the top valence band, we found that this leads to
only minor differences in the results. 

The trigonal phase of GeTe has two atoms in its primitive
cell, there are therefore six phonons, with a single LO mode, highlighted in Fig.\,\ref{fig:GeTe_bands}b. The layered structure of GeTe makes this system anisotropic, therefore the
precise energy of the LO phonon depends on the direction of $\mathbf{q}$, as $\mathbf{q}\rightarrow \mathbf{0}$. This anisotropy is reflected in the values of
the Born effective charges associated with
the Ge and Te atoms, as reported in Table\,\ref{table:GeTe_Born}, which assume different in-plane and out-of-plane values, and similarly for the high-frequency and static
dielectric constants reported in Table\,\ref{table:GeTe_results}. Given the anisotropy of the system, the long-range phonon-mediated attraction is obtained as $-\gamma^2/(\epsilon_{\rm eff}r)$, where $\epsilon_{\rm eff}$ is the spherical average of the dielectric tensor following Eq.\,\eqref{eq:eps_eff}, and similarly $\gamma^2$ is obtained from Eq.\,\eqref{eq:gamma_general_anisotropic}, where an orientational average over the unit sphere is taken. Table\,\ref{table:GeTe_results} reports the values for $\epsilon_{\rm eff}$ and $\gamma^2$ that we obtain here. 

\begin{table*}[tb]
\centering
  \setlength{\tabcolsep}{6pt} 
\begin{tabular}{ccc}
\hline
atom & $Z^*_{\parallel}$ & $Z^*_{\perp}$  \\
\hline
Ge & $7.139$ & $5.081$  \\
Te & $-7.019$ & $-4.995$  \\
\hline
\end{tabular}
\caption{Born effective charges of GeTe.}
\label{table:GeTe_Born}
\end{table*}

\begin{table*}[tb]
\centering
  \setlength{\tabcolsep}{6pt} 
\begin{tabular}{ccccccccccc}
\hline
$\epsilon^{\parallel}_{\infty}$ & $\epsilon^{\parallel}_o$ & $\epsilon^{\perp}_{\infty}$ & $\epsilon^{\perp}_o$ & $\epsilon^{\rm eff}_{\infty}$ & $\gamma^2$ &$U^{el}$ (eV) & $V^{el}$ (eV) & $U^{ph,st}$ (eV)  & $V^{ph,st,\mathcal{S}}$ (eV) & $V^{ph,st,\mathcal{L}}$ (eV) \\
\hline
$61.90$ & $279.33$ & $51.29$ & $93.71$ & $57.91$ & $0.649$ & $1.102$ & $0.118$ & $-0.873$ & $-0.128$ & $-0.038$ \\
\hline
\end{tabular}
\caption{Numerical results for GeTe: High-frequency and static dielectric constants, $\gamma^2$ following Eq.\,\eqref{eq:gamma_general_anisotropic}, static on-site and nearest neighbor Coulomb repulsion, static on-site and nearest-neighbor phonon-induced attraction due to short-range electron-phonon coupling, and static nearest-neighbor phonon-induced attraction due to long-range electron-phonon coupling.}
\label{table:GeTe_results}
\end{table*}

The magnitude of the on-site and nearest-neighbor electron-electron repulsion is given
in Table\,\ref{table:GeTe_results}, as is the phonon-induced
renormalization of these. Here we give the intra-band values for the top valence band within the active space, however, similar to the case of MgO, the inter-band phonon-mediated interactions are found to be of comparable magnitude. Despite the anisotropic character of
the system, the values of these quantities do not significantly change depending on the spatial direction, 
we therefore only give the values for the in-plane interactions for brevity. The first striking observation
from the results of Table\,\ref{table:GeTe_results} is that
the on-site Coulomb repulsion $U^{el}=1.102$\,eV is strongly
renormalized by phonons, leading to a total interaction of $U^{tot}=U^{el}+U^{ph}=0.229$\,eV in the static limit, \emph{i.e.} amounting to a
$79$\% reduction of the electronic repulsion due to phonon screening. 
The second interesting observation is that the total phonon-induced attraction between nearest neighbor sites, due to the combined effects of short-range and long-range electron-phonon coupling, is equal to $V^{ph,st,tot}=V^{ph,st,\mathcal{S}}+V^{ph,st,\mathcal{L}}=-0.166$\,eV, which will not only completely screen out the repulsive
electron-electron term $V^{el}=0.118$\,eV, but will even
result in an overall attractive interaction of $V^{el}+V^{ph,st,tot}=-0.048$\,eV in the static limit.

\begin{figure}[tb]
    \centering
    \includegraphics[width=\linewidth]{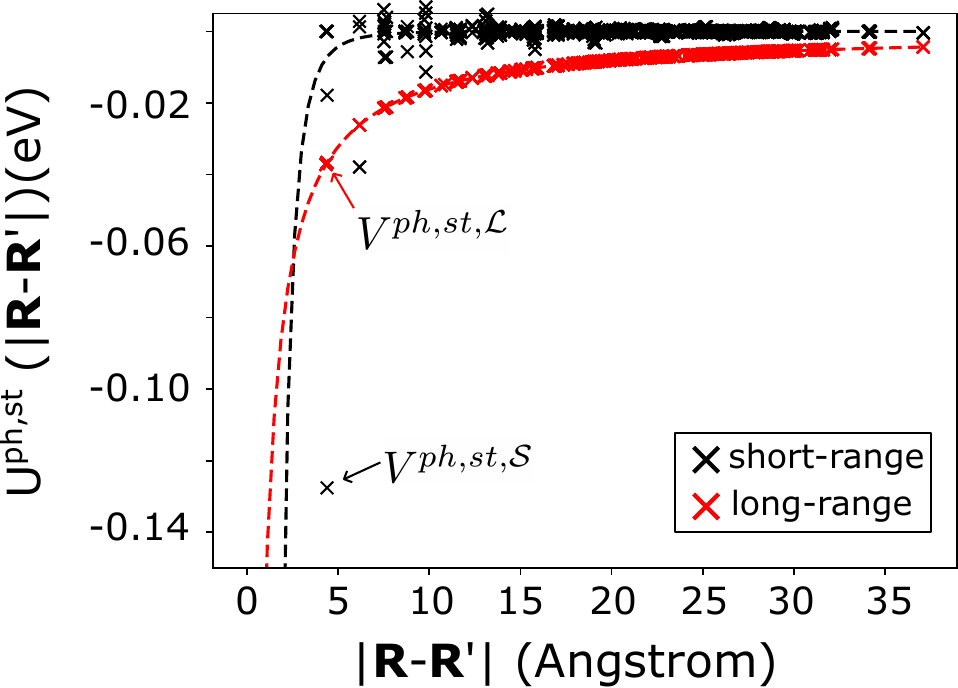}
    \caption{Spatial decay of the static phonon-induced electron-electron attraction in GeTe, due to short-range and long-range electron-phonon coupling. The dashed lines serve as a guide to the eye.}
    \label{fig:GeTe_Wph_vs_R}
\end{figure}

Interestingly,
unlike the case of MgO where the nearest-neighbor phonon-induced attractive interaction is almost entirely due to
long-range electron-phonon coupling, we see in Table\,\ref{table:GeTe_results}, that for GeTe the short-range
term $V^{ph,st,\mathcal{S}}$ is in fact more negative than
the long-range part $V^{ph,st,\mathcal{L}}$. To better understand this result, we plot in Fig.\,\ref{fig:GeTe_Wph_vs_R} the spatial decay of the
static phonon-induced attraction, for short-range and long-range electron-phonon coupling. As expected, the short-range part decays
rapidly, quickly reducing to a negligible contribution with
increasing distance between the lattice sites $|\mathbf{R}-\mathbf{R}'|$. Its most important contributions are found for
small distances, primarily for the on-site interaction, where
as discussed previously, it induces a $79$\% reduction of the electronic repulsion. Despite the fast decay of this term
with increasing distance, its strong on-site value
leads to a still appreciable contribution for nearest neighbors, which is in fact more attractive than that of the 
generalized Fr\"ohlich vertex. Nevertheless, here too
the long-range mechanism of the electron-phonon interaction leads to an appreciable renormalization of
the electron-electron interaction, and incorporating it
in the downfolded representation of GeTe is important
towards achieving predictive accuracy for the physics of this system. 
Moreover, the Fr\"ohlich term remains finite
at large values of $|\mathbf{R}-\mathbf{R}'|$, given its long-range character.

\begin{figure}[tb]
    \centering
    \includegraphics[width=\linewidth]{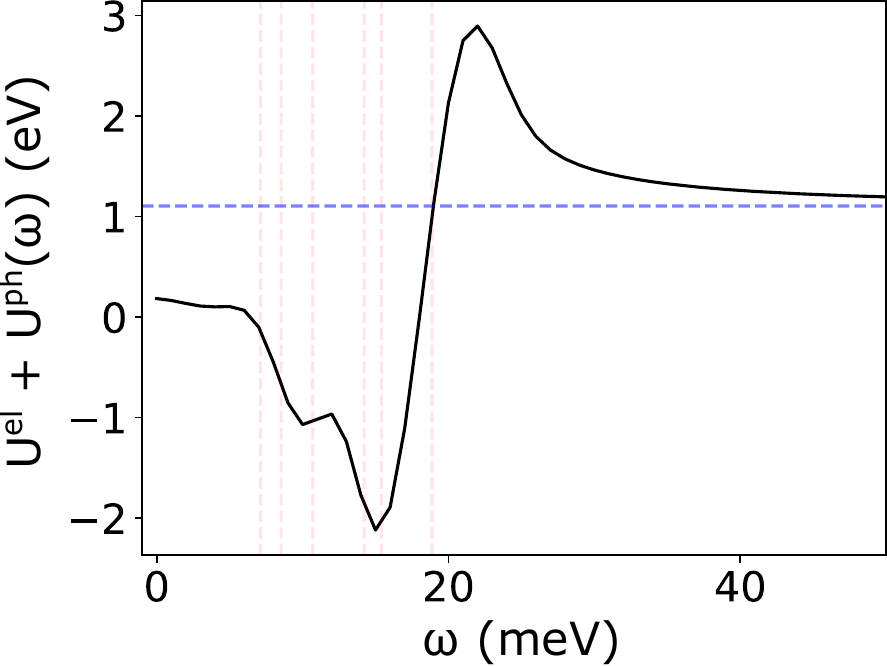}
    \caption{Frequency-dependence of the total on-site Hubbard term for the highest-lying valence band of GeTe. The red vertical lines indicate the average phonon frequencies on the $20\times20\times20$ $\mathbf{q}$-grid where $U^{ph}$ is computed, while
    the blue horizontal line marks the value of the electronic Hubbard term of $1.102$\,eV for this band, in the $\omega=0$ limit.}
    \label{fig:GeTe_frequency}
\end{figure}

Now let us consider the frequency dependence of the on-site electron-electron interactions, once phonon screening is accounted for, visualized in Fig.\,\ref{fig:GeTe_frequency}. We
see that already for small values of the electronic frequency and as we are approaching the phonon poles, the on-site
Coulomb interaction assumes negative values. Moreover, for
larger frequency values in the plotted range, the total electronic interaction $U^{el}+U^{ph}(\omega)$ approaches the static limit of the purely electronic contribution, as expected. 

\begin{figure}[tb]
    \centering
    \includegraphics[width=\linewidth]{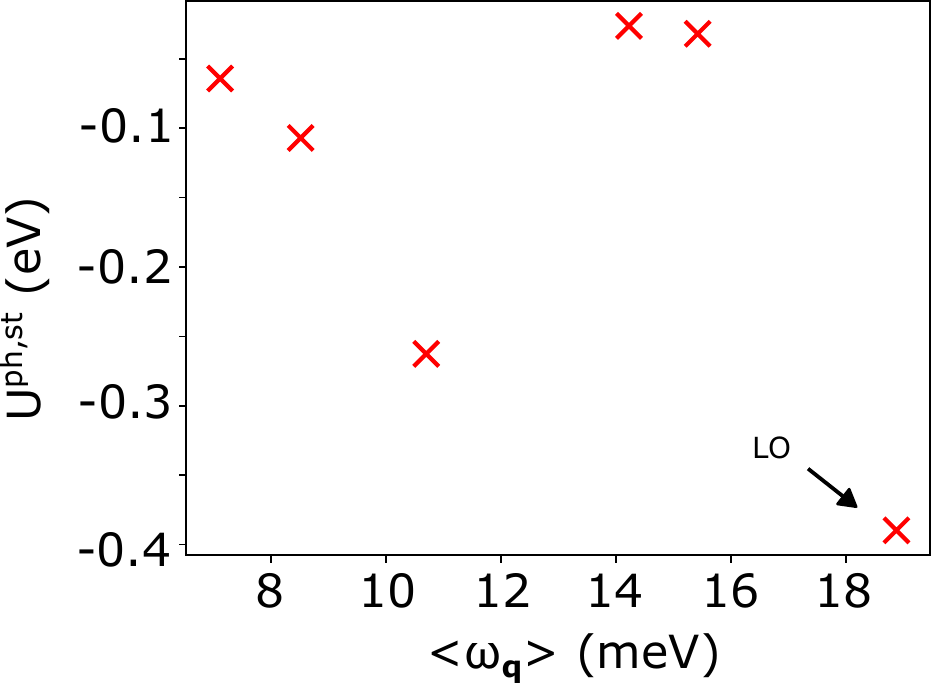}
    \caption{Contributions of individual phonon modes to
    the renormalization of the on-site Hubbard term of GeTe.}
    \label{fig:GeTe_modes}
\end{figure}

The strong variations of the frequency-dependent interaction strength in Fig.\,\ref{fig:GeTe_frequency} as one approaches the poles corresponding to acoustic and optical
phonons suggests that the screening from several modes renormalize the electron-electron interactions strongly. In Fig.\,\ref{fig:GeTe_modes} we plot the attractive contribution of different phonons in the static limit. Similar to MgO, the LO phonon again dominates this short-range contribution to the attractive
interaction. However, it is worth noting that acoustic modes also provide a very significant attraction
in this case. This is attributed to the strong piezoelectric electron-phonon interaction in GeTe~\cite{PhysRevB.78.205203}, which also manifests as strong phonon screening from acoustic modes in other piezoelectrics, such as CdS~\cite{Alvertis2024}. 

\begin{figure}[tb]
    \centering
    \includegraphics[width=\linewidth]{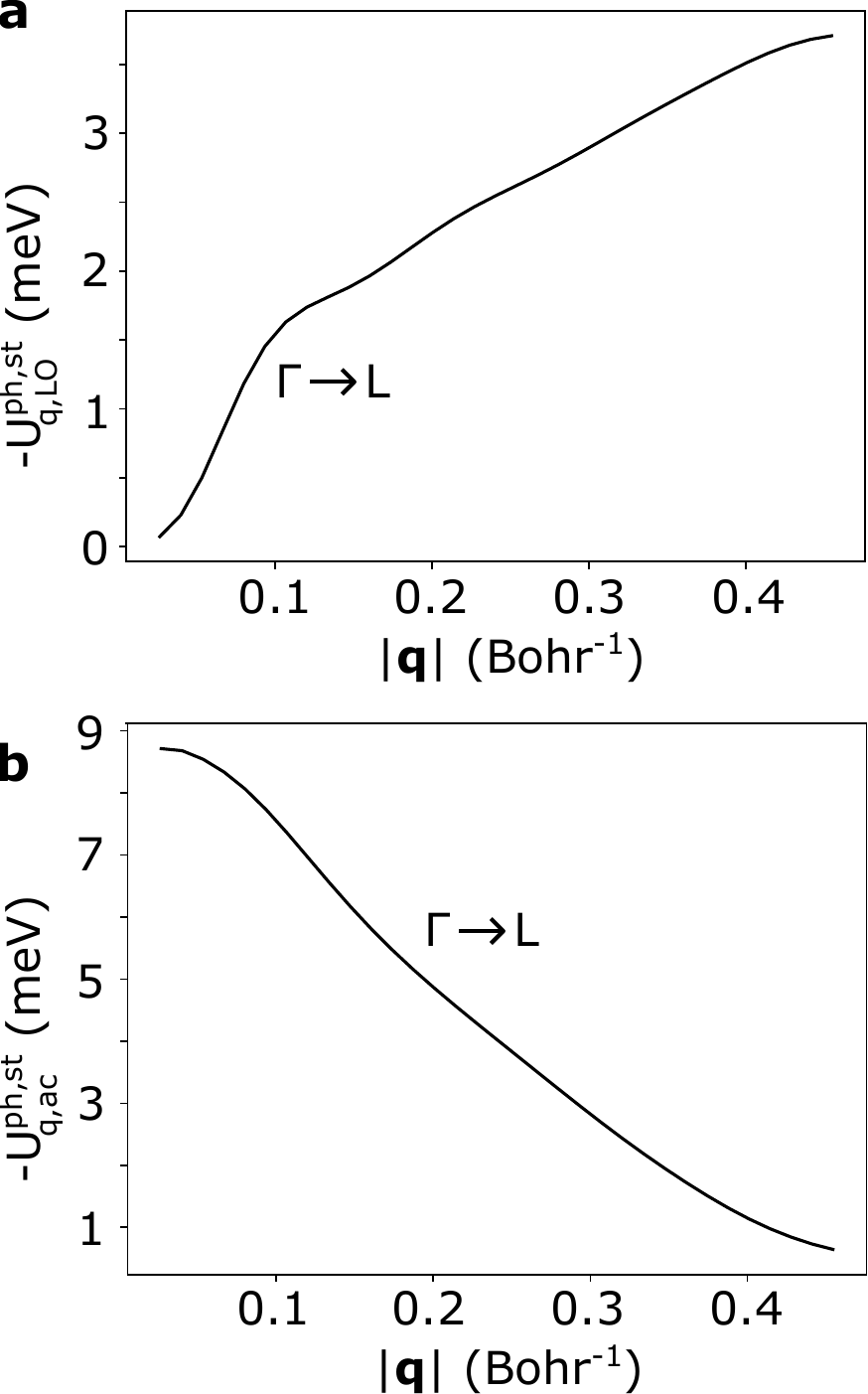}
    \caption{Dependence of the short-range phonon-induced
    correction of the on-site Coulomb interaction on the phonon momentum, in the static limit. This is visualized for the LO phonon of GeTe, along the $\Gamma \rightarrow L$ path in reciprocal space (panel $\textbf{a}$), and for the strongest coupled acoustic phonon along the $\Gamma \rightarrow L$ path (panel $\textbf{b}$).}
    \label{fig:GeTe_q}
\end{figure}

In order to gain more insights into the microscopic mechanism of the phonon-induced electron-electron
attraction due to short-range interactions, we plot in Fig.\,\ref{fig:GeTe_q} the dependence of $U^{ph,st}$ in the static limit on the phonon momentum $|\mathbf{q}|$, for the LO phonon (panel \textbf{a}) and for the
dominant acoustic mode (panel \textbf{b}). For the LO phonon we see that $U^{ph}$ is roughly proportional to $|\mathbf{q}|$ and vanishes at $\Gamma$, in accordance with a deformation potential mechanism. This channel is particularly strong in GeTe, due to the large values of the Born effective charges in this material (Table\,\ref{table:GeTe_Born}), which constitute the character of the
Ge-Te bond mixed ionic-covalent; a feature which is commonly associated with strong
deformation potential~\cite{Cohen_Louie_2016}. Intuitively, this can be understood as having a significant charge localization
on the atoms, making distortions of the atomic configuration particularly impactful on the properties
of the electronic wave function. This is similar to the mechanism of electron-phonon coupling in molecular crystals and molecules in vacuum, where strong electronic localization leads to stronger coupling to vibrations associated with
local atomic distortions~\cite{PhysRevB.102.081122,10.1063/5.0052247}. The acoustic phonon on the other hand, does not demonstrate behavior
consistent with a deformation potential mechanism as in the case of MgO, with its contribution growing
as we approach the $\Gamma$ point. As already stated previously, this is consistent with the piezoelectric mechanism of the electron-phonon interaction. These terms arise from
long-range quadrupole contributions, which are captured
within DFPT calculations of the electron-phonon matrix elements~\cite{poncé2025verificationvalidationzeropointelectronphonon}. 
Similar to the dipole terms, the quadrupole terms should be subtracted from
the electron-phonon matrix elements prior to Wannier-Fourier interpolation~\cite{PhysRevB.102.125203,Lee2023}, in order to avoid irregular behavior, particularly in the $\mathbf{q}\rightarrow \mathbf{0}$ region. 
However, for
GeTe we do not perform any interpolation of the electron-phonon matrix elements, as outlined in Section\,\ref{computational_details}, it is therefore
not necessary to subtract the quadrupole terms from our short-range 
electron-phonon matrix elements computed from DFPT, and
hence the piezoelectric phonon-mediated electron-electron interaction in GeTe is captured within our approach. 

Our first-principles formalism predicts the emergence of
overall attractive nearest-neighbor interactions in the valence bands of GeTe, in the static limit, which arise from the
interplay between the Fr\"ohlich, deformation
potential and piezoelectric mechanisms of
electron-phonon coupling. Such an attractive
nearest neighbor attraction has commonly
been connected to the emergence of superconductivity~\cite{doi:10.1126/science.abf5174,PhysRevB.105.024510,PhysRevB.107.L201102,PhysRevB.108.064514}. Additionally, the
frequency dependence of the on-site Hubbard
term due to phonon contributions, visualized in Fig.\,\ref{fig:GeTe_frequency}, demonstrates that this term also becomes attractive at relatively low electronic
energies, which could be relevant already
at relatively low doping levels, and which
could further contribute to superconductivity~\cite{PhysRevB.108.064514}. Indeed it
has previously been shown in the literature that a purely on-site phonon-mediated attraction between electrons, can alone induce superconductivity, with appropriate retardation and strength~\cite{PhysRevLett.125.167001}. 
Therefore, our findings of a strongly attractive on-site interaction between electrons, which overcomes the repulsion already at low frequencies, as well as the existence of a net attractive nearest neighbor interaction, already in the static limit, are strongly suggestive of the emergence of phonon-mediated superconductivity in GeTe, through the combination of the Fr\"ohlich, deformation potential, and piezoelectric mechanisms. 
GeTe
was already proposed to be a superconductor in 1969 by Allen and Cohen~\cite{PhysRev.177.704} based on a deformation potential mechanism, which we indeed find to provide a key contribution
to the overall electron-electron attraction. Recent experiments
have confirmed GeTe to be superconducting
upon hole doping~\cite{PhysRevLett.124.047002}. 

It is worth
considering what makes GeTe unique within our formalism in terms of
its ability to generate attractive electron-electron interactions
compared to other semiconductors. The long-range Fr\"ohlich 
coupling due to the single LO phonon in
this material is not sufficient to lead to
an overall attractive nearest-neighbor electron-electron interaction. As discussed for Eq.\,\eqref{eq:Uph_long_range_Frohlich_static}, LO phonons can, at most, cancel
the electron-electron repulsion in the limit of $\epsilon_o \rightarrow \infty$, however
cannot result in an overall attractive
interaction, in agreement with the prediction of Ref.~\cite{RevModPhys.53.81}. Nevertheless, the strong
dipolar electron-phonon coupling effectively screens out
long-range electron-electron repulsion, and
other mechanisms of electrons coupling to
phonons, including the deformation
potential and piezoelectric mechanisms here,
contribute to an additional attraction, which
can flip the overall sign of the interaction
between electrons. GeTe is known to exhibit
significant piezoelectricity~\cite{PhysRevB.78.205203}, resulting in important contributions
from this mechanism of electron-phonon coupling. Additionally, the large values of its Born
effective charges indicate a 
significant mixed ionic-covalent bonding character for
the Ge-Te bond, and the covalent character is associated with
strong deformation potential coupling~\cite{Cohen_Louie_2016}. Overall, 
it is the combination of the Fr\"ohlich, piezoelectric, and deformation potential
mechanisms for the electron-phonon interaction
of GeTe, which are all substantial in this
material, that results in an overall attraction between electrons.

\section{Discussion and conclusions}
\label{discussion}
In this work, we have developed a theoretical and computational framework to account for phonon screening
on electron-electron interactions, entirely from first
principles. We account for short-range and long-range electron-phonon coupling on equal
footing, and we have demonstrated simplifications of our
expression in different limits, \emph{e.g.}, in the static
limit, in the case where long-range electron-phonon interactions are described by the traditional Fr\"ohlich
model, and beyond. Moreover, we have demonstrated the connections of
our formalism to applying a polaron transformation to a
many-body electron-phonon Hamiltonian. Our theoretical scheme for including phonon effects in
electronic interactions is directly compatible with \emph{ab initio} downfolding schemes, which have grown in
popularity for obtaining the properties of strongly-correlated materials from first principles~\cite{Nakamura2008,Nakamura2010,Arita2015,Zheng2018}. 

We have applied our computational framework to MgO and GeTe,
and in both cases we find significant phonon-induced 
renormalization of the electron-electron interactions, particularly for on-site Coulomb terms. This results mostly from short-range electron-phonon interactions based on the deformation potential mechanism, which, despite their fast decay, still lead
to substantial nearest-neighbor attraction in GeTe. The
interplay of this mechanism with long-range electron-phonon coupling screens out repulsive electron-electron interactions, and leads to an overall attraction, which
could provide an explanation for superconductivity in this
system. 

Our theoretical work could have important implications
for accurately predicting the electronic phase diagrams of diverse materials, and interpreting experiments. \emph{Ab initio} downfolding casts the problem of
describing the electronic structure of a material in
the form of an extended Hubbard model (see Eq.\,\eqref{eq:Hamiltonian}), which generally allows
capturing complex phases such as charge density waves, spin density waves, and $d$-wave superconductivity depending on the values of the parameters $t,U,V$~\cite{PhysRevB.70.094523,PhysRevLett.99.216403}. Therefore, our approach to obtain the potentially
strong phonon-induced renormalization of the Coulomb
terms could significantly reshape theoretically predicted phase diagrams. The impact of long-range electron-phonon coupling, which our approach has incorporated into \emph{ab initio} downfolding for the first time, would be particularly pronounced here, as
this interaction can greatly renormalize the nearest-neighbor Coulomb interaction $V$ (see our examples of MgO and GeTe), which can often determine the electronic phase of a material~\cite{PhysRevB.70.094523,PhysRevLett.99.216403}. An implication of our work relevant to experiment includes
the potential emergence of superconductivity in GeTe, due to static nearest-neighbor and retarded on-site attractive electron-electron interactions mediated by phonons. The renormalization of the Coulomb terms will also affect the transport properties of materials, particularly in the Mott insulator regime, where the conductivity can depend sensitively on the $U,V$ values~\cite{PhysRevB.79.125116,PhysRevB.94.235115}.

\section{Acknowledgments}
The authors are thankful to Marvin L. Cohen for
helpful and inspiring discussions. The authors also express their gratitude to the anonymous referees for 
pointing out the regime of validity of eq.\,\eqref{eq:Born_LO_TO}, which led to modifications in Appendix~\ref{derivation_long_range} and the addition of Appendix~\ref{gamma_LO_TO}. 
This material is based upon work supported by the U.S. Department of Energy, Office of Science, National Quantum Information Science Research Centers, Superconducting Quantum Materials and Systems Center (SQMS) under contract No. DE-AC02-07CH11359.  We are grateful for support from NASA Ames Research Center. A.M.A. thanks the Physics Department and the Oden Institute for Computational Engineering and Sciences at The University of Texas at Austin for their support. C. J. N. C. and M. R. F. acknowledge support from the UK Engineering and Physical Sciences Research Council (EPSRC). Z.L. acknowledges the support by the U.S. National Science Foundation (NSF) CAREER Award under Grant No. DMR-2440763. VV was supported by the National Science Foundation (NSF) CAREER award through grant No. DMR-1945098.  This research used resources of the National Energy Research
Scientific Computing Center, a DOE Office of Science User Facility supported by the Office of Science of the U.S. Department of Energy under Contract No. DE-AC02-05CH11231 using NERSC awards HEP-ERCAP0029167 and DDR-ERCAP0029710.

\section{Data Availability}
The data underlying this publication are available from the authors upon reasonable request. 

\appendix

\section{Detailed derivation of the phonon correction to the Coulomb interaction}
\label{derivation_Uph}

The general $4$-center Coulomb interaction
appearing in downfolded Hamiltonians can
be written as
\begin{align}
    \label{eq:coulomb_4_center}
    U_{i\mathbf{R}j\mathbf{R}'k\mathbf{R}''l\mathbf{R}'''}(\omega)=\nonumber \\ \int_V  d\mathbf{r}_1 d\mathbf{r}_2 \phi_{i\mathbf{R}}^*(\mathbf{r}_1)\phi_{i\mathbf{R}'}(\mathbf{r}_2)W(\mathbf{r}_1,\mathbf{r}_2;\omega)\phi_{k\mathbf{R}''}^*(\mathbf{r}_1)\phi_{l\mathbf{R}'''}(\mathbf{r}_2).
\end{align}
Taking $W=W^{el}+W^{ph}$ the integral becomes 
\begin{equation}
    U_{i\mathbf{R}j\mathbf{R}'k\mathbf{R}''l\mathbf{R}'''}(\omega)=U^{el}_{i\mathbf{R}j\mathbf{R}'k\mathbf{R}''l\mathbf{R}'''}(\omega)+U^{ph}_{i\mathbf{R}j\mathbf{R}'k\mathbf{R}''l\mathbf{R}'''}(\omega),
\end{equation}
where
the electronically screened Coulomb interaction is given by 
\begin{align}
        U^{el}_{i\mathbf{R}j\mathbf{R'},k\mathbf{R''}l\mathbf{R'''}}(\omega)
        &= \langle{\phi_{i\mathbf{R}}\phi_{j\mathbf{R}'}|W^{el}(\omega)|\phi_{k\mathbf{R''}}\phi_{l\mathbf{R'''}}\rangle},
\end{align}
and the phonon-screened term is
\begin{align}
        U^{ph}_{i\mathbf{R}j\mathbf{R'},k\mathbf{R''}l\mathbf{R'''}}(\omega)
        = \langle{\phi_{i\mathbf{R}}\phi_{j\mathbf{R}'}|W^{ph}(\omega)|\phi_{k\mathbf{R''}}\phi_{l\mathbf{R'''}}\rangle}.
\end{align}
We now focus on developing the expression for $U^{ph}_{i\mathbf{R}j\mathbf{R'},k\mathbf{R''}l\mathbf{R'''}}(\omega)$. Using Eq.\,\eqref{eq:phonon_screening} for $W^{ph}$ and resolving the identity in the position basis, we have 
\begin{align}
        U^{ph}_{i\mathbf{R}j\mathbf{R'},k\mathbf{R''}l\mathbf{R'''}}(\omega) \nonumber \\
         = \int d\mathbf{r}_1d\mathbf{r}_2 \phi^*_{i\mathbf{R}}(\mathbf{r}_1)\phi^*_{j\mathbf{R'}}(\mathbf{r}_2)W^{ph}(\mathbf{r}_1,\mathbf{r}_2;\omega) \nonumber \\ \times \phi_{k\mathbf{R''}}(\mathbf{r}_1)\phi_{l\mathbf{R'''}}(\mathbf{r}_2),
\end{align}
which gives 
\begin{align}
        U^{ph}_{i\mathbf{R}j\mathbf{R'},k\mathbf{R''}l\mathbf{R'''}}(\omega) \nonumber \\
        = \sum_{\mathbf{q}\nu}D_{\mathbf{q}\nu}(\omega)\int d\mathbf{r}_1d\mathbf{r}_2 \phi^*_{i\mathbf{R}}(\mathbf{r}_1)\phi^*_{j\mathbf{R'}}(\mathbf{r}_2)g_{\mathbf{q}\nu}(\mathbf{r}_1)g^*_{\mathbf{q}\nu}(\mathbf{r}_2)\nonumber \\
        \times\phi_{k\mathbf{R''}}(\mathbf{r}_1)\phi_{l\mathbf{R'''}}(\mathbf{r}_2)
\end{align}
The integral factorizes and so we have 
\begin{align}
        U^{ph}_{i\mathbf{R}j\mathbf{R'},k\mathbf{R''}l\mathbf{R'''}}(\omega)  \nonumber \\
        = \sum_{\mathbf{q}\nu}D_{\mathbf{q}\nu}(\omega)\int d\mathbf{r}_1 \phi^*_{i\mathbf{R}}(\mathbf{r}_1)g_{\mathbf{q}\nu}(\mathbf{r}_1)\phi_{k\mathbf{R''}}(\mathbf{r}_1)\nonumber \\
        \int d\mathbf{r}_2\phi^*_{j\mathbf{R'}}(\mathbf{r}_2)g^*_{\mathbf{q}\nu}(\mathbf{r}_2)\phi_{l\mathbf{R'''}}(\mathbf{r}_2)
\end{align}
The typical electron-phonon vertex integral is given by 
\begin{align}
        \int d\mathbf{r}_1 \phi^*_{i\mathbf{R}}(\mathbf{r}_1)g_{\mathbf{q}\nu}(\mathbf{r}_1)\phi_{k\mathbf{R''}}(\mathbf{r}_1) = \nonumber \\
        \int d\mathbf{r}_1 \phi^*_{i\mathbf{0}}(\mathbf{r}_1-\mathbf{R})g_{\mathbf{q}\nu}(\mathbf{r}_1)\phi_{k\mathbf{0}}(\mathbf{r}_1-\mathbf{R''})
\end{align}
where we have used the property that $\phi_{i\mathbf{R}}(\mathbf{r}_1)=\phi_{i\mathbf{0}}(\mathbf{r}_1-\mathbf{R})$. Further, changing variables to $\mathbf{r}=\mathbf{r}_1-\mathbf{R}$, we have 
\begin{align}
        \int d\mathbf{r}_1 \phi^*_{i\mathbf{0}}(\mathbf{r}_1-\mathbf{R})g_{\mathbf{q}\nu}(\mathbf{r}_1)\phi_{i\mathbf{0}}(\mathbf{r}_1-\mathbf{R''}) = \nonumber \\
        \int d\mathbf{r} \phi^*_{i\mathbf{0}}(\mathbf{r})g_{\mathbf{q}\nu}(\mathbf{r}+\mathbf{R})\phi_{i\mathbf{0}}(\mathbf{r}-(\mathbf{R''}-\mathbf{R})) = g_{ii\mathbf{q}\nu}(\mathbf{R},\mathbf{R''}),
\end{align}
where 
\begin{align}
\label{eq:g_dR}
        g_{mn\mathbf{q}\nu}(\mathbf{R},\mathbf{R''}) = \bra{\phi_{m\mathbf{R}}}g_{\mathbf{q}\nu}\ket{\phi_{n\mathbf{R''}}}\nonumber \\
        = \sum_{n'm'\mathbf{k}}e^{-i\mathbf{k}\cdot(\mathbf{R''}-\mathbf{R})}e^{i\mathbf{q}\cdot\mathbf{R}}w^\dag_{mn'}(\mathbf{k+q})g_{n'm'\nu}(\mathbf{k},\mathbf{q})w_{m'n}(\mathbf{k}),
\end{align}
$w_{mn}$ being Wannier rotation matrices,
and $g_{nm\nu}(\mathbf{k},\mathbf{q})$ the
electron-phonon matrix element in the Bloch
basis.
Therefore, we have 
\begin{align}    U^{ph}_{i\mathbf{R}j\mathbf{R'},k\mathbf{R''}l\mathbf{R'''}}(\omega)  \nonumber \\
        =\sum_{\mathbf{q}\nu}D_{\mathbf{q}\nu}(\omega)g_{ik\mathbf{q}\nu}(\mathbf{R},\mathbf{R''})g^*_{lj\mathbf{q}\nu}(\mathbf{R'''},\mathbf{R'})
\end{align}
and finally we have 
\begin{align}
        U^{ph}_{i\mathbf{R}j\mathbf{R'},k\mathbf{R''}l\mathbf{R'''}}(\omega) = \sum_{\mathbf{q}\nu}g_{ik\mathbf{q}\nu}(\mathbf{R},\mathbf{R''})g^*_{lj\mathbf{q}\nu}(\mathbf{R'''},\mathbf{R'})\nonumber \\
        \times\left(\frac{1}{\omega-\omega_{\mathbf{q}\nu}+i\eta}-\frac{1}{\omega+\omega_{\mathbf{q}\nu}-i\eta}\right).
\end{align}
The density-density interaction is given by the following matrix element
\begin{align}
    U^{ph}_{i\mathbf{R}j\mathbf{R'}}(\omega)\equiv U^{ph}_{i\mathbf{R}j\mathbf{R'},i\mathbf{R}j\mathbf{R'}}(\omega)
\end{align}
which in the static limit $\omega=0$ reduces to 
\begin{equation}
    \label{eq:Uph_st_appendix}
        U^{ph,st}_{i\mathbf{R}j\mathbf{R'}} = -2\sum_{\mathbf{q}\nu}\frac{g_{ii\mathbf{q}\nu}(\mathbf{R})g^*_{jj\mathbf{q}\nu}(\mathbf{R'})}{\omega_{\mathbf{q}\nu}} \ ,
\end{equation}
where here we define 
\begin{align}
\label{eq:g_R}
        g_{mn\mathbf{q}\nu}(\mathbf{R}) \equiv g_{mn\mathbf{q}\nu}(\mathbf{R},\mathbf{R})\nonumber \\
        = \sum_{n'm'\mathbf{k}}e^{i\mathbf{q}\cdot\mathbf{R}}w^\dag_{mn'}(\mathbf{k+q})g_{n'm'\nu}(\mathbf{k},\mathbf{q})w_{m'n}(\mathbf{k}).
\end{align}
Using the translational symmetry of the lattice, we have 
\begin{equation}
        U_{i\mathbf{R}j\mathbf{R'}} = U_{i\mathbf{0}j\mathbf{R'-R}}
\end{equation}
and so 
\begin{equation}
        U^{ph,st}_{i\mathbf{R}j\mathbf{R'}} = -2\sum_{\mathbf{q}\nu}\frac{g_{ii\mathbf{q}\nu}(\mathbf{0})g^*_{jj\mathbf{q}\nu}(\mathbf{R'-R})}{\omega_{\mathbf{q}\nu}} \ .
\end{equation}
By focusing on the on-site interaction we find
\begin{equation}
        U^{ph,st}_{i\mathbf{R}i\mathbf{R}} = -2\sum_{\mathbf{q}\nu}\frac{|g_{ii\mathbf{q}\nu}(\mathbf{0})|^2}{\omega_{\mathbf{q}\nu}} \ .
\end{equation}

\section{Derivation of the long-range contribution to
$U^{ph}$}
\label{derivation_long_range}

For brevity, below we drop band indices and we work in the static limit $\omega=0$, however, the results immediately generalize to the case of finite $\omega$. The long-range attractive electron-electron interaction due to phonons is written as (Eq.\,\eqref{eq:Uph_st_appendix}):
\begin{equation}
        U_{\mathbf{R}\mathbf{R}'}^{ph,\mathcal{L}}(\omega=0) = -2\sum_{\mathbf{q}\nu}\frac{g^{\mathcal{L}}_{\mathbf{q}\nu}(\mathbf{R})g^{\mathcal{L}*}_{\mathbf{q}\nu}(\mathbf{R}')}{\omega_{\mathbf{q}\nu}}.
\end{equation}
Using the expression for the generalized Fröhlich vertex Eq.\,\eqref{eq:generalized_Frohlich} gives 
\begin{align}
        U_{\mathbf{R}\mathbf{R}'}^{ph,\mathcal{L}}(\omega=0) = -2(\frac{4\pi}{V})^2\sum_{\mathbf{q}\nu}\sum_{\kappa\kappa'}(\frac{1}{2N\omega^2_{\mathbf{q}\nu}})\frac{1}{\sqrt{M_{\kappa}M_{\kappa'}}}\nonumber \\
        \times \frac{(\mathbf{q}\cdot\mathbf{Z}_{\kappa}\cdot\mathbf{e}_{\kappa\nu}(\mathbf{q}))(\mathbf{q}\cdot\mathbf{Z}_{\kappa'}\cdot\mathbf{e}_{\kappa'\nu}(\mathbf{q}))}{(\mathbf{q}\cdot\mathbf{\epsilon}_{\infty}\cdot\mathbf{q})^2}\nonumber \\ \times\bra{\phi_{i\mathbf{R}}(\mathbf{r})\phi_{j\mathbf{R}'}(\mathbf{r}')}e^{i\mathbf{q}\cdot(\mathbf{r-r'})}\ket{\phi_{i\mathbf{R}}(\mathbf{r})\phi_{j\mathbf{R}'}(\mathbf{r}')}.
\end{align}
First let us focus on the matrix element $\bra{\phi_{i\mathbf{R}}(\mathbf{r})\phi_{j\mathbf{R}'}(\mathbf{r}')}e^{i\mathbf{q}\cdot(\mathbf{r-r'})}\ket{\phi_{i\mathbf{R}}(\mathbf{r})\phi_{j\mathbf{R}'}(\mathbf{r}')}$. Written
out in integral form, this factorizes as:
\begin{align}
    \bra{\phi_{i\mathbf{R}}(\mathbf{r})\phi_{j\mathbf{R}'}(\mathbf{r}')}e^{i\mathbf{q}\cdot(\mathbf{r-r'})}\ket{\phi_{i\mathbf{R}}(\mathbf{r})\phi_{j\mathbf{R}'}(\mathbf{r}')}\nonumber \\ = \int d\mathbf{r}e^{i\mathbf{q}\cdot \mathbf{r}}|\phi_{i\mathbf{R}}(\mathbf{r})|^2\int d\mathbf{r}'e^{-i\mathbf{q}\cdot \mathbf{r}'}|\phi_{j\mathbf{R}'}(\mathbf{r}')|^2.
\end{align}
For well-localized Wannier functions, the corresponding form factors
$F_i(\mathbf q)=\int d\mathbf r\,e^{i\mathbf q\cdot\mathbf r}|\phi_{i\mathbf R}(\mathbf r)|^2$
are smooth and satisfy $F_i(\mathbf q)=e^{i\mathbf q\cdot\mathbf R}[1+\mathcal O(q^2)]$.
Thus, to leading order in the long-wavelength regime, the matrix element reduces to
$e^{i\mathbf q\cdot(\mathbf R-\mathbf R')}$.

We can thus write:
\begin{align}
        U_{\mathbf{R}\mathbf{R}'}^{ph,\mathcal{L}}(\omega=0) = -2(\frac{4\pi}{V})^2\sum_{\mathbf{q}\nu}\sum_{\kappa\kappa'}(\frac{1}{2N\omega^2_{\mathbf{q}\nu}})\frac{1}{\sqrt{M_{\kappa}M_{\kappa'}}}\nonumber \\
        \times \frac{(\mathbf{q}\cdot\mathbf{Z}_{\kappa}\cdot\mathbf{e}_{\kappa\nu}(\mathbf{q}))(\mathbf{q}\cdot\mathbf{Z}_{\kappa'}\cdot\mathbf{e}_{\kappa'\nu}(\mathbf{q}))}{(\mathbf{q}\cdot\mathbf{\epsilon}_{\infty}\cdot\mathbf{q})^2}e^{i\mathbf{q}\cdot(\mathbf{R-R'})}.
\end{align}
This expression is amenable to first-principles calculation, as everything that enters it may be obtained through DFPT calculations. However, we are now going to make a few well-justified approximations, which will greatly simplify its form. 

The long-range Fröhlich interaction is controlled by the long-wavelength region of reciprocal space. In this regime, the optical phonon eigenvectors and frequencies are smooth functions of $|\mathbf q|$ (for fixed direction $\hat{\mathbf q}$), and may be expanded as
\begin{equation}
\mathbf e_{\kappa\nu}(\mathbf q)
=
\mathbf e_{\kappa\nu}(\mathbf{q\rightarrow 0})
+ \mathcal{O}(q).
\end{equation}
To leading order in the long-wavelength limit we therefore replace the polarization vectors by their $\mathbf{q\rightarrow 0}$ values,
\begin{equation}
\mathbf e_{\kappa\nu}(\mathbf q)\approx \mathbf e_{\kappa\nu}(\mathbf{q\rightarrow 0}),
\end{equation}
while retaining the explicit $q$-dependence of the Fröhlich kernel. The neglected terms are higher order in $q$ and contribute only subleading corrections to $U^{ph,\mathcal L}$.

We therefore have:
\begin{align}
        U_{\mathbf{R}\mathbf{R}'}^{ph,\mathcal{L}}(\omega=0) = -2(\frac{4\pi}{V})^2\sum_{\mathbf{q}\nu}\sum_{\kappa\kappa'}(\frac{1}{2N\omega^2_{\mathbf{q}\nu}})\frac{1}{\sqrt{M_{\kappa}M_{\kappa'}}}\nonumber \\
        \times \frac{(\mathbf{q}\cdot\mathbf{Z}_{\kappa}\cdot\mathbf{e}_{\kappa\nu}(\mathbf{q}\rightarrow\mathbf{0}))(\mathbf{q}\cdot\mathbf{Z}_{\kappa'}\cdot\mathbf{e}_{\kappa'\nu}(\mathbf{q}\rightarrow\mathbf{0}))}{(\mathbf{q}\cdot\mathbf{\epsilon}_{\infty}\cdot\mathbf{q})^2}e^{i\mathbf{q}\cdot(\mathbf{R-R'})}.
\end{align}

We now introduce the (mass-weighted) mode effective charge vector
\begin{equation}
\label{eq:mode_effective_charge}
\mathcal Z_{\nu,\alpha}
\equiv
\sum_{\kappa\beta}\frac{1}{\sqrt{M_\kappa}}\,
Z_{\kappa,\alpha\beta}\,e_{\kappa\beta\nu}(\mathbf{q}\rightarrow\mathbf{0}),
\end{equation}
so that
\begin{equation}
\mathbf q\cdot \boldsymbol{\mathcal Z}_\nu
=
\sum_{\kappa}\frac{1}{\sqrt{M_\kappa}}\,
\mathbf q\cdot \mathbf Z_\kappa\cdot \mathbf e_{\kappa\nu}(\mathbf{q}\rightarrow\mathbf{0}).
\end{equation}

We therefore obtain
\begin{align}
U_{\mathbf{R}\mathbf{R}'}^{ph,\mathcal{L}}(\omega=0)
=
-\left(\frac{4\pi}{V}\right)^2
\sum_{\mathbf q\nu}
\frac{1}{N\omega_{\mathbf q\nu}^2}\,\nonumber \\
\times
\frac{\left|\mathbf q\cdot \boldsymbol{\mathcal Z}_\nu\right|^2}
{\left(\mathbf q\cdot \boldsymbol\epsilon_\infty\cdot \mathbf q\right)^2}
\,e^{i\mathbf q\cdot(\mathbf R-\mathbf R')}.
\end{align}

Writing $\mathbf q = |\mathbf q|\,\hat{\mathbf q}$, the quadratic contractions scale as
\begin{align}
\left|\mathbf q\cdot \boldsymbol{\mathcal Z}_\nu\right|^2
&\to |\mathbf q|^2\,\left|\hat{\mathbf q}\cdot \boldsymbol{\mathcal Z}_\nu\right|^2,\\
\left(\mathbf q\cdot \boldsymbol\epsilon_\infty\cdot \mathbf q\right)^2
&\to |\mathbf q|^4\,
\left(\hat{\mathbf q}\cdot \boldsymbol\epsilon_\infty\cdot \hat{\mathbf q}\right)^2.
\end{align}

Thus we are left with an overall factor $1/|\mathbf q|^2$ in the denominator. 
For simplicity, we denote the directional components $\hat q_\alpha$ simply by 
$q_\alpha$ in what follows. We therefore have:
\begin{align}
\label{eq:full_ph_mediated_Coulomb}
U_{\mathbf{R}\mathbf{R}'}^{ph,\mathcal{L}}(\omega=0)
=
-\left(\frac{4\pi}{V}\right)^2
\sum_{\mathbf q\nu}
\frac{1}{N\omega_{\mathbf q\nu}^2}\cdot
\frac{e^{i\mathbf q\cdot(\mathbf R-\mathbf R')}}{|\mathbf q|^2}\nonumber \\
\times
\frac{\left|q_\alpha \mathcal Z_{\nu,\alpha}\right|^2}
{\left(q_\alpha \epsilon^\infty_{\alpha\beta} q_\beta\right)^2}.
\end{align}
This expression is amenable to first-principles computation, and is consistent with the ion part of the total dielectric function that was derived in the seminal review of Dolgov, Kirzhnits and Maksymov ~\cite{RevModPhys.53.81}, which the authors of Ref.~\cite{RevModPhys.53.81} analyzed for the case of a two-atom unit cell.
Here we will not use Eq.\,\eqref{eq:full_ph_mediated_Coulomb} itself, but instead the limiting 
case of an isotropic system and of a system with mild anisotropy, which we derive in the following sections. Finally, we will show how for certain systems Eq.\,\eqref{eq:full_ph_mediated_Coulomb} reduces to a simpler form that only requires knowledge of the dielectric constant and LO-TO splittings.

\subsection{Long-range phonon-mediated interaction for an isotropic system}
To make further progress from Eq.\,\eqref{eq:full_ph_mediated_Coulomb}, we assume that the long-wavelength interaction is isotropic, 
so that the dielectric tensor may be written as 
$\epsilon^\infty_{\alpha\beta}=\epsilon_\infty\delta_{\alpha\beta}$.
In this limit,
\begin{equation}
(q_\alpha \epsilon^\infty_{\alpha\beta} q_\beta)^2
=
\epsilon_\infty^2.
\end{equation}
Moreover, in an isotropic medium the long-wavelength LO modes are characterized
by a polarization parallel to $\mathbf q$, so that the associated mode effective
charge is aligned with $\mathbf q$. The contraction then becomes independent of
direction and reduces to
\begin{equation}
|q_\alpha \mathcal Z_{\nu,\alpha}|^2
=
|\boldsymbol{\mathcal Z}_{\nu}|^2.
\end{equation}
The mode effective charge contributes to the long-range interaction only for
LO modes. We therefore restrict the mode index to the LO
branches, labeled by $m$.
Given the smooth variation of the LO phonon frequencies in the long-wavelength 
limit, we substitute them by their $\mathbf q\rightarrow\mathbf 0$ values,
\emph{i.e.}, $\omega_{\mathbf q m}=\omega_{LO,m}$, which is also independent of 
the direction of $\mathbf q$ under the isotropy assumption.
Using the identity
\(
\frac{1}{r}=\frac{4\pi}{NV}\sum_{\mathbf q}\frac{e^{i\mathbf q\cdot\mathbf r}}{|\mathbf q|^2},
\)
we arrive at
\begin{equation}
U_{\mathbf{R}\mathbf{R}'}^{ph,\mathcal{L}}(\omega=0)
=
-\sum_{m}
\left(\frac{4\pi}{V}\right)
\frac{|\boldsymbol{\mathcal Z}_m|^2}
{\epsilon_\infty\,\omega_{LO,m}^2}
\cdot
\frac{1}{\epsilon_\infty|\mathbf{R-R'}|}.
\end{equation}

Finally, defining
\begin{equation}
\label{eq:gamma_general}
\gamma^2
=
\left(\frac{4\pi}{V}\right)
\sum_{m}
\frac{|\boldsymbol{\mathcal Z}_m|^2}
{\epsilon_\infty\,\omega_{LO,m}^2},
\end{equation}
we can write this in the compact form
\begin{equation}
\label{eq:ph_mediated_interaction}
U_{\mathbf{R}\mathbf{R}'}^{ph,\mathcal{L}}(\omega=0)
=
-\frac{\gamma^2}{\epsilon_{\infty}|\mathbf{R-R'}|}.
\end{equation}

\subsection{Long-range phonon-mediated interaction for a mildly anisotropic system}

For an anisotropic system we retain the full dielectric tensor
$\epsilon^\infty_{\alpha\beta}$ in Eq.~\eqref{eq:full_ph_mediated_Coulomb}. In this case the long-range
interaction acquires a directional dependence through the factor
$\left|q_\alpha \mathcal Z_{\nu,\alpha}\right|^2/\left(q_\alpha\epsilon^\infty_{\alpha\beta}q_\beta\right)^2$.
For use in an effective isotropic model, we define an isotropized long-range
interaction by replacing this directional dependence by its orientational average
over the unit sphere. This is a good approximation for systems with modest dielectric anisotropy. Restricting again to the LO branches labeled by $m$ and
using $\omega_{\mathbf q m}\simeq \omega_{LO,m}$ in the long-wavelength limit, Eq.\,\eqref{eq:full_ph_mediated_Coulomb} becomes:
\begin{align}
U_{\mathbf{R}\mathbf{R}'}^{ph,\mathcal{L}}(\omega=0)
\simeq
-\left(\frac{4\pi}{V}\right)^2
\sum_{m}
\frac{1}{N\omega_{LO,m}^2}
\nonumber \\ \times\left[\int\frac{d\Omega}{4\pi}\,
\frac{\left|q_\alpha \mathcal Z_{m,\alpha}\right|^2}
{\left(q_\alpha \epsilon^\infty_{\alpha\beta} q_\beta\right)^2}\right]
\sum_{\mathbf q}\frac{e^{i\mathbf q\cdot(\mathbf R-\mathbf R')}}{|\mathbf q|^2}.
\end{align}
Using
\(
\frac{1}{r}=\frac{4\pi}{NV}\sum_{\mathbf q}\frac{e^{i\mathbf q\cdot\mathbf r}}{|\mathbf q|^2}
\),
this yields the effective isotropic form
\begin{align}
U_{\mathbf{R}\mathbf{R}'}^{ph,\mathcal{L}}(\omega=0)
\simeq
-\left(\frac{4\pi}{V}\right)\sum_{m}\frac{1}{\omega_{LO,m}^2}
\nonumber \\ \times\left[
\int\frac{d\Omega}{4\pi}\,
\frac{\left|q_\alpha \mathcal Z_{m,\alpha}\right|^2}
{\left(q_\alpha \epsilon^\infty_{\alpha\beta} q_\beta\right)^2}\right]
\frac{1}{|\mathbf R-\mathbf R'|}.
\end{align}

To retain a Coulomb-like screened form, we introduce an effective dielectric
constant
\begin{equation}
\label{eq:eps_eff_def}
\frac{1}{\epsilon_{\rm eff}}
\equiv
\int\frac{d\Omega}{4\pi}\,
\frac{1}{q_\alpha \epsilon^\infty_{\alpha\beta} q_\beta}.
\end{equation}
We then define $\gamma^2$ as
\begin{equation}
\label{eq:gamma_general_aniso_eps_eff}
\gamma^2
=
\epsilon_{\rm eff}\left(\frac{4\pi}{V}\right)
\sum_{m}\frac{1}{\omega_{LO,m}^2}
\int\frac{d\Omega}{4\pi}\,
\frac{\left|q_\alpha \mathcal Z_{m,\alpha}\right|^2}
{\left(q_\alpha \epsilon^\infty_{\alpha\beta} q_\beta\right)^2},
\end{equation}
so that the long-range interaction retains the compact Coulomb-like form
\begin{equation}
U_{\mathbf{R}\mathbf{R}'}^{ph,\mathcal{L}}(\omega=0)
\simeq
-\frac{\gamma^2}{\epsilon_{\rm eff}\,|\mathbf R-\mathbf R'|}.
\end{equation}

\subsection{Simplification to an expression involving the LO-TO splittings}

In some cases, it is possible to make an additional simplification in the definition
of $\gamma^2$ and the final result for the phonon-mediated electron-electron
interaction. We demonstrate this in the case of an isotropic material, however
it is straightforward to generalize the discussion to an anisotropic system.

For certain materials, symmetry guarantees that the phonon eigenvectors at
$\mathbf{q}\rightarrow \mathbf{0}$ are identical to those at $\mathbf{q}=\mathbf{0}$,
even if the frequencies differ. In those cases, the longitudinal projection of
the mode effective charge is related to the LO–TO splitting via~\cite{PhysRevB.55.10355}
\begin{align}
\left(\frac{4\pi}{V}\right)
\frac{|q_\alpha \mathcal Z_{\nu,\alpha}|^2}
{q_\alpha \epsilon^{\infty}_{\alpha\beta} q_\beta}
=
\omega^{2}_{LO,\nu}-\omega^{2}_{TO,\nu}.
\end{align}

Using this relation in the definition of $\gamma^2$ yields
\begin{equation}
\label{eq:gamma_simple_appendix}
\gamma^2
=
\sum_m
\frac{\omega^{2}_{LO,m}-\omega^{2}_{TO,m}}{\omega^{2}_{LO,m}},
\end{equation}
and the phonon-mediated electron-electron interaction takes the simple form
\begin{equation}
\label{eq:ph_mediated_interaction_simpler}
U_{\mathbf{R}\mathbf{R}'}^{ph,\mathcal{L}}(\omega=0)
=
-\sum_{m}
\left(
\frac{\omega^{2}_{LO,m}-\omega^{2}_{TO,m}}{\omega^{2}_{LO,m}}
\right)
\frac{1}{\epsilon_{\infty}|\mathbf{R-R'}|}.
\end{equation}
While this is a very convenient form for the Coulomb interaction, we emphasize
again that it is only strictly valid when
$\mathbf{e}(\mathbf{q}\rightarrow \mathbf{0})=\mathbf{e}(\mathbf{q}=\mathbf{0})$.
For MgO and GeTe this condition holds to excellent accuracy, as
discussed in the following Appendix~\ref{gamma_LO_TO}, and hence
Eq.~\eqref{eq:ph_mediated_interaction_simpler} may be used. 
In the more general case,  $\gamma^2$ is computed
directly from the Born effective charges and phonon eigenvectors through the
mode effective charge defined in Eq.~\eqref{eq:mode_effective_charge}.

\section{Validity of Eq.~\eqref{eq:ph_mediated_interaction_simple}/Eq.~\eqref{eq:ph_mediated_interaction_simpler} for MgO and GeTe}
\label{gamma_LO_TO}

The simplified form of the long-range interaction,
Eqs.~\eqref{eq:ph_mediated_interaction_simple}/\eqref{eq:ph_mediated_interaction_simpler},
relies on the condition that the phonon eigenvectors at long wavelength coincide
with those at the Brillouin-zone center,
\begin{equation}
\mathbf e_\nu(\mathbf q\rightarrow 0)=\mathbf e_\nu(\mathbf q=0),
\end{equation}
up to an overall phase and possible unitary rotations within degenerate
subspaces. We verify this explicitly in MgO and GeTe by computing the overlap between
mass-weighted phonon eigenvectors obtained from DFPT. 

For a mode $\nu$ at $\mathbf q\rightarrow 0$ and a mode $\nu'$ at $\mathbf q=0$,
we define the squared overlap
\begin{equation}
\mathcal O_{\nu\nu'} =
\left|
\sum_{\kappa\alpha}
e^*_{\kappa\alpha,\nu}(\mathbf q\rightarrow 0)\,
e_{\kappa\alpha,\nu'}(\mathbf q=0)
\right|^2,
\end{equation}
where $\kappa$ labels atoms and $\alpha=x,y,z$ Cartesian components.
The eigenvectors are normalized according to the convention used in the
DFPT output. This quantity is invariant under overall phase changes and
provides a direct measure of the similarity between the two modes.

We also explicitly compare the values of $\gamma^2$ 
that determine the phonon-mediated interaction $-\gamma^2/(\epsilon \cdot r)$, when obtained 
from the general isotropic expression that involves the mode effective charges (Eq.\,\eqref{eq:gamma_general}), and the simpler expression involving the LO-TO splitting (Eq.\,\eqref{eq:gamma_simple_appendix}), which relies on 
the approximation:
\begin{align}
\left(\frac{4\pi}{V}\right)
\frac{|q_\alpha \mathcal Z_{\nu,\alpha}|^2}
{q_\alpha \epsilon^{\infty}_{\alpha\beta} q_\beta}
=
\omega^{2}_{LO,\nu}-\omega^{2}_{TO,\nu}.
\end{align}

While the simplified expression of Eq.~\eqref{eq:ph_mediated_interaction_simple}/Eq.~\eqref{eq:ph_mediated_interaction_simpler} holds for the systems we studied here, we emphasize that this should not be used without explicitly verifying that the phonon eigenvectors do not change between the $\mathbf{q}=\mathbf{0}$ point and the $\mathbf{q} \rightarrow \mathbf{0}$ region. This is only generally true for materials of certain symmetries, and is not expected to hold in more complex cases, such as for example in cubic perovskites. In those cases the Coulomb interaction will still be of the form $-\gamma^2/(\epsilon \cdot r)$, however the strength of the electron-phonon coupling $\gamma^2$ should be evaluated 
using the more general expressions we derived for isotropic and mildly anisotropic materials, which involve the mode effective charges rather than the LO-TO splitting
(Eq.\,\eqref{eq:gamma_general} and Eq.\,\eqref{eq:gamma_general_aniso_eps_eff} for isotropic and anisotropic systems respectively).

\begin{table*}[t]
\centering
\caption{Mass-weighted phonon eigenvectors for MgO used in the overlap
analysis. Components are listed in the order
$(\mathrm{Mg}_x,\mathrm{Mg}_y,\mathrm{Mg}_z,\mathrm{O}_x,\mathrm{O}_y,\mathrm{O}_z)$,
as read from the DFPT eigenvector output files.}
\label{table:mgo_evecs}
\begin{tabular}{lrrrrrr}
\hline\hline
Mode & Mg$_x$ & Mg$_y$ & Mg$_z$ & O$_x$ & O$_y$ & O$_z$ \\
\hline
$q\!\to\!0$ LO  & $-0.549839$ & $0$ & $0$ & $0.835271$ & $0$ & $0$ \\
$q=0$ opt.\ 1   & $0.411871$ & $-0.256155$ & $-0.026550$ 
                & $-0.741143$ & $0.460940$ & $0.047776$ \\
$q=0$ opt.\ 2   & $0.203254$ & $0.354090$ & $-0.263184$ 
                & $-0.365747$ & $-0.637170$ & $0.473587$ \\
$q=0$ opt.\ 3   & $-0.158140$ & $-0.212043$ & $-0.407416$ 
                & $0.284566$ & $0.381563$ & $0.733127$ \\
\hline\hline
\end{tabular}
\end{table*}

\begin{table*}[t]
\centering
\caption{Mass-weighted phonon eigenvectors for GeTe used in the overlap analysis
along $\mathbf q\parallel x$. Components are listed in the order
$(\mathrm{Ge}_x,\mathrm{Ge}_y,\mathrm{Ge}_z;\mathrm{Te}_x,\mathrm{Te}_y,\mathrm{Te}_z)$,
as read from the DFPT eigenvector output files. Tiny $\sim 10^{-6}$ components reflect
printing precision.}
\label{tab:gete_evecs_x}
\begin{tabular}{lcrrrrr}
\hline\hline
& Ge$_x$ & Ge$_y$ & Ge$_z$ & Te$_x$ & Te$_y$ & Te$_z$ \\
\hline
$q\!\to\!0$ LO ($\mathbf q\parallel x$) 
& $0.869047$ & $0$ & $0$ & $-0.494730$ & $0$ & $0$ \\
$q=0$ opt. 1 
& $1\times 10^{-6}$ & $0.870151$ & $0$ & $0$ & $-0.492785$ & $0$ \\
$q=0$ opt. 2 
& $0.870151$ & $-1\times 10^{-6}$ & $0$ & $-0.492785$ & $0$ & $0$ \\
\hline\hline
\end{tabular}
\end{table*}

\subsection{Direct comparison of phonon eigenvectors at $\mathbf{q}\rightarrow 0$ and $\mathbf{q}=0$ for MgO}
\label{MgO_evec_overlap}

In MgO, the optical modes at $\mathbf q=0$ form a triply-degenerate
subspace. In this case, individual eigenvectors at $\mathbf q=0$ are not
unique and may differ by unitary rotations within this subspace.
A meaningful comparison is therefore obtained by projecting the
long-wavelength eigenvector onto the entire degenerate subspace,
\begin{equation}
\sum_{\nu'\in \mathrm{deg.\ subspace}}
\mathcal O_{\nu\nu'} \simeq 1.
\end{equation}
Numerically, we find that this sum is equal to $0.994$ for the LO mode of MgO,
demonstrating that the eigenvector at $\mathbf q\rightarrow 0$ lies almost
entirely within the optical subspace at $\mathbf q=0$. This confirms that
$\mathbf e(\mathbf q\rightarrow 0)$ and $\mathbf e(\mathbf q=0)$ coincide to
excellent accuracy.

To make the procedure fully reproducible, we list in
Table~\ref{table:mgo_evecs} the DFPT eigenvector components used in the
comparison. These correspond to the LO mode at $\mathbf q\rightarrow 0$
and the three degenerate optical modes at $\mathbf q=0$.

Using these values, the individual overlaps between the long-wavelength LO
eigenvector and the three optical modes at $\mathbf q=0$ are found to be
$0.715$, $0.174$, and $0.105$, respectively. Their sum,
$0.994$, is very close to unity, confirming that the long-wavelength
eigenvector is essentially identical to the corresponding zone-center
optical eigenvector, up to a rotation within the degenerate subspace.

This numerical result justifies the use of the relation
\begin{equation}
\left(\frac{4\pi}{V}\right)
\frac{|q_\alpha \mathcal Z_{\nu,\alpha}|^2}
{q_\alpha \epsilon^{\infty}_{\alpha\beta} q_\beta}
=
\omega^{2}_{LO,\nu}-\omega^{2}_{TO,\nu}
\end{equation}
for MgO. Using the more general isotropic expression that involves the mode-effective charges (Eq.\,\eqref{eq:gamma_general}) for $\gamma^2$, we find $\gamma^2=0.660$, whereas the simpler expression of Eq.\,\eqref{eq:gamma_simple_appendix} involving the LO-TO splitting yields $\gamma^2=0.667$. Therefore the two values are in excellent agreement.  

\subsection{Direct comparison of phonon eigenvectors at $\mathbf{q}\rightarrow 0$ and $\mathbf{q}=0$ for GeTe}
\label{sec:evec_overlap_gete}

For GeTe we verify the condition $\mathbf e(\mathbf q\rightarrow 0)\simeq \mathbf e(\mathbf q=0)$
by explicitly comparing DFPT phonon eigenvectors. Since GeTe is mildly anisotropic,
$\gamma^2$ is evaluated using the orientational averaging procedure outlined above.
However, the validity of the LO--TO simplification depends only on the continuity of the
eigenvectors between $\mathbf q\to 0$ and $\mathbf q=0$. For brevity, we present the check
for $\mathbf q\parallel x$; the same conclusion is
obtained for other directions.

Here too we quantify the agreement using the squared overlap
\begin{equation}
\mathcal O_{\nu\nu'}=
\left|
\sum_{\kappa\alpha}
e^*_{\kappa\alpha,\nu}(\mathbf q\rightarrow 0)\,
e_{\kappa\alpha,\nu'}(\mathbf q=0)
\right|^2,
\end{equation}
with components listed in the order
$(\mathrm{Ge}_x,\mathrm{Ge}_y,\mathrm{Ge}_z;\mathrm{Te}_x,\mathrm{Te}_y,\mathrm{Te}_z)$.

At $\mathbf q=0$, the optical modes at $8.6$\,meV are two-fold degenerate, corresponding to two orthogonal polarizations.
Therefore, as with MgO, we compute the total projection of the eigenvectors onto the degenerate subspace:
\begin{equation}
\sum_{\nu'\in \mathrm{deg.\ subspace}} \mathcal O_{\nu\nu'} \simeq 1.
\end{equation}

Table~\ref{tab:gete_evecs_x} lists the DFPT eigenvectors relevant for the comparison along
$\mathbf q\parallel x$: the LO mode at $\mathbf q\rightarrow 0$ (mode 6 at $18.3$\,meV)
and the two degenerate optical modes at $\mathbf q=0$.
Using these values, we find
\begin{align}
\mathcal O_{(q\to 0)\mathrm{LO},\,q=0\,\mathrm{opt.}\,1} &\simeq 7.6\times 10^{-13},\\
\mathcal O_{(q\to 0)\mathrm{LO},\,q=0\,\mathrm{opt.}\,2} &\simeq 0.999995,
\end{align}
so that the projection onto the degenerate $q=0$ subspace is
$0.999995\simeq 1$.
This confirms that $\mathbf e(\mathbf q\rightarrow 0)\approx \mathbf e(\mathbf q=0)$ to
excellent accuracy for GeTe as well, thereby justifying the use of the LO--TO-based
simplification when evaluating the long-range interaction. Using the anisotropic expression that involves the mode-effective charges (Eq.\,\eqref{eq:gamma_general_aniso_eps_eff}) for $\gamma^2$, we find $\gamma^2=0.649$, whereas the simpler expression of Eq.\,\eqref{eq:gamma_simple_appendix} involving the LO-TO splitting yields $\gamma^2=0.78$, therefore slightly overestimating the long-range interaction.

\section{Connection between downfolding $W^{ph}$ and the Lang-Firsov transformation}
\label{lang_firsov_derivation}

Here, we demonstrate the equivalence between the Lang-Firsov transformation and the static expression obtained for $U^{ph}$ when downfolded in an extended Hubbard model. We work in a local basis and write the Hubbard model including an electron-phonon and a phonon term as 
\begin{align}
        H = \sum_{i\mathbf{R},j\mathbf{R'}}t_{i\mathbf{R},j\mathbf{R'}}a^\dag_{i\mathbf{R}}a_{j\mathbf{R'}}  \nonumber \\ +\frac{1}{2}\sum_{i\mathbf{R},j\mathbf{R'}} U_{i\mathbf{R},j\mathbf{R'}} a^\dag_{i\mathbf{R}}a^\dag_{j\mathbf{R'}}a_{j\mathbf{R'}}a_{i\mathbf{R}} \nonumber \\
        +\sum_{\mathbf{q}\nu}\omega_{\mathbf{q}\nu}b^\dag_{\mathbf{q}\nu}b_{\mathbf{q}\nu} \nonumber \\ + \sum_{i\mathbf{R}j\mathbf{R'}\mathbf{q}\nu} g_{ij\mathbf{q}\nu}(\mathbf{R,R'})(b_{\mathbf{q}\nu}+b^\dag_{-\mathbf{q}\nu})a^\dag_{i\mathbf{R}}a_{j\mathbf{R'}},
\end{align}
where we have dropped the spin indices for brevity, however these
are given explicitly in the formulas appearing in the main part of the manuscript. 
This general Hamiltonian includes non-local
scattering between $\mathbf{R}$ and $\mathbf{R}'$, caused by coupling to a single
phonon, described by the matrix element of Eq.\,\eqref{eq:g_dR}.
For $\mathbf{R}=\mathbf{R}'$ this expression
yields $g(\mathbf{R})$ (Eq.\,\eqref{eq:g_R}).

Let us first consider the non-local electron-phonon coupling $\mathbf{R}\neq\mathbf{R}'$, which results
in a modification of the hopping term 
\begin{equation}
\nonumber
    \sum_{i\mathbf{R},j\mathbf{R'}}(t_{i\mathbf{R},j\mathbf{R'}}+g_{ij\mathbf{q}\nu}(\mathbf{R,R'}))(b_{\mathbf{q}\nu}+b^\dag_{-\mathbf{q}\nu}))a^\dag_{i\mathbf{R}}a_{j\mathbf{R'}},
\end{equation}
which describes a Su-Schrieffer-Heeger (SSH) mechanism~\cite{PhysRevLett.42.1698,RevModPhys.60.781} of electron-phonon coupling. Given the strongly localized
nature of Wannier functions, which is the basis we
employ here, we expect these non-local terms $\bra{\phi_{m\mathbf{R}}}g_{\mathbf{q}\nu}\ket{\phi_{n\mathbf{R'}}}$ to be small compared to the local terms $g_{mn\mathbf{q}\nu}(\mathbf{R})=\bra{\phi_{m\mathbf{R}}}g_{\mathbf{q}\nu}\ket{\phi_{n\mathbf{R}}}$. We will therefore ignore them moving forward, however, it will be interesting to examine the impact of these terms from first principles
within a downfolding context, as a part of a future work. 

We now restrict the electron-phonon interaction to the local form 
 \begin{equation}
        V_{ep}=\sum_{i\mathbf{R}\mathbf{q}\nu} g_{ii\mathbf{q}\nu}(\mathbf{R})(b_{\mathbf{q}\nu}+b^\dag_{-\mathbf{q}\nu})a^\dag_{i\mathbf{R}}a_{i\mathbf{R}}.
\end{equation}
We introduce the Lang-Firsov transformation on the Hamiltonian as 
\begin{equation}
    \Bar{H} = e^{S}He^{-S}
\end{equation}
where 
\begin{equation}
        S = \sum_{i\mathbf{R}\mathbf{q}\nu}a^\dag_{i\mathbf{R}}a_{i\mathbf{R}}\frac{g_{ii\mathbf{q}\nu}(\mathbf{R})}{\omega_{\mathbf{q}\nu}}(b^\dag_{\mathbf{q}\nu}-b_{-\mathbf{q}\nu}) \ .
\end{equation}
Application of this transformation results in the electronic operators as 
\begin{subequations}
    \begin{equation}
            \Bar{a}_{i\mathbf{R}} = e^{S}a_{i\mathbf{R}}e^{-S} = a_{i\mathbf{R}}X_{i\mathbf{R}}
    \end{equation}
    with
    \begin{equation}
            \Bar{a}^\dag_{i\mathbf{R}} = e^{S}a^\dag_{i\mathbf{R}}e^{-S} = a^\dag_{i\mathbf{R}}X^\dag_{i\mathbf{R}}
        \end{equation}
        where 
        \begin{equation}
                    X_{i\mathbf{R}}=\exp\left(-\sum_{\mathbf{q}\nu}\frac{g_{ii\mathbf{q}\nu}(\mathbf{R)}}{\omega_{\mathbf{q}\nu}}(b^\dag_{\mathbf{q}\nu}-b_{-\mathbf{q}\nu})\right) \ .
        \end{equation}
\end{subequations}
The phonon operators transform as 
\begin{subequations}
    \begin{equation}
            \Bar{b}_{\mathbf{q}\nu} = b_{\mathbf{q}\nu}-\sum_{j\mathbf{R'}}\frac{g^{*}_{jj\mathbf{q}\nu}(\mathbf{R'})}{\omega_{\mathbf{q}\nu}}a^\dag_{j\mathbf{R'}}a_{j\mathbf{R'}}
    \end{equation}
    with 
    \begin{equation}
            \Bar{b}^\dag_{\mathbf{q}\nu} = b^\dag_{\mathbf{q}\nu}-\sum_{j\mathbf{R'}}\frac{g_{jj\mathbf{q}\nu}(\mathbf{R'})}{\omega_{\mathbf{q}\nu}}a^\dag_{j\mathbf{R'}}a_{j\mathbf{R'}} \ .
    \end{equation}
\end{subequations}
The number operator is conserved since $X_{i}$ is a unitary operator, \emph{i.e.}, $X^\dag_{i\mathbf{R}}X_{i\mathbf{R}} = 1$,
which means that 
\begin{equation}
    \Bar{a}^\dag_{i\mathbf{R}}\Bar{a}_{i\mathbf{R}} =a^\dag_{i\mathbf{R}}a_{i\mathbf{R}}.
\end{equation}
Additionally, we have
\begin{equation}     
X^\dag_{i\mathbf{R}}=\exp\left(\sum_{\mathbf{q}\nu}\frac{g_{ii\mathbf{q}\nu}(\mathbf{R)}}{\omega_{\mathbf{q}\nu}}(b^\dag_{\mathbf{q}\nu}-b_{-\mathbf{q}\nu})\right) \ .
\end{equation}
Thus, the similarity transformed Hamiltonian becomes 
\begin{align}
        \Bar{H} = \sum_{i\mathbf{R},j\mathbf{R'}}t_{i\mathbf{R},j\mathbf{R'}}X^\dag_{i}X_{j}a^\dag_{i\mathbf{R}}a_{j\mathbf{R'}} \nonumber \\ + \frac{1}{2}\sum_{i\mathbf{R},j\mathbf{R'}} U_{i\mathbf{R},j\mathbf{R'}} a^\dag_{i\mathbf{R}}a^\dag_{j\mathbf{R'}}a_{j\mathbf{R'}}a_{i\mathbf{R}}\nonumber \\
        +\sum_{\mathbf{q}\nu}\omega_{\mathbf{q}\nu}\Big(b^\dag_{\mathbf{q}\nu}-\sum_{i\mathbf{R}}\frac{g_{ii\mathbf{q}\nu}(\mathbf{R})}{\omega_{\mathbf{q}\nu}}a^\dag_{i\mathbf{R}}a_{i\mathbf{R}}\Big)\nonumber \\
        \times\Big(b_{\mathbf{q}\nu}-\sum_{j\mathbf{R'}}\frac{g^{*}_{jj\mathbf{q}\nu}(\mathbf{R'})}{\omega_{\mathbf{q}\nu}}a^\dag_{j\mathbf{R'}}a_{j\mathbf{R'}}\Big)\nonumber \\
        + \sum_{i\mathbf{R}\mathbf{q}\nu} g_{ii\mathbf{q}\nu}(\mathbf{R})\Bigg(b_{\mathbf{q}\nu}+b^\dag_{-\mathbf{q}\nu}-2\sum_{j\mathbf{R'}}\frac{g^{*}_{jj\mathbf{q}\nu}(\mathbf{R'})}{\omega_{\mathbf{q}\nu}}a^\dag_{j\mathbf{R'}}a_{j\mathbf{R'}}\Bigg)\nonumber \\
        \times a^\dag_{i\mathbf{R}}a_{i\mathbf{R}}
\end{align}
The transformation results in a cancellation of the electron-phonon interaction, thereby giving 
\begin{align}
        \Bar{H} = \sum_{i\mathbf{R},j\mathbf{R'}}t_{i\mathbf{R},j\mathbf{R'}}X^\dag_{i}X_{j}a^\dag_{i\mathbf{R}}a_{j\mathbf{R'}} \nonumber \\
        + \frac{1}{2}\sum_{i\mathbf{R},j\mathbf{R'}} U_{i\mathbf{R},j\mathbf{R'}} a^\dag_{i\mathbf{R}}a^\dag_{j\mathbf{R'}}a_{j\mathbf{R'}}a_{i\mathbf{R}}\nonumber \\
        +\sum_{\mathbf{q}\nu}\omega_{\mathbf{q}\nu}b^\dag_{\mathbf{q}\nu}b_{\mathbf{q}\nu}\nonumber \\+\sum_{i\mathbf{R}j\mathbf{R'}\mathbf{q}\nu}\frac{g_{ii\mathbf{q}\nu}(\mathbf{R})g^*_{jj\mathbf{q}\nu}(\mathbf{R'})}{\omega_{\mathbf{q}\nu}}a^\dag_{i\mathbf{R}}a_{i\mathbf{R}}a^\dag_{j\mathbf{R'}}a_{j\mathbf{R'}}\nonumber \\
        -2\sum_{j\mathbf{R'}i\mathbf{R}\mathbf{q}\nu} \frac{g_{ii\mathbf{q}\nu}(\mathbf{R})g^{*}_{jj\mathbf{q}\nu}(\mathbf{R'})}{\omega_{\mathbf{q}\nu}}a^\dag_{j\mathbf{R'}}a_{j\mathbf{R'}}a^\dag_{i\mathbf{R}}a_{i\mathbf{R}}
\end{align}
Now, using the anti-commutation relations 
\begin{equation}
        \{a_{i\mathbf{R}},a^\dag_{j\mathbf{R'}}\} = \delta_{ij}\delta_{\mathbf{RR'}}
\end{equation}
we may re-write the quartic terms generated by the Lang-Firsov transformation to give 
\begin{align}
        \Bar{H} = \sum_{i\mathbf{R}}\left(t_{i\mathbf{R},i\mathbf{R}}-\sum_{\mathbf{q}\nu}\frac{|g_{ii\mathbf{q}\nu}(\mathbf{0})|^2}{\omega_{\mathbf{q}\nu}}\right)a^\dag_{i\mathbf{R}}a_{i\mathbf{R}}\nonumber \\
        +\sum_{i\neq j\mathbf{R}\mathbf{R'}}t_{i\mathbf{R},j\mathbf{R'}}\nonumber \\ \times \exp\left(-\sum_{\mathbf{q}\nu}\frac{[g_{jj\mathbf{q}\nu}(\mathbf{R'})-g_{ii\mathbf{q}\nu}(\mathbf{R})]}{\omega_{\mathbf{q}\nu}}(b^\dag_{\mathbf{q}\nu}-b_{-\mathbf{q}\nu})\right)a^\dag_{i\mathbf{R}}a_{j\mathbf{R'}} \nonumber \\
        + \frac{1}{2}\sum_{i\mathbf{R},j\mathbf{R'}} \left(U_{i\mathbf{R},j\mathbf{R'}}-2\sum_{\mathbf{q}\nu}\frac{g_{ii\mathbf{q}\nu}(\mathbf{R})g^{*}_{jj\mathbf{q}\nu}(\mathbf{R'})}{\omega_{\mathbf{q}\nu}} \right)\nonumber \\ \times a^\dag_{i\mathbf{R}}a^\dag_{j\mathbf{R'}}a_{j\mathbf{R'}}a_{i\mathbf{R}} 
        +\sum_{\mathbf{q}\nu}\omega_{\mathbf{q}\nu}b^\dag_{\mathbf{q}\nu}b_{\mathbf{q}\nu}
\end{align}
where we have used $|g_{ii\mathbf{q}\nu}(\mathbf{R})|^2=|g_{ii\mathbf{q}\nu}(\mathbf{0})|^2$. The result of the transformation are the effective interactions 
\begin{subequations}
    \begin{equation}
            \tilde{t}_{i\mathbf{R},i\mathbf{R}} = t_{i\mathbf{R},i\mathbf{R}}-\sum_{\mathbf{q}\nu}\frac{|g_{ii\mathbf{q}\nu}(\mathbf{0})|^2}{\omega_{\mathbf{q}\nu}}
    \end{equation}
    with the off-site interaction modulated by 
    \begin{align}
            \tilde{t}_{i\mathbf{R},j\mathbf{R'}} = t_{i\mathbf{R},j\mathbf{R'}}\nonumber \\ \times \exp[-\sum_{\mathbf{q}\nu}\frac{[g_{jj\mathbf{q}\nu}(\mathbf{R'})-g_{ii\mathbf{q}\nu}(\mathbf{R})]}{\omega_{\mathbf{q}\nu}}(b^\dag_{\mathbf{q}\nu}-b_{-\mathbf{q}\nu})]
    \end{align}
    and the two-body interaction is renormalized as 
    \begin{equation}
            \tilde{U}_{i\mathbf{R},j\mathbf{R'}} = U_{i\mathbf{R},j\mathbf{R'}}-2\sum_{\mathbf{q}\nu}\frac{g_{ii\mathbf{q}\nu}(\mathbf{R})g^{*}_{jj\mathbf{q}\nu}(\mathbf{R'})}{\omega_{\mathbf{q}\nu}} 
    \end{equation}
\end{subequations}
As we can see, the contribution to the two-body term is exactly the static phonon-screened contribution so we may write 
\begin{equation}
            \tilde{U}_{i\mathbf{R},j\mathbf{R'}} = U_{i\mathbf{R},j\mathbf{R'}}+U^{ph,st}_{i\mathbf{R},j\mathbf{R'}}  \ .
\end{equation}
Using these effective interactions, the similarity transformed Hamiltonian takes the form
\begin{align}
        \Bar{H} = \sum_{i\mathbf{R},j\mathbf{R'}}\tilde{t}_{i\mathbf{R},j\mathbf{R'}}a^\dag_{i\mathbf{R}}a_{j\mathbf{R'}}\nonumber \\+ \frac{1}{2}\sum_{i\mathbf{R},j\mathbf{R'}} \tilde{U}_{i\mathbf{R},j\mathbf{R'}}a^\dag_{i\mathbf{R}}a^\dag_{j\mathbf{R'}}a_{j\mathbf{R'}}a_{i\mathbf{R}}
        +\sum_{\mathbf{q}\nu}\omega_{\mathbf{q}\nu}b^\dag_{\mathbf{q}\nu}b_{\mathbf{q}\nu}
\end{align}
Assuming a coherent decoupling of the phonon and electronic states, we may evaluate the temperature dependent trace of the off-site hopping term by 
\begin{align}
            \tilde{t}_{i\mathbf{R},j\mathbf{R'}}(T) \approx t_{i\mathbf{R},j\mathbf{R'}}\nonumber \\ \times \exp[-\sum_{\mathbf{q}\nu}\frac{|g_{jj\mathbf{q}\nu}(\mathbf{R'})-g_{ii\mathbf{q}\nu}(\mathbf{R})|^2}{\omega^2_{\mathbf{q}\nu}}(N_{\mathbf{q}\nu}(T)+\frac{1}{2})]
\end{align}


\begin{thebibliography}{92}%
\makeatletter
\providecommand \@ifxundefined [1]{%
 \@ifx{#1\undefined}
}%
\providecommand \@ifnum [1]{%
 \ifnum #1\expandafter \@firstoftwo
 \else \expandafter \@secondoftwo
 \fi
}%
\providecommand \@ifx [1]{%
 \ifx #1\expandafter \@firstoftwo
 \else \expandafter \@secondoftwo
 \fi
}%
\providecommand \natexlab [1]{#1}%
\providecommand \enquote  [1]{``#1''}%
\providecommand \bibnamefont  [1]{#1}%
\providecommand \bibfnamefont [1]{#1}%
\providecommand \citenamefont [1]{#1}%
\providecommand \href@noop [0]{\@secondoftwo}%
\providecommand \href [0]{\begingroup \@sanitize@url \@href}%
\providecommand \@href[1]{\@@startlink{#1}\@@href}%
\providecommand \@@href[1]{\endgroup#1\@@endlink}%
\providecommand \@sanitize@url [0]{\catcode `\\12\catcode `\$12\catcode `\&12\catcode `\#12\catcode `\^12\catcode `\_12\catcode `\%12\relax}%
\providecommand \@@startlink[1]{}%
\providecommand \@@endlink[0]{}%
\providecommand \url  [0]{\begingroup\@sanitize@url \@url }%
\providecommand \@url [1]{\endgroup\@href {#1}{\urlprefix }}%
\providecommand \urlprefix  [0]{URL }%
\providecommand \Eprint [0]{\href }%
\providecommand \doibase [0]{https://doi.org/}%
\providecommand \selectlanguage [0]{\@gobble}%
\providecommand \bibinfo  [0]{\@secondoftwo}%
\providecommand \bibfield  [0]{\@secondoftwo}%
\providecommand \translation [1]{[#1]}%
\providecommand \BibitemOpen [0]{}%
\providecommand \bibitemStop [0]{}%
\providecommand \bibitemNoStop [0]{.\EOS\space}%
\providecommand \EOS [0]{\spacefactor3000\relax}%
\providecommand \BibitemShut  [1]{\csname bibitem#1\endcsname}%
\let\auto@bib@innerbib\@empty
\bibitem [{\citenamefont {Nakamura}\ \emph {et~al.}(2008)\citenamefont {Nakamura}, \citenamefont {Yoshimoto}, \citenamefont {Arita}, \citenamefont {Tsuneyuki},\ and\ \citenamefont {Imada}}]{Nakamura2008}%
  \BibitemOpen
  \bibfield  {author} {\bibinfo {author} {\bibfnamefont {K.}~\bibnamefont {Nakamura}}, \bibinfo {author} {\bibfnamefont {Y.}~\bibnamefont {Yoshimoto}}, \bibinfo {author} {\bibfnamefont {R.}~\bibnamefont {Arita}}, \bibinfo {author} {\bibfnamefont {S.}~\bibnamefont {Tsuneyuki}},\ and\ \bibinfo {author} {\bibfnamefont {M.}~\bibnamefont {Imada}},\ }\bibfield  {title} {\bibinfo {title} {{Optical absorption study by ab initio downfolding approach: Application to GaAs}},\ }\href {https://doi.org/10.1103/PhysRevB.77.195126} {\bibfield  {journal} {\bibinfo  {journal} {Physical Review B - Condensed Matter and Materials Physics}\ }\textbf {\bibinfo {volume} {77}},\ \bibinfo {pages} {1} (\bibinfo {year} {2008})},\ \Eprint {https://arxiv.org/abs/0710.4371} {arXiv:0710.4371} \BibitemShut {NoStop}%
\bibitem [{\citenamefont {Nakamura}\ \emph {et~al.}(2010)\citenamefont {Nakamura}, \citenamefont {Yoshimoto}, \citenamefont {Nohara},\ and\ \citenamefont {Imada}}]{Nakamura2010}%
  \BibitemOpen
  \bibfield  {author} {\bibinfo {author} {\bibfnamefont {K.}~\bibnamefont {Nakamura}}, \bibinfo {author} {\bibfnamefont {Y.}~\bibnamefont {Yoshimoto}}, \bibinfo {author} {\bibfnamefont {Y.}~\bibnamefont {Nohara}},\ and\ \bibinfo {author} {\bibfnamefont {M.}~\bibnamefont {Imada}},\ }\bibfield  {title} {\bibinfo {title} {{Ab initio low-dimensional physics opened up by dimensional downfolding: Application to LaFeAsO}},\ }\href {https://doi.org/10.1143/JPSJ.79.123708} {\bibfield  {journal} {\bibinfo  {journal} {Journal of the Physical Society of Japan}\ }\textbf {\bibinfo {volume} {79}},\ \bibinfo {pages} {12} (\bibinfo {year} {2010})},\ \Eprint {https://arxiv.org/abs/1007.4429} {1007.4429} \BibitemShut {NoStop}%
\bibitem [{\citenamefont {Arita}\ \emph {et~al.}(2015)\citenamefont {Arita}, \citenamefont {Ikeda}, \citenamefont {Sakai},\ and\ \citenamefont {Suzuki}}]{Arita2015}%
  \BibitemOpen
  \bibfield  {author} {\bibinfo {author} {\bibfnamefont {R.}~\bibnamefont {Arita}}, \bibinfo {author} {\bibfnamefont {H.}~\bibnamefont {Ikeda}}, \bibinfo {author} {\bibfnamefont {S.}~\bibnamefont {Sakai}},\ and\ \bibinfo {author} {\bibfnamefont {M.~T.}\ \bibnamefont {Suzuki}},\ }\bibfield  {title} {\bibinfo {title} {{Ab initio downfolding study of the iron-based ladder superconductor $\text{BaFe}_2\text{S}_3$}},\ }\href {https://doi.org/10.1103/PhysRevB.92.054515} {\bibfield  {journal} {\bibinfo  {journal} {Physical Review B - Condensed Matter and Materials Physics}\ }\textbf {\bibinfo {volume} {92}},\ \bibinfo {pages} {1} (\bibinfo {year} {2015})},\ \Eprint {https://arxiv.org/abs/1507.05715} {1507.05715} \BibitemShut {NoStop}%
\bibitem [{\citenamefont {Zheng}\ \emph {et~al.}(2018)\citenamefont {Zheng}, \citenamefont {Changlani}, \citenamefont {Williams}, \citenamefont {Busemeyer},\ and\ \citenamefont {Wagner}}]{Zheng2018}%
  \BibitemOpen
  \bibfield  {author} {\bibinfo {author} {\bibfnamefont {H.}~\bibnamefont {Zheng}}, \bibinfo {author} {\bibfnamefont {H.~J.}\ \bibnamefont {Changlani}}, \bibinfo {author} {\bibfnamefont {K.~T.}\ \bibnamefont {Williams}}, \bibinfo {author} {\bibfnamefont {B.}~\bibnamefont {Busemeyer}},\ and\ \bibinfo {author} {\bibfnamefont {L.~K.}\ \bibnamefont {Wagner}},\ }\bibfield  {title} {\bibinfo {title} {{From real materials to model Hamiltonians with density matrix downfolding}},\ }\href {https://doi.org/10.3389/fphy.2018.00043} {\bibfield  {journal} {\bibinfo  {journal} {Frontiers in Physics}\ }\textbf {\bibinfo {volume} {6}},\ \bibinfo {pages} {1} (\bibinfo {year} {2018})},\ \Eprint {https://arxiv.org/abs/1712.00477} {1712.00477} \BibitemShut {NoStop}%
\bibitem [{\citenamefont {Pham}\ \emph {et~al.}(2020)\citenamefont {Pham}, \citenamefont {Hermes},\ and\ \citenamefont {Gagliardi}}]{Pham2020}%
  \BibitemOpen
  \bibfield  {author} {\bibinfo {author} {\bibfnamefont {H.~Q.}\ \bibnamefont {Pham}}, \bibinfo {author} {\bibfnamefont {M.~R.}\ \bibnamefont {Hermes}},\ and\ \bibinfo {author} {\bibfnamefont {L.}~\bibnamefont {Gagliardi}},\ }\bibfield  {title} {\bibinfo {title} {{Periodic Electronic Structure Calculations with the Density Matrix Embedding Theory}},\ }\href {https://doi.org/10.1021/acs.jctc.9b00939} {\bibfield  {journal} {\bibinfo  {journal} {Journal of Chemical Theory and Computation}\ }\textbf {\bibinfo {volume} {16}},\ \bibinfo {pages} {130} (\bibinfo {year} {2020})}\BibitemShut {NoStop}%
\bibitem [{\citenamefont {Bauman}\ and\ \citenamefont {Kowalski}(2022)}]{Bauman2022}%
  \BibitemOpen
  \bibfield  {author} {\bibinfo {author} {\bibfnamefont {N.~P.}\ \bibnamefont {Bauman}}\ and\ \bibinfo {author} {\bibfnamefont {K.}~\bibnamefont {Kowalski}},\ }\bibfield  {title} {\bibinfo {title} {{Coupled Cluster Downfolding Theory: towards universal many-body algorithms for dimensionality reduction of composite quantum systems in chemistry and materials science}},\ }\href {https://doi.org/10.1186/s41313-022-00046-8} {\bibfield  {journal} {\bibinfo  {journal} {Materials Theory}\ }\textbf {\bibinfo {volume} {6}},\ \bibinfo {pages} {17} (\bibinfo {year} {2022})}\BibitemShut {NoStop}%
\bibitem [{\citenamefont {Weisburn}\ \emph {et~al.}(2024)\citenamefont {Weisburn}, \citenamefont {Cho}, \citenamefont {Bensberg}, \citenamefont {Meitei}, \citenamefont {Reiher},\ and\ \citenamefont {Voorhis}}]{weisburn2024multiscaleembeddingquantumcomputing}%
  \BibitemOpen
  \bibfield  {author} {\bibinfo {author} {\bibfnamefont {L.~P.}\ \bibnamefont {Weisburn}}, \bibinfo {author} {\bibfnamefont {M.}~\bibnamefont {Cho}}, \bibinfo {author} {\bibfnamefont {M.}~\bibnamefont {Bensberg}}, \bibinfo {author} {\bibfnamefont {O.~R.}\ \bibnamefont {Meitei}}, \bibinfo {author} {\bibfnamefont {M.}~\bibnamefont {Reiher}},\ and\ \bibinfo {author} {\bibfnamefont {T.~V.}\ \bibnamefont {Voorhis}},\ }\href {https://arxiv.org/abs/2409.06813} {\bibinfo {title} {Multiscale embedding for quantum computing}} (\bibinfo {year} {2024}),\ \Eprint {https://arxiv.org/abs/2409.06813} {arXiv:2409.06813 [physics.chem-ph]} \BibitemShut {NoStop}%
\bibitem [{\citenamefont {Aryasetiawan}\ \emph {et~al.}(2004)\citenamefont {Aryasetiawan}, \citenamefont {Imada}, \citenamefont {Georges}, \citenamefont {Kotliar}, \citenamefont {Biermann},\ and\ \citenamefont {Lichtenstein}}]{PhysRevB.70.195104}%
  \BibitemOpen
  \bibfield  {author} {\bibinfo {author} {\bibfnamefont {F.}~\bibnamefont {Aryasetiawan}}, \bibinfo {author} {\bibfnamefont {M.}~\bibnamefont {Imada}}, \bibinfo {author} {\bibfnamefont {A.}~\bibnamefont {Georges}}, \bibinfo {author} {\bibfnamefont {G.}~\bibnamefont {Kotliar}}, \bibinfo {author} {\bibfnamefont {S.}~\bibnamefont {Biermann}},\ and\ \bibinfo {author} {\bibfnamefont {A.~I.}\ \bibnamefont {Lichtenstein}},\ }\bibfield  {title} {\bibinfo {title} {Frequency-dependent local interactions and low-energy effective models from electronic structure calculations},\ }\href {https://doi.org/10.1103/PhysRevB.70.195104} {\bibfield  {journal} {\bibinfo  {journal} {Phys. Rev. B}\ }\textbf {\bibinfo {volume} {70}},\ \bibinfo {pages} {195104} (\bibinfo {year} {2004})}\BibitemShut {NoStop}%
\bibitem [{\citenamefont {Muechler}\ \emph {et~al.}(2022)\citenamefont {Muechler}, \citenamefont {Badrtdinov}, \citenamefont {Hampel}, \citenamefont {Cano}, \citenamefont {R\"osner},\ and\ \citenamefont {Dreyer}}]{PhysRevB.105.235104}%
  \BibitemOpen
  \bibfield  {author} {\bibinfo {author} {\bibfnamefont {L.}~\bibnamefont {Muechler}}, \bibinfo {author} {\bibfnamefont {D.~I.}\ \bibnamefont {Badrtdinov}}, \bibinfo {author} {\bibfnamefont {A.}~\bibnamefont {Hampel}}, \bibinfo {author} {\bibfnamefont {J.}~\bibnamefont {Cano}}, \bibinfo {author} {\bibfnamefont {M.}~\bibnamefont {R\"osner}},\ and\ \bibinfo {author} {\bibfnamefont {C.~E.}\ \bibnamefont {Dreyer}},\ }\bibfield  {title} {\bibinfo {title} {Quantum embedding methods for correlated excited states of point defects: Case studies and challenges},\ }\href {https://doi.org/10.1103/PhysRevB.105.235104} {\bibfield  {journal} {\bibinfo  {journal} {Phys. Rev. B}\ }\textbf {\bibinfo {volume} {105}},\ \bibinfo {pages} {235104} (\bibinfo {year} {2022})}\BibitemShut {NoStop}%
\bibitem [{\citenamefont {Chang}\ \emph {et~al.}(2024)\citenamefont {Chang}, \citenamefont {van Loon}, \citenamefont {Eskridge}, \citenamefont {Busemeyer}, \citenamefont {Morales}, \citenamefont {Dreyer}, \citenamefont {Millis}, \citenamefont {Zhang}, \citenamefont {Wehling}, \citenamefont {Wagner},\ and\ \citenamefont {R{\"{o}}sner}}]{Chang2024}%
  \BibitemOpen
  \bibfield  {author} {\bibinfo {author} {\bibfnamefont {Y.}~\bibnamefont {Chang}}, \bibinfo {author} {\bibfnamefont {E.~G.}\ \bibnamefont {van Loon}}, \bibinfo {author} {\bibfnamefont {B.}~\bibnamefont {Eskridge}}, \bibinfo {author} {\bibfnamefont {B.}~\bibnamefont {Busemeyer}}, \bibinfo {author} {\bibfnamefont {M.~A.}\ \bibnamefont {Morales}}, \bibinfo {author} {\bibfnamefont {C.~E.}\ \bibnamefont {Dreyer}}, \bibinfo {author} {\bibfnamefont {A.~J.}\ \bibnamefont {Millis}}, \bibinfo {author} {\bibfnamefont {S.}~\bibnamefont {Zhang}}, \bibinfo {author} {\bibfnamefont {T.~O.}\ \bibnamefont {Wehling}}, \bibinfo {author} {\bibfnamefont {L.~K.}\ \bibnamefont {Wagner}},\ and\ \bibinfo {author} {\bibfnamefont {M.}~\bibnamefont {R{\"{o}}sner}},\ }\bibfield  {title} {\bibinfo {title} {{Downfolding from ab initio to interacting model Hamiltonians: comprehensive analysis and benchmarking of the DFT+cRPA approach}},\ }\href {https://doi.org/10.1038/s41524-024-01314-6} {\bibfield  {journal} {\bibinfo  {journal} {npj
  Computational Materials}\ }\textbf {\bibinfo {volume} {10}},\ \bibinfo {pages} {129} (\bibinfo {year} {2024})}\BibitemShut {NoStop}%
\bibitem [{\citenamefont {Yoshimi}\ \emph {et~al.}(2021)\citenamefont {Yoshimi}, \citenamefont {Tsumuraya},\ and\ \citenamefont {Misawa}}]{Yoshimi2021}%
  \BibitemOpen
  \bibfield  {author} {\bibinfo {author} {\bibfnamefont {K.}~\bibnamefont {Yoshimi}}, \bibinfo {author} {\bibfnamefont {T.}~\bibnamefont {Tsumuraya}},\ and\ \bibinfo {author} {\bibfnamefont {T.}~\bibnamefont {Misawa}},\ }\bibfield  {title} {\bibinfo {title} {{Ab initio derivation and exact diagonalization analysis of low-energy effective Hamiltonians for $\beta$- X[Pd(dmit)2]2}},\ }\href {https://doi.org/10.1103/PhysRevResearch.3.043224} {\bibfield  {journal} {\bibinfo  {journal} {Physical Review Research}\ }\textbf {\bibinfo {volume} {3}},\ \bibinfo {pages} {1} (\bibinfo {year} {2021})},\ \Eprint {https://arxiv.org/abs/2109.10542} {2109.10542} \BibitemShut {NoStop}%
\bibitem [{\citenamefont {Botzung}\ and\ \citenamefont {Nataf}(2024)}]{Botzung2024}%
  \BibitemOpen
  \bibfield  {author} {\bibinfo {author} {\bibfnamefont {T.}~\bibnamefont {Botzung}}\ and\ \bibinfo {author} {\bibfnamefont {P.}~\bibnamefont {Nataf}},\ }\bibfield  {title} {\bibinfo {title} {{Exact Diagonalization of SU (N) Fermi-Hubbard Models}},\ }\href {https://doi.org/10.1103/PhysRevLett.132.153001} {\bibfield  {journal} {\bibinfo  {journal} {Physical Review Letters}\ }\textbf {\bibinfo {volume} {132}},\ \bibinfo {pages} {153001} (\bibinfo {year} {2024})}\BibitemShut {NoStop}%
\bibitem [{\citenamefont {Ma}\ \emph {et~al.}(2015)\citenamefont {Ma}, \citenamefont {Purwanto}, \citenamefont {Zhang},\ and\ \citenamefont {Krakauer}}]{Ma2015}%
  \BibitemOpen
  \bibfield  {author} {\bibinfo {author} {\bibfnamefont {F.}~\bibnamefont {Ma}}, \bibinfo {author} {\bibfnamefont {W.}~\bibnamefont {Purwanto}}, \bibinfo {author} {\bibfnamefont {S.}~\bibnamefont {Zhang}},\ and\ \bibinfo {author} {\bibfnamefont {H.}~\bibnamefont {Krakauer}},\ }\bibfield  {title} {\bibinfo {title} {{Quantum monte carlo calculations in solids with downfolded hamiltonians}},\ }\href {https://doi.org/10.1103/PhysRevLett.114.226401} {\bibfield  {journal} {\bibinfo  {journal} {Physical Review Letters}\ }\textbf {\bibinfo {volume} {114}},\ \bibinfo {pages} {1} (\bibinfo {year} {2015})},\ \Eprint {https://arxiv.org/abs/1412.0322} {arXiv:1412.0322} \BibitemShut {NoStop}%
\bibitem [{\citenamefont {Yoshida}\ \emph {et~al.}(2024)\citenamefont {Yoshida}, \citenamefont {Takemori},\ and\ \citenamefont {Mizukami}}]{10.1063/5.0213525}%
  \BibitemOpen
  \bibfield  {author} {\bibinfo {author} {\bibfnamefont {Y.}~\bibnamefont {Yoshida}}, \bibinfo {author} {\bibfnamefont {N.}~\bibnamefont {Takemori}},\ and\ \bibinfo {author} {\bibfnamefont {W.}~\bibnamefont {Mizukami}},\ }\bibfield  {title} {\bibinfo {title} {{{Ab initio extended Hubbard model of short polyenes for efficient quantum computing}}},\ }\href {https://doi.org/10.1063/5.0213525} {\bibfield  {journal} {\bibinfo  {journal} {The Journal of Chemical Physics}\ }\textbf {\bibinfo {volume} {161}},\ \bibinfo {pages} {084303} (\bibinfo {year} {2024})}\BibitemShut {NoStop}%
\bibitem [{\citenamefont {Alvertis}\ \emph {et~al.}(2024{\natexlab{a}})\citenamefont {Alvertis}, \citenamefont {Khan},\ and\ \citenamefont {Tubman}}]{alvertis2024downfolding}%
  \BibitemOpen
  \bibfield  {author} {\bibinfo {author} {\bibfnamefont {A.~M.}\ \bibnamefont {Alvertis}}, \bibinfo {author} {\bibfnamefont {A.}~\bibnamefont {Khan}},\ and\ \bibinfo {author} {\bibfnamefont {N.~M.}\ \bibnamefont {Tubman}},\ }\href {https://arxiv.org/abs/2409.12237} {\bibinfo {title} {Ground states of strongly-correlated materials on quantum computers using ab initio downfolding}} (\bibinfo {year} {2024}{\natexlab{a}}),\ \Eprint {https://arxiv.org/abs/2409.12237} {arXiv:2409.12237 [quant-ph]} \BibitemShut {NoStop}%
\bibitem [{\citenamefont {Hirayama}\ \emph {et~al.}(2019)\citenamefont {Hirayama}, \citenamefont {Misawa}, \citenamefont {Ohgoe}, \citenamefont {Yamaji},\ and\ \citenamefont {Imada}}]{Hirayama2019}%
  \BibitemOpen
  \bibfield  {author} {\bibinfo {author} {\bibfnamefont {M.}~\bibnamefont {Hirayama}}, \bibinfo {author} {\bibfnamefont {T.}~\bibnamefont {Misawa}}, \bibinfo {author} {\bibfnamefont {T.}~\bibnamefont {Ohgoe}}, \bibinfo {author} {\bibfnamefont {Y.}~\bibnamefont {Yamaji}},\ and\ \bibinfo {author} {\bibfnamefont {M.}~\bibnamefont {Imada}},\ }\bibfield  {title} {\bibinfo {title} {{Effective Hamiltonian for cuprate superconductors derived from multiscale ab initio scheme with level renormalization}},\ }\href {https://doi.org/10.1103/PhysRevB.99.245155} {\bibfield  {journal} {\bibinfo  {journal} {Physical Review B}\ }\textbf {\bibinfo {volume} {99}},\ \bibinfo {pages} {1} (\bibinfo {year} {2019})},\ \Eprint {https://arxiv.org/abs/1901.00763} {1901.00763} \BibitemShut {NoStop}%
\bibitem [{\citenamefont {Ohgoe}\ \emph {et~al.}(2020)\citenamefont {Ohgoe}, \citenamefont {Hirayama}, \citenamefont {Misawa}, \citenamefont {Ido}, \citenamefont {Yamaji},\ and\ \citenamefont {Imada}}]{Ohgoe2020}%
  \BibitemOpen
  \bibfield  {author} {\bibinfo {author} {\bibfnamefont {T.}~\bibnamefont {Ohgoe}}, \bibinfo {author} {\bibfnamefont {M.}~\bibnamefont {Hirayama}}, \bibinfo {author} {\bibfnamefont {T.}~\bibnamefont {Misawa}}, \bibinfo {author} {\bibfnamefont {K.}~\bibnamefont {Ido}}, \bibinfo {author} {\bibfnamefont {Y.}~\bibnamefont {Yamaji}},\ and\ \bibinfo {author} {\bibfnamefont {M.}~\bibnamefont {Imada}},\ }\bibfield  {title} {\bibinfo {title} {{Ab initio study of superconductivity and inhomogeneity in a Hg-based cuprate superconductor}},\ }\href {https://doi.org/10.1103/PhysRevB.101.045124} {\bibfield  {journal} {\bibinfo  {journal} {Physical Review B}\ }\textbf {\bibinfo {volume} {101}},\ \bibinfo {pages} {1} (\bibinfo {year} {2020})},\ \Eprint {https://arxiv.org/abs/1902.00122} {arXiv:1902.00122} \BibitemShut {NoStop}%
\bibitem [{\citenamefont {Been}\ \emph {et~al.}(2021)\citenamefont {Been}, \citenamefont {Lee}, \citenamefont {Hwang}, \citenamefont {Cui}, \citenamefont {Zaanen}, \citenamefont {Devereaux}, \citenamefont {Moritz},\ and\ \citenamefont {Jia}}]{Been2021}%
  \BibitemOpen
  \bibfield  {author} {\bibinfo {author} {\bibfnamefont {E.}~\bibnamefont {Been}}, \bibinfo {author} {\bibfnamefont {W.~S.}\ \bibnamefont {Lee}}, \bibinfo {author} {\bibfnamefont {H.~Y.}\ \bibnamefont {Hwang}}, \bibinfo {author} {\bibfnamefont {Y.}~\bibnamefont {Cui}}, \bibinfo {author} {\bibfnamefont {J.}~\bibnamefont {Zaanen}}, \bibinfo {author} {\bibfnamefont {T.}~\bibnamefont {Devereaux}}, \bibinfo {author} {\bibfnamefont {B.}~\bibnamefont {Moritz}},\ and\ \bibinfo {author} {\bibfnamefont {C.}~\bibnamefont {Jia}},\ }\bibfield  {title} {\bibinfo {title} {{Electronic Structure Trends across the Rare-Earth Series in Superconducting Infinite-Layer Nickelates}},\ }\href {https://doi.org/10.1103/PhysRevX.11.011050} {\bibfield  {journal} {\bibinfo  {journal} {Physical Review X}\ }\textbf {\bibinfo {volume} {11}},\ \bibinfo {pages} {11050} (\bibinfo {year} {2021})},\ \Eprint {https://arxiv.org/abs/2002.12300} {arXiv:2002.12300} \BibitemShut {NoStop}%
\bibitem [{\citenamefont {Schmid}\ \emph {et~al.}(2023)\citenamefont {Schmid}, \citenamefont {Mor{\'{e}}e}, \citenamefont {Kaneko}, \citenamefont {Yamaji},\ and\ \citenamefont {Imada}}]{Schmid2023}%
  \BibitemOpen
  \bibfield  {author} {\bibinfo {author} {\bibfnamefont {M.~T.}\ \bibnamefont {Schmid}}, \bibinfo {author} {\bibfnamefont {J.~B.}\ \bibnamefont {Mor{\'{e}}e}}, \bibinfo {author} {\bibfnamefont {R.}~\bibnamefont {Kaneko}}, \bibinfo {author} {\bibfnamefont {Y.}~\bibnamefont {Yamaji}},\ and\ \bibinfo {author} {\bibfnamefont {M.}~\bibnamefont {Imada}},\ }\bibfield  {title} {\bibinfo {title} {{Superconductivity Studied by Solving Ab Initio Low-Energy Effective Hamiltonians for Carrier Doped $\text{CaCuO}_2$, $\text{Bi}_2\text{Sr}_2\text{CuO}_6$, $\text{Bi}_2\text{Sr}_2\text{CaCu}_2\text{O}_8$, and $\text{HgBa}_2\text{CuO}_4$}},\ }\href {https://doi.org/10.1103/PhysRevX.13.041036} {\bibfield  {journal} {\bibinfo  {journal} {Physical Review X}\ }\textbf {\bibinfo {volume} {13}},\ \bibinfo {pages} {1} (\bibinfo {year} {2023})}\BibitemShut {NoStop}%
\bibitem [{\citenamefont {Mazza}\ \emph {et~al.}(2020)\citenamefont {Mazza}, \citenamefont {R\"osner}, \citenamefont {Windg\"atter}, \citenamefont {Latini}, \citenamefont {H\"ubener}, \citenamefont {Millis}, \citenamefont {Rubio},\ and\ \citenamefont {Georges}}]{PhysRevLett.124.197601}%
  \BibitemOpen
  \bibfield  {author} {\bibinfo {author} {\bibfnamefont {G.}~\bibnamefont {Mazza}}, \bibinfo {author} {\bibfnamefont {M.}~\bibnamefont {R\"osner}}, \bibinfo {author} {\bibfnamefont {L.}~\bibnamefont {Windg\"atter}}, \bibinfo {author} {\bibfnamefont {S.}~\bibnamefont {Latini}}, \bibinfo {author} {\bibfnamefont {H.}~\bibnamefont {H\"ubener}}, \bibinfo {author} {\bibfnamefont {A.~J.}\ \bibnamefont {Millis}}, \bibinfo {author} {\bibfnamefont {A.}~\bibnamefont {Rubio}},\ and\ \bibinfo {author} {\bibfnamefont {A.}~\bibnamefont {Georges}},\ }\bibfield  {title} {\bibinfo {title} {{Nature of Symmetry Breaking at the Excitonic Insulator Transition: ${\mathrm{Ta}}_{2}{\mathrm{NiSe}}_{5}$}},\ }\href {https://doi.org/10.1103/PhysRevLett.124.197601} {\bibfield  {journal} {\bibinfo  {journal} {Phys. Rev. Lett.}\ }\textbf {\bibinfo {volume} {124}},\ \bibinfo {pages} {197601} (\bibinfo {year} {2020})}\BibitemShut {NoStop}%
\bibitem [{\citenamefont {Yoshimi}\ \emph {et~al.}(2023)\citenamefont {Yoshimi}, \citenamefont {Misawa}, \citenamefont {Tsumuraya},\ and\ \citenamefont {Seo}}]{Yoshimi2023}%
  \BibitemOpen
  \bibfield  {author} {\bibinfo {author} {\bibfnamefont {K.}~\bibnamefont {Yoshimi}}, \bibinfo {author} {\bibfnamefont {T.}~\bibnamefont {Misawa}}, \bibinfo {author} {\bibfnamefont {T.}~\bibnamefont {Tsumuraya}},\ and\ \bibinfo {author} {\bibfnamefont {H.}~\bibnamefont {Seo}},\ }\bibfield  {title} {\bibinfo {title} {{Comprehensive Ab Initio Investigation of the Phase Diagram of Quasi-One-Dimensional Molecular Solids}},\ }\href {https://doi.org/10.1103/PhysRevLett.131.036401} {\bibfield  {journal} {\bibinfo  {journal} {Physical Review Letters}\ }\textbf {\bibinfo {volume} {131}},\ \bibinfo {pages} {36401} (\bibinfo {year} {2023})}\BibitemShut {NoStop}%
\bibitem [{\citenamefont {Schobert}\ \emph {et~al.}(2024{\natexlab{a}})\citenamefont {Schobert}, \citenamefont {Berges}, \citenamefont {van Loon}, \citenamefont {Sentef}, \citenamefont {Brener}, \citenamefont {Rossi},\ and\ \citenamefont {Wehling}}]{Schobert2024}%
  \BibitemOpen
  \bibfield  {author} {\bibinfo {author} {\bibfnamefont {A.}~\bibnamefont {Schobert}}, \bibinfo {author} {\bibfnamefont {J.}~\bibnamefont {Berges}}, \bibinfo {author} {\bibfnamefont {E.~G.}\ \bibnamefont {van Loon}}, \bibinfo {author} {\bibfnamefont {M.~A.}\ \bibnamefont {Sentef}}, \bibinfo {author} {\bibfnamefont {S.}~\bibnamefont {Brener}}, \bibinfo {author} {\bibfnamefont {M.}~\bibnamefont {Rossi}},\ and\ \bibinfo {author} {\bibfnamefont {T.~O.}\ \bibnamefont {Wehling}},\ }\bibfield  {title} {\bibinfo {title} {{Ab initio electron-lattice downfolding: Potential energy landscapes, anharmonicity, and molecular dynamics in charge density wave materials}},\ }\href {https://doi.org/10.21468/SciPostPhys.16.2.046} {\bibfield  {journal} {\bibinfo  {journal} {SciPost Physics}\ }\textbf {\bibinfo {volume} {16}},\ \bibinfo {pages} {1} (\bibinfo {year} {2024}{\natexlab{a}})},\ \Eprint {https://arxiv.org/abs/2303.07261} {arXiv:2303.07261} \BibitemShut {NoStop}%
\bibitem [{\citenamefont {Giustino}\ \emph {et~al.}(2010)\citenamefont {Giustino}, \citenamefont {Louie},\ and\ \citenamefont {Cohen}}]{PhysRevLett.105.265501}%
  \BibitemOpen
  \bibfield  {author} {\bibinfo {author} {\bibfnamefont {F.}~\bibnamefont {Giustino}}, \bibinfo {author} {\bibfnamefont {S.~G.}\ \bibnamefont {Louie}},\ and\ \bibinfo {author} {\bibfnamefont {M.~L.}\ \bibnamefont {Cohen}},\ }\bibfield  {title} {\bibinfo {title} {Electron-phonon renormalization of the direct band gap of diamond},\ }\href {https://doi.org/10.1103/PhysRevLett.105.265501} {\bibfield  {journal} {\bibinfo  {journal} {Phys. Rev. Lett.}\ }\textbf {\bibinfo {volume} {105}},\ \bibinfo {pages} {265501} (\bibinfo {year} {2010})}\BibitemShut {NoStop}%
\bibitem [{\citenamefont {Antonius}\ \emph {et~al.}(2014)\citenamefont {Antonius}, \citenamefont {Ponc\'e}, \citenamefont {Boulanger}, \citenamefont {C\^ot\'e},\ and\ \citenamefont {Gonze}}]{PhysRevLett.112.215501}%
  \BibitemOpen
  \bibfield  {author} {\bibinfo {author} {\bibfnamefont {G.}~\bibnamefont {Antonius}}, \bibinfo {author} {\bibfnamefont {S.}~\bibnamefont {Ponc\'e}}, \bibinfo {author} {\bibfnamefont {P.}~\bibnamefont {Boulanger}}, \bibinfo {author} {\bibfnamefont {M.}~\bibnamefont {C\^ot\'e}},\ and\ \bibinfo {author} {\bibfnamefont {X.}~\bibnamefont {Gonze}},\ }\bibfield  {title} {\bibinfo {title} {Many-body effects on the zero-point renormalization of the band structure},\ }\href {https://doi.org/10.1103/PhysRevLett.112.215501} {\bibfield  {journal} {\bibinfo  {journal} {Phys. Rev. Lett.}\ }\textbf {\bibinfo {volume} {112}},\ \bibinfo {pages} {215501} (\bibinfo {year} {2014})}\BibitemShut {NoStop}%
\bibitem [{\citenamefont {Monserrat}\ \emph {et~al.}(2015)\citenamefont {Monserrat}, \citenamefont {Engel},\ and\ \citenamefont {Needs}}]{PhysRevB.92.140302}%
  \BibitemOpen
  \bibfield  {author} {\bibinfo {author} {\bibfnamefont {B.}~\bibnamefont {Monserrat}}, \bibinfo {author} {\bibfnamefont {E.~A.}\ \bibnamefont {Engel}},\ and\ \bibinfo {author} {\bibfnamefont {R.~J.}\ \bibnamefont {Needs}},\ }\bibfield  {title} {\bibinfo {title} {Giant electron-phonon interactions in molecular crystals and the importance of nonquadratic coupling},\ }\href {https://doi.org/10.1103/PhysRevB.92.140302} {\bibfield  {journal} {\bibinfo  {journal} {Phys. Rev. B}\ }\textbf {\bibinfo {volume} {92}},\ \bibinfo {pages} {140302} (\bibinfo {year} {2015})}\BibitemShut {NoStop}%
\bibitem [{\citenamefont {Marini}(2008)}]{Marini2008}%
  \BibitemOpen
  \bibfield  {author} {\bibinfo {author} {\bibfnamefont {A.}~\bibnamefont {Marini}},\ }\bibfield  {title} {\bibinfo {title} {{Ab initio finite-temperature excitons}},\ }\href {https://doi.org/10.1103/PhysRevLett.101.106405} {\bibfield  {journal} {\bibinfo  {journal} {Physical Review Letters}\ }\textbf {\bibinfo {volume} {101}},\ \bibinfo {pages} {1} (\bibinfo {year} {2008})},\ \Eprint {https://arxiv.org/abs/0712.3365} {0712.3365} \BibitemShut {NoStop}%
\bibitem [{\citenamefont {Alvertis}\ \emph {et~al.}(2020)\citenamefont {Alvertis}, \citenamefont {Pandya}, \citenamefont {Muscarella}, \citenamefont {Sawhney}, \citenamefont {Nguyen}, \citenamefont {Ehrler}, \citenamefont {Rao}, \citenamefont {Friend}, \citenamefont {Chin},\ and\ \citenamefont {Monserrat}}]{PhysRevB.102.081122}%
  \BibitemOpen
  \bibfield  {author} {\bibinfo {author} {\bibfnamefont {A.~M.}\ \bibnamefont {Alvertis}}, \bibinfo {author} {\bibfnamefont {R.}~\bibnamefont {Pandya}}, \bibinfo {author} {\bibfnamefont {L.~A.}\ \bibnamefont {Muscarella}}, \bibinfo {author} {\bibfnamefont {N.}~\bibnamefont {Sawhney}}, \bibinfo {author} {\bibfnamefont {M.}~\bibnamefont {Nguyen}}, \bibinfo {author} {\bibfnamefont {B.}~\bibnamefont {Ehrler}}, \bibinfo {author} {\bibfnamefont {A.}~\bibnamefont {Rao}}, \bibinfo {author} {\bibfnamefont {R.~H.}\ \bibnamefont {Friend}}, \bibinfo {author} {\bibfnamefont {A.~W.}\ \bibnamefont {Chin}},\ and\ \bibinfo {author} {\bibfnamefont {B.}~\bibnamefont {Monserrat}},\ }\bibfield  {title} {\bibinfo {title} {Impact of exciton delocalization on exciton-vibration interactions in organic semiconductors},\ }\href {https://doi.org/10.1103/PhysRevB.102.081122} {\bibfield  {journal} {\bibinfo  {journal} {Phys. Rev. B}\ }\textbf {\bibinfo {volume} {102}},\ \bibinfo {pages} {081122} (\bibinfo {year} {2020})}\BibitemShut
  {NoStop}%
\bibitem [{\citenamefont {Huang}\ \emph {et~al.}(2021)\citenamefont {Huang}, \citenamefont {Zacharias}, \citenamefont {Lewis}, \citenamefont {Giustino},\ and\ \citenamefont {Sharifzadeh}}]{Huang2021}%
  \BibitemOpen
  \bibfield  {author} {\bibinfo {author} {\bibfnamefont {T.~A.}\ \bibnamefont {Huang}}, \bibinfo {author} {\bibfnamefont {M.}~\bibnamefont {Zacharias}}, \bibinfo {author} {\bibfnamefont {D.~K.}\ \bibnamefont {Lewis}}, \bibinfo {author} {\bibfnamefont {F.}~\bibnamefont {Giustino}},\ and\ \bibinfo {author} {\bibfnamefont {S.}~\bibnamefont {Sharifzadeh}},\ }\bibfield  {title} {\bibinfo {title} {{Exciton-Phonon Interactions in Monolayer Germanium Selenide from First Principles}},\ }\href {https://doi.org/10.1021/acs.jpclett.1c00264} {\bibfield  {journal} {\bibinfo  {journal} {The Journal of Physical Chemistry Letters}\ }\textbf {\bibinfo {volume} {12}},\ \bibinfo {pages} {3802} (\bibinfo {year} {2021})}\BibitemShut {NoStop}%
\bibitem [{\citenamefont {Bishop}\ and\ \citenamefont {Overhauser}(1981)}]{PhysRevB.23.3627}%
  \BibitemOpen
  \bibfield  {author} {\bibinfo {author} {\bibfnamefont {M.~F.}\ \bibnamefont {Bishop}}\ and\ \bibinfo {author} {\bibfnamefont {A.~W.}\ \bibnamefont {Overhauser}},\ }\bibfield  {title} {\bibinfo {title} {Phonon-mediated electron-electron interaction in real space},\ }\href {https://doi.org/10.1103/PhysRevB.23.3627} {\bibfield  {journal} {\bibinfo  {journal} {Phys. Rev. B}\ }\textbf {\bibinfo {volume} {23}},\ \bibinfo {pages} {3627} (\bibinfo {year} {1981})}\BibitemShut {NoStop}%
\bibitem [{\citenamefont {Tse}\ and\ \citenamefont {Das~Sarma}(2007)}]{PhysRevLett.99.236802}%
  \BibitemOpen
  \bibfield  {author} {\bibinfo {author} {\bibfnamefont {W.-K.}\ \bibnamefont {Tse}}\ and\ \bibinfo {author} {\bibfnamefont {S.}~\bibnamefont {Das~Sarma}},\ }\bibfield  {title} {\bibinfo {title} {Phonon-induced many-body renormalization of the electronic properties of graphene},\ }\href {https://doi.org/10.1103/PhysRevLett.99.236802} {\bibfield  {journal} {\bibinfo  {journal} {Phys. Rev. Lett.}\ }\textbf {\bibinfo {volume} {99}},\ \bibinfo {pages} {236802} (\bibinfo {year} {2007})}\BibitemShut {NoStop}%
\bibitem [{\citenamefont {Nomura}\ and\ \citenamefont {Arita}(2015)}]{Nomura2015}%
  \BibitemOpen
  \bibfield  {author} {\bibinfo {author} {\bibfnamefont {Y.}~\bibnamefont {Nomura}}\ and\ \bibinfo {author} {\bibfnamefont {R.}~\bibnamefont {Arita}},\ }\bibfield  {title} {\bibinfo {title} {{Ab initio downfolding for electron-phonon-coupled systems: Constrained density-functional perturbation theory}},\ }\href {https://doi.org/10.1103/PhysRevB.92.245108} {\bibfield  {journal} {\bibinfo  {journal} {Physical Review B - Condensed Matter and Materials Physics}\ }\textbf {\bibinfo {volume} {92}},\ \bibinfo {pages} {1} (\bibinfo {year} {2015})},\ \Eprint {https://arxiv.org/abs/1509.01138} {arXiv:1509.01138} \BibitemShut {NoStop}%
\bibitem [{\citenamefont {Giovannetti}\ \emph {et~al.}(2014)\citenamefont {Giovannetti}, \citenamefont {Casula}, \citenamefont {Werner}, \citenamefont {Mauri},\ and\ \citenamefont {Capone}}]{PhysRevB.90.115435}%
  \BibitemOpen
  \bibfield  {author} {\bibinfo {author} {\bibfnamefont {G.}~\bibnamefont {Giovannetti}}, \bibinfo {author} {\bibfnamefont {M.}~\bibnamefont {Casula}}, \bibinfo {author} {\bibfnamefont {P.}~\bibnamefont {Werner}}, \bibinfo {author} {\bibfnamefont {F.}~\bibnamefont {Mauri}},\ and\ \bibinfo {author} {\bibfnamefont {M.}~\bibnamefont {Capone}},\ }\bibfield  {title} {\bibinfo {title} {{Downfolding electron-phonon Hamiltonians from ab initio calculations: Application to ${\mathrm{K}}_{3}$ picene}},\ }\href {https://doi.org/10.1103/PhysRevB.90.115435} {\bibfield  {journal} {\bibinfo  {journal} {Phys. Rev. B}\ }\textbf {\bibinfo {volume} {90}},\ \bibinfo {pages} {115435} (\bibinfo {year} {2014})}\BibitemShut {NoStop}%
\bibitem [{\citenamefont {van Loon}\ \emph {et~al.}(2021)\citenamefont {van Loon}, \citenamefont {Berges},\ and\ \citenamefont {Wehling}}]{PhysRevB.103.205103}%
  \BibitemOpen
  \bibfield  {author} {\bibinfo {author} {\bibfnamefont {E.~G. C.~P.}\ \bibnamefont {van Loon}}, \bibinfo {author} {\bibfnamefont {J.}~\bibnamefont {Berges}},\ and\ \bibinfo {author} {\bibfnamefont {T.~O.}\ \bibnamefont {Wehling}},\ }\bibfield  {title} {\bibinfo {title} {Downfolding approaches to electron-ion coupling: Constrained density-functional perturbation theory for molecules},\ }\href {https://doi.org/10.1103/PhysRevB.103.205103} {\bibfield  {journal} {\bibinfo  {journal} {Phys. Rev. B}\ }\textbf {\bibinfo {volume} {103}},\ \bibinfo {pages} {205103} (\bibinfo {year} {2021})}\BibitemShut {NoStop}%
\bibitem [{\citenamefont {Schobert}\ \emph {et~al.}(2024{\natexlab{b}})\citenamefont {Schobert}, \citenamefont {Berges}, \citenamefont {van Loon}, \citenamefont {Sentef}, \citenamefont {Brener}, \citenamefont {Rossi},\ and\ \citenamefont {Wehling}}]{10.21468/SciPostPhys.16.2.046}%
  \BibitemOpen
  \bibfield  {author} {\bibinfo {author} {\bibfnamefont {A.}~\bibnamefont {Schobert}}, \bibinfo {author} {\bibfnamefont {J.}~\bibnamefont {Berges}}, \bibinfo {author} {\bibfnamefont {E.~G. C.~P.}\ \bibnamefont {van Loon}}, \bibinfo {author} {\bibfnamefont {M.~A.}\ \bibnamefont {Sentef}}, \bibinfo {author} {\bibfnamefont {S.}~\bibnamefont {Brener}}, \bibinfo {author} {\bibfnamefont {M.}~\bibnamefont {Rossi}},\ and\ \bibinfo {author} {\bibfnamefont {T.~O.}\ \bibnamefont {Wehling}},\ }\bibfield  {title} {\bibinfo {title} {{Ab initio electron-lattice downfolding: Potential energy landscapes, anharmonicity, and molecular dynamics in charge density wave materials}},\ }\href {https://doi.org/10.21468/SciPostPhys.16.2.046} {\bibfield  {journal} {\bibinfo  {journal} {SciPost Phys.}\ }\textbf {\bibinfo {volume} {16}},\ \bibinfo {pages} {046} (\bibinfo {year} {2024}{\natexlab{b}})}\BibitemShut {NoStop}%
\bibitem [{\citenamefont {Berges}\ \emph {et~al.}(2023)\citenamefont {Berges}, \citenamefont {Girotto}, \citenamefont {Wehling}, \citenamefont {Marzari},\ and\ \citenamefont {Ponc\'e}}]{PhysRevX.13.041009}%
  \BibitemOpen
  \bibfield  {author} {\bibinfo {author} {\bibfnamefont {J.}~\bibnamefont {Berges}}, \bibinfo {author} {\bibfnamefont {N.}~\bibnamefont {Girotto}}, \bibinfo {author} {\bibfnamefont {T.}~\bibnamefont {Wehling}}, \bibinfo {author} {\bibfnamefont {N.}~\bibnamefont {Marzari}},\ and\ \bibinfo {author} {\bibfnamefont {S.}~\bibnamefont {Ponc\'e}},\ }\bibfield  {title} {\bibinfo {title} {Phonon self-energy corrections: To screen, or not to screen},\ }\href {https://doi.org/10.1103/PhysRevX.13.041009} {\bibfield  {journal} {\bibinfo  {journal} {Phys. Rev. X}\ }\textbf {\bibinfo {volume} {13}},\ \bibinfo {pages} {041009} (\bibinfo {year} {2023})}\BibitemShut {NoStop}%
\bibitem [{\citenamefont {Yue}\ \emph {et~al.}(2023)\citenamefont {Yue}, \citenamefont {Nomura}, \citenamefont {Prassides},\ and\ \citenamefont {Werner}}]{PhysRevB.108.L220508}%
  \BibitemOpen
  \bibfield  {author} {\bibinfo {author} {\bibfnamefont {C.}~\bibnamefont {Yue}}, \bibinfo {author} {\bibfnamefont {Y.}~\bibnamefont {Nomura}}, \bibinfo {author} {\bibfnamefont {K.}~\bibnamefont {Prassides}},\ and\ \bibinfo {author} {\bibfnamefont {P.}~\bibnamefont {Werner}},\ }\bibfield  {title} {\bibinfo {title} {{Instability of the $Pa\overline{3}$ fulleride ${\mathrm{Cs}}_{3}{\mathrm{C}}_{60}$ at ambient pressure and superconducting state of the fcc phase}},\ }\href {https://doi.org/10.1103/PhysRevB.108.L220508} {\bibfield  {journal} {\bibinfo  {journal} {Phys. Rev. B}\ }\textbf {\bibinfo {volume} {108}},\ \bibinfo {pages} {L220508} (\bibinfo {year} {2023})}\BibitemShut {NoStop}%
\bibitem [{\citenamefont {Verdi}\ and\ \citenamefont {Giustino}(2015)}]{Verdi2015}%
  \BibitemOpen
  \bibfield  {author} {\bibinfo {author} {\bibfnamefont {C.}~\bibnamefont {Verdi}}\ and\ \bibinfo {author} {\bibfnamefont {F.}~\bibnamefont {Giustino}},\ }\bibfield  {title} {\bibinfo {title} {{Fr{\"{o}}hlich electron-phonon vertex from first principles}},\ }\href {https://doi.org/10.1103/PhysRevLett.115.176401} {\bibfield  {journal} {\bibinfo  {journal} {Physical Review Letters}\ }\textbf {\bibinfo {volume} {115}},\ \bibinfo {pages} {1} (\bibinfo {year} {2015})},\ \Eprint {https://arxiv.org/abs/1510.06373} {1510.06373} \BibitemShut {NoStop}%
\bibitem [{\citenamefont {Sio}\ \emph {et~al.}(2019)\citenamefont {Sio}, \citenamefont {Verdi}, \citenamefont {Ponc\'e},\ and\ \citenamefont {Giustino}}]{PhysRevB.99.235139}%
  \BibitemOpen
  \bibfield  {author} {\bibinfo {author} {\bibfnamefont {W.~H.}\ \bibnamefont {Sio}}, \bibinfo {author} {\bibfnamefont {C.}~\bibnamefont {Verdi}}, \bibinfo {author} {\bibfnamefont {S.}~\bibnamefont {Ponc\'e}},\ and\ \bibinfo {author} {\bibfnamefont {F.}~\bibnamefont {Giustino}},\ }\bibfield  {title} {\bibinfo {title} {Ab initio theory of polarons: Formalism and applications},\ }\href {https://doi.org/10.1103/PhysRevB.99.235139} {\bibfield  {journal} {\bibinfo  {journal} {Phys. Rev. B}\ }\textbf {\bibinfo {volume} {99}},\ \bibinfo {pages} {235139} (\bibinfo {year} {2019})}\BibitemShut {NoStop}%
\bibitem [{\citenamefont {Marzari}\ \emph {et~al.}(2012)\citenamefont {Marzari}, \citenamefont {Mostofi}, \citenamefont {Yates}, \citenamefont {Souza},\ and\ \citenamefont {Vanderbilt}}]{RevModPhys.84.1419}%
  \BibitemOpen
  \bibfield  {author} {\bibinfo {author} {\bibfnamefont {N.}~\bibnamefont {Marzari}}, \bibinfo {author} {\bibfnamefont {A.~A.}\ \bibnamefont {Mostofi}}, \bibinfo {author} {\bibfnamefont {J.~R.}\ \bibnamefont {Yates}}, \bibinfo {author} {\bibfnamefont {I.}~\bibnamefont {Souza}},\ and\ \bibinfo {author} {\bibfnamefont {D.}~\bibnamefont {Vanderbilt}},\ }\bibfield  {title} {\bibinfo {title} {Maximally localized wannier functions: Theory and applications},\ }\href {https://doi.org/10.1103/RevModPhys.84.1419} {\bibfield  {journal} {\bibinfo  {journal} {Rev. Mod. Phys.}\ }\textbf {\bibinfo {volume} {84}},\ \bibinfo {pages} {1419} (\bibinfo {year} {2012})}\BibitemShut {NoStop}%
\bibitem [{\citenamefont {Romanova}\ \emph {et~al.}(2023)\citenamefont {Romanova}, \citenamefont {Weng}, \citenamefont {Apelian},\ and\ \citenamefont {Vl{\v{c}}ek}}]{Romanova2023}%
  \BibitemOpen
  \bibfield  {author} {\bibinfo {author} {\bibfnamefont {M.}~\bibnamefont {Romanova}}, \bibinfo {author} {\bibfnamefont {G.}~\bibnamefont {Weng}}, \bibinfo {author} {\bibfnamefont {A.}~\bibnamefont {Apelian}},\ and\ \bibinfo {author} {\bibfnamefont {V.}~\bibnamefont {Vl{\v{c}}ek}},\ }\bibfield  {title} {\bibinfo {title} {{Dynamical downfolding for localized quantum states}},\ }\bibfield  {journal} {\bibinfo  {journal} {npj Computational Materials}\ }\textbf {\bibinfo {volume} {9}},\ \href {https://doi.org/10.1038/s41524-023-01078-5} {10.1038/s41524-023-01078-5} (\bibinfo {year} {2023})\BibitemShut {NoStop}%
\bibitem [{\citenamefont {Strinati}(1988)}]{Strinati1988}%
  \BibitemOpen
  \bibfield  {author} {\bibinfo {author} {\bibfnamefont {G.}~\bibnamefont {Strinati}},\ }\bibfield  {title} {\bibinfo {title} {{Application of the Green's functions method to the study of the optical properties of semiconductors}},\ }\href {https://doi.org/10.1007/BF02725962} {\bibfield  {journal} {\bibinfo  {journal} {La Rivista Del Nuovo Cimento Series 3}\ }\textbf {\bibinfo {volume} {11}},\ \bibinfo {pages} {1} (\bibinfo {year} {1988})}\BibitemShut {NoStop}%
\bibitem [{\citenamefont {Rohlfing}\ and\ \citenamefont {Louie}(1998)}]{Rohlfing1998}%
  \BibitemOpen
  \bibfield  {author} {\bibinfo {author} {\bibfnamefont {M.}~\bibnamefont {Rohlfing}}\ and\ \bibinfo {author} {\bibfnamefont {S.~G.}\ \bibnamefont {Louie}},\ }\bibfield  {title} {\bibinfo {title} {{Electron-hole excitations in semiconductors and insulators}},\ }\href {https://doi.org/10.1103/PhysRevLett.81.2312} {\bibfield  {journal} {\bibinfo  {journal} {Physical Review Letters}\ }\textbf {\bibinfo {volume} {81}},\ \bibinfo {pages} {2312} (\bibinfo {year} {1998})}\BibitemShut {NoStop}%
\bibitem [{\citenamefont {Rohlfing}\ and\ \citenamefont {Louie}(2000)}]{Rohlfing2000}%
  \BibitemOpen
  \bibfield  {author} {\bibinfo {author} {\bibfnamefont {M.}~\bibnamefont {Rohlfing}}\ and\ \bibinfo {author} {\bibfnamefont {S.~G.}\ \bibnamefont {Louie}},\ }\bibfield  {title} {\bibinfo {title} {{Electron-hole excitations and optical spectra from first principles}},\ }\href {https://doi.org/10.1103/PhysRevB.62.4927} {\bibfield  {journal} {\bibinfo  {journal} {Physical Review B - Condensed Matter and Materials Physics}\ }\textbf {\bibinfo {volume} {62}},\ \bibinfo {pages} {4927} (\bibinfo {year} {2000})},\ \Eprint {https://arxiv.org/abs/0406203v3} {0406203v3 [arXiv:cond-mat]} \BibitemShut {NoStop}%
\bibitem [{\citenamefont {Deslippe}\ \emph {et~al.}(2012)\citenamefont {Deslippe}, \citenamefont {Samsonidze}, \citenamefont {Strubbe}, \citenamefont {Jain}, \citenamefont {Cohen},\ and\ \citenamefont {Louie}}]{Deslippe2012}%
  \BibitemOpen
  \bibfield  {author} {\bibinfo {author} {\bibfnamefont {J.}~\bibnamefont {Deslippe}}, \bibinfo {author} {\bibfnamefont {G.}~\bibnamefont {Samsonidze}}, \bibinfo {author} {\bibfnamefont {D.~A.}\ \bibnamefont {Strubbe}}, \bibinfo {author} {\bibfnamefont {M.}~\bibnamefont {Jain}}, \bibinfo {author} {\bibfnamefont {M.~L.}\ \bibnamefont {Cohen}},\ and\ \bibinfo {author} {\bibfnamefont {S.~G.}\ \bibnamefont {Louie}},\ }\bibfield  {title} {\bibinfo {title} {{BerkeleyGW: A massively parallel computer package for the calculation of the quasiparticle and optical properties of materials and nanostructures}},\ }\href {https://doi.org/10.1016/j.cpc.2011.12.006} {\bibfield  {journal} {\bibinfo  {journal} {Computer Physics Communications}\ }\textbf {\bibinfo {volume} {183}},\ \bibinfo {pages} {1269} (\bibinfo {year} {2012})},\ \Eprint {https://arxiv.org/abs/1111.4429} {1111.4429} \BibitemShut {NoStop}%
\bibitem [{\citenamefont {Canestraight}\ \emph {et~al.}(2025)\citenamefont {Canestraight}, \citenamefont {Huang}, \citenamefont {Lin},\ and\ \citenamefont {Vlcek}}]{canestraight2024renormalizationstatesquasiparticlesmanybody}%
  \BibitemOpen
  \bibfield  {author} {\bibinfo {author} {\bibfnamefont {A.}~\bibnamefont {Canestraight}}, \bibinfo {author} {\bibfnamefont {Z.}~\bibnamefont {Huang}}, \bibinfo {author} {\bibfnamefont {L.}~\bibnamefont {Lin}},\ and\ \bibinfo {author} {\bibfnamefont {V.}~\bibnamefont {Vlcek}},\ }\bibfield  {title} {\bibinfo {title} {Renormalization of states and quasiparticles in many-body downfolding},\ }\href {https://doi.org/10.1063/5.0276233} {\bibfield  {journal} {\bibinfo  {journal} {The Journal of Chemical Physics}\ }\textbf {\bibinfo {volume} {163}},\ \bibinfo {pages} {024114} (\bibinfo {year} {2025})}\BibitemShut {NoStop}%
\bibitem [{\citenamefont {Scott}\ and\ \citenamefont {Booth}(2024)}]{PhysRevLett.132.076401}%
  \BibitemOpen
  \bibfield  {author} {\bibinfo {author} {\bibfnamefont {C.~J.~C.}\ \bibnamefont {Scott}}\ and\ \bibinfo {author} {\bibfnamefont {G.~H.}\ \bibnamefont {Booth}},\ }\bibfield  {title} {\bibinfo {title} {Rigorous screened interactions for realistic correlated electron systems},\ }\href {https://doi.org/10.1103/PhysRevLett.132.076401} {\bibfield  {journal} {\bibinfo  {journal} {Phys. Rev. Lett.}\ }\textbf {\bibinfo {volume} {132}},\ \bibinfo {pages} {076401} (\bibinfo {year} {2024})}\BibitemShut {NoStop}%
\bibitem [{\citenamefont {Baym}(1961)}]{Baym1961}%
  \BibitemOpen
  \bibfield  {author} {\bibinfo {author} {\bibfnamefont {G.}~\bibnamefont {Baym}},\ }\bibfield  {title} {\bibinfo {title} {{Field-theoretic approach to the properties of the solid state}},\ }\bibfield  {journal} {\bibinfo  {journal} {Annals of Physics}\ }\textbf {\bibinfo {volume} {14}},\ \href {https://doi.org/10.1006/aphy.2000.6009} {10.1006/aphy.2000.6009} (\bibinfo {year} {1961})\BibitemShut {NoStop}%
\bibitem [{\citenamefont {Hedin}(1965)}]{Hedin1965}%
  \BibitemOpen
  \bibfield  {author} {\bibinfo {author} {\bibfnamefont {L.}~\bibnamefont {Hedin}},\ }\bibfield  {title} {\bibinfo {title} {{New Method for Calculating the One-Particle Green's Function with Application to the Electron-Gas Problem}},\ }\href {https://doi.org/10.1007/BF01115969} {\bibfield  {journal} {\bibinfo  {journal} {Physical Review}\ }\textbf {\bibinfo {volume} {139}},\ \bibinfo {pages} {796} (\bibinfo {year} {1965})}\BibitemShut {NoStop}%
\bibitem [{\citenamefont {Giustino}(2017)}]{Giustino2017}%
  \BibitemOpen
  \bibfield  {author} {\bibinfo {author} {\bibfnamefont {F.}~\bibnamefont {Giustino}},\ }\bibfield  {title} {\bibinfo {title} {{Electron-phonon interactions from first principles}},\ }\href {https://doi.org/10.1103/RevModPhys.89.015003} {\bibfield  {journal} {\bibinfo  {journal} {Reviews of Modern Physics}\ }\textbf {\bibinfo {volume} {89}},\ \bibinfo {pages} {1} (\bibinfo {year} {2017})}\BibitemShut {NoStop}%
\bibitem [{\citenamefont {Ruhman}(2025)}]{ruhman2025commentarxiv250117230arxiv250200103phononmediated}%
  \BibitemOpen
  \bibfield  {author} {\bibinfo {author} {\bibfnamefont {J.}~\bibnamefont {Ruhman}},\ }\href {https://arxiv.org/abs/2503.00041} {\bibinfo {title} {{Comment on arXiv:2501.17230 and arXiv:2502.00103 "Phonon-mediated electron attraction in SrTiO3 via the generalized Frohlich and deformation potential mechanisms" and "Theory of ab initio downfolding with arbitrary range electron-phonon coupling"}}} (\bibinfo {year} {2025}),\ \Eprint {https://arxiv.org/abs/2503.00041} {arXiv:2503.00041 [cond-mat.supr-con]} \BibitemShut {NoStop}%
\bibitem [{\citenamefont {Gonze}\ and\ \citenamefont {Lee}(1997)}]{PhysRevB.55.10355}%
  \BibitemOpen
  \bibfield  {author} {\bibinfo {author} {\bibfnamefont {X.}~\bibnamefont {Gonze}}\ and\ \bibinfo {author} {\bibfnamefont {C.}~\bibnamefont {Lee}},\ }\bibfield  {title} {\bibinfo {title} {Dynamical matrices, born effective charges, dielectric permittivity tensors, and interatomic force constants from density-functional perturbation theory},\ }\href {https://doi.org/10.1103/PhysRevB.55.10355} {\bibfield  {journal} {\bibinfo  {journal} {Phys. Rev. B}\ }\textbf {\bibinfo {volume} {55}},\ \bibinfo {pages} {10355} (\bibinfo {year} {1997})}\BibitemShut {NoStop}%
\bibitem [{\citenamefont {Mahan}(2013)}]{MahanGeraldD2013MP}%
  \BibitemOpen
  \bibfield  {author} {\bibinfo {author} {\bibfnamefont {G.~D.}\ \bibnamefont {Mahan}},\ }\href@noop {} {\emph {\bibinfo {title} {{Many-Particle Physics}}}},\ Physics of Solids and Liquids\ (\bibinfo  {publisher} {Springer},\ \bibinfo {year} {2013})\BibitemShut {NoStop}%
\bibitem [{\citenamefont {Alvertis}\ \emph {et~al.}(2024{\natexlab{b}})\citenamefont {Alvertis}, \citenamefont {Haber}, \citenamefont {Li}, \citenamefont {Coveney}, \citenamefont {Louie}, \citenamefont {Filip},\ and\ \citenamefont {Neaton}}]{Alvertis2024}%
  \BibitemOpen
  \bibfield  {author} {\bibinfo {author} {\bibfnamefont {A.~M.}\ \bibnamefont {Alvertis}}, \bibinfo {author} {\bibfnamefont {J.~B.}\ \bibnamefont {Haber}}, \bibinfo {author} {\bibfnamefont {Z.}~\bibnamefont {Li}}, \bibinfo {author} {\bibfnamefont {C.~J.~N.}\ \bibnamefont {Coveney}}, \bibinfo {author} {\bibfnamefont {S.~G.}\ \bibnamefont {Louie}}, \bibinfo {author} {\bibfnamefont {M.~R.}\ \bibnamefont {Filip}},\ and\ \bibinfo {author} {\bibfnamefont {J.~B.}\ \bibnamefont {Neaton}},\ }\bibfield  {title} {\bibinfo {title} {{Phonon screening and dissociation of excitons at finite temperatures from first principles}},\ }\href {https://doi.org/10.1073/pnas.2403434121} {\bibfield  {journal} {\bibinfo  {journal} {Proceedings of the National Academy of Sciences}\ }\textbf {\bibinfo {volume} {121}},\ \bibinfo {pages} {e2403434121} (\bibinfo {year} {2024}{\natexlab{b}})}\BibitemShut {NoStop}%
\bibitem [{\citenamefont {Filip}\ \emph {et~al.}(2021)\citenamefont {Filip}, \citenamefont {Haber},\ and\ \citenamefont {Neaton}}]{PhysRevLett.127.067401}%
  \BibitemOpen
  \bibfield  {author} {\bibinfo {author} {\bibfnamefont {M.~R.}\ \bibnamefont {Filip}}, \bibinfo {author} {\bibfnamefont {J.~B.}\ \bibnamefont {Haber}},\ and\ \bibinfo {author} {\bibfnamefont {J.~B.}\ \bibnamefont {Neaton}},\ }\bibfield  {title} {\bibinfo {title} {Phonon screening of excitons in semiconductors: Halide perovskites and beyond},\ }\href {https://doi.org/10.1103/PhysRevLett.127.067401} {\bibfield  {journal} {\bibinfo  {journal} {Phys. Rev. Lett.}\ }\textbf {\bibinfo {volume} {127}},\ \bibinfo {pages} {067401} (\bibinfo {year} {2021})}\BibitemShut {NoStop}%
\bibitem [{\citenamefont {Su}\ \emph {et~al.}(1979)\citenamefont {Su}, \citenamefont {Schrieffer},\ and\ \citenamefont {Heeger}}]{PhysRevLett.42.1698}%
  \BibitemOpen
  \bibfield  {author} {\bibinfo {author} {\bibfnamefont {W.~P.}\ \bibnamefont {Su}}, \bibinfo {author} {\bibfnamefont {J.~R.}\ \bibnamefont {Schrieffer}},\ and\ \bibinfo {author} {\bibfnamefont {A.~J.}\ \bibnamefont {Heeger}},\ }\bibfield  {title} {\bibinfo {title} {Solitons in polyacetylene},\ }\href {https://doi.org/10.1103/PhysRevLett.42.1698} {\bibfield  {journal} {\bibinfo  {journal} {Phys. Rev. Lett.}\ }\textbf {\bibinfo {volume} {42}},\ \bibinfo {pages} {1698} (\bibinfo {year} {1979})}\BibitemShut {NoStop}%
\bibitem [{\citenamefont {Heeger}\ \emph {et~al.}(1988)\citenamefont {Heeger}, \citenamefont {Kivelson}, \citenamefont {Schrieffer},\ and\ \citenamefont {Su}}]{RevModPhys.60.781}%
  \BibitemOpen
  \bibfield  {author} {\bibinfo {author} {\bibfnamefont {A.~J.}\ \bibnamefont {Heeger}}, \bibinfo {author} {\bibfnamefont {S.}~\bibnamefont {Kivelson}}, \bibinfo {author} {\bibfnamefont {J.~R.}\ \bibnamefont {Schrieffer}},\ and\ \bibinfo {author} {\bibfnamefont {W.~P.}\ \bibnamefont {Su}},\ }\bibfield  {title} {\bibinfo {title} {Solitons in conducting polymers},\ }\href {https://doi.org/10.1103/RevModPhys.60.781} {\bibfield  {journal} {\bibinfo  {journal} {Rev. Mod. Phys.}\ }\textbf {\bibinfo {volume} {60}},\ \bibinfo {pages} {781} (\bibinfo {year} {1988})}\BibitemShut {NoStop}%
\bibitem [{\citenamefont {Giustino}\ \emph {et~al.}(2007)\citenamefont {Giustino}, \citenamefont {Cohen},\ and\ \citenamefont {Louie}}]{Giustino2007}%
  \BibitemOpen
  \bibfield  {author} {\bibinfo {author} {\bibfnamefont {F.}~\bibnamefont {Giustino}}, \bibinfo {author} {\bibfnamefont {M.~L.}\ \bibnamefont {Cohen}},\ and\ \bibinfo {author} {\bibfnamefont {S.~G.}\ \bibnamefont {Louie}},\ }\bibfield  {title} {\bibinfo {title} {{Electron-phonon interaction using Wannier functions}},\ }\href {https://doi.org/10.1103/PhysRevB.76.165108} {\bibfield  {journal} {\bibinfo  {journal} {Physical Review B - Condensed Matter and Materials Physics}\ }\textbf {\bibinfo {volume} {76}},\ \bibinfo {pages} {1} (\bibinfo {year} {2007})}\BibitemShut {NoStop}%
\bibitem [{\citenamefont {Dolgov}\ \emph {et~al.}(1981)\citenamefont {Dolgov}, \citenamefont {Kirzhnits},\ and\ \citenamefont {Maksimov}}]{RevModPhys.53.81}%
  \BibitemOpen
  \bibfield  {author} {\bibinfo {author} {\bibfnamefont {O.~V.}\ \bibnamefont {Dolgov}}, \bibinfo {author} {\bibfnamefont {D.~A.}\ \bibnamefont {Kirzhnits}},\ and\ \bibinfo {author} {\bibfnamefont {E.~G.}\ \bibnamefont {Maksimov}},\ }\bibfield  {title} {\bibinfo {title} {On an admissible sign of the static dielectric function of matter},\ }\href {https://doi.org/10.1103/RevModPhys.53.81} {\bibfield  {journal} {\bibinfo  {journal} {Rev. Mod. Phys.}\ }\textbf {\bibinfo {volume} {53}},\ \bibinfo {pages} {81} (\bibinfo {year} {1981})}\BibitemShut {NoStop}%
\bibitem [{\citenamefont {Fröhlich}\ \emph {et~al.}(1950)\citenamefont {Fröhlich}, \citenamefont {Pelzer},\ and\ \citenamefont {Zienau}}]{doi:10.1080/14786445008521794}%
  \BibitemOpen
  \bibfield  {author} {\bibinfo {author} {\bibfnamefont {H.}~\bibnamefont {Fröhlich}}, \bibinfo {author} {\bibfnamefont {H.}~\bibnamefont {Pelzer}},\ and\ \bibinfo {author} {\bibfnamefont {S.}~\bibnamefont {Zienau}},\ }\bibfield  {title} {\bibinfo {title} {Xx. properties of slow electrons in polar materials},\ }\href {https://doi.org/10.1080/14786445008521794} {\bibfield  {journal} {\bibinfo  {journal} {The London, Edinburgh, and Dublin Philosophical Magazine and Journal of Science}\ }\textbf {\bibinfo {volume} {41}},\ \bibinfo {pages} {221} (\bibinfo {year} {1950})},\ \Eprint {https://arxiv.org/abs/https://doi.org/10.1080/14786445008521794} {https://doi.org/10.1080/14786445008521794} \BibitemShut {NoStop}%
\bibitem [{\citenamefont {Sio}\ and\ \citenamefont {Giustino}(2022)}]{PhysRevB.105.115414}%
  \BibitemOpen
  \bibfield  {author} {\bibinfo {author} {\bibfnamefont {W.~H.}\ \bibnamefont {Sio}}\ and\ \bibinfo {author} {\bibfnamefont {F.}~\bibnamefont {Giustino}},\ }\bibfield  {title} {\bibinfo {title} {Unified ab initio description of fr\"ohlich electron-phonon interactions in two-dimensional and three-dimensional materials},\ }\href {https://doi.org/10.1103/PhysRevB.105.115414} {\bibfield  {journal} {\bibinfo  {journal} {Phys. Rev. B}\ }\textbf {\bibinfo {volume} {105}},\ \bibinfo {pages} {115414} (\bibinfo {year} {2022})}\BibitemShut {NoStop}%
\bibitem [{\citenamefont {Hannewald}\ \emph {et~al.}(2004)\citenamefont {Hannewald}, \citenamefont {Stojanovi\ifmmode~\acute{c}\else \'{c}\fi{}}, \citenamefont {Schellekens}, \citenamefont {Bobbert}, \citenamefont {Kresse},\ and\ \citenamefont {Hafner}}]{PhysRevB.69.075211}%
  \BibitemOpen
  \bibfield  {author} {\bibinfo {author} {\bibfnamefont {K.}~\bibnamefont {Hannewald}}, \bibinfo {author} {\bibfnamefont {V.~M.}\ \bibnamefont {Stojanovi\ifmmode~\acute{c}\else \'{c}\fi{}}}, \bibinfo {author} {\bibfnamefont {J.~M.~T.}\ \bibnamefont {Schellekens}}, \bibinfo {author} {\bibfnamefont {P.~A.}\ \bibnamefont {Bobbert}}, \bibinfo {author} {\bibfnamefont {G.}~\bibnamefont {Kresse}},\ and\ \bibinfo {author} {\bibfnamefont {J.}~\bibnamefont {Hafner}},\ }\bibfield  {title} {\bibinfo {title} {Theory of polaron bandwidth narrowing in organic molecular crystals},\ }\href {https://doi.org/10.1103/PhysRevB.69.075211} {\bibfield  {journal} {\bibinfo  {journal} {Phys. Rev. B}\ }\textbf {\bibinfo {volume} {69}},\ \bibinfo {pages} {075211} (\bibinfo {year} {2004})}\BibitemShut {NoStop}%
\bibitem [{\citenamefont {Bostr\"om}\ \emph {et~al.}(2019)\citenamefont {Bostr\"om}, \citenamefont {Helmer}, \citenamefont {Werner},\ and\ \citenamefont {Verdozzi}}]{PhysRevResearch.1.013017}%
  \BibitemOpen
  \bibfield  {author} {\bibinfo {author} {\bibfnamefont {E.~V.~n.}\ \bibnamefont {Bostr\"om}}, \bibinfo {author} {\bibfnamefont {P.}~\bibnamefont {Helmer}}, \bibinfo {author} {\bibfnamefont {P.}~\bibnamefont {Werner}},\ and\ \bibinfo {author} {\bibfnamefont {C.}~\bibnamefont {Verdozzi}},\ }\bibfield  {title} {\bibinfo {title} {Electron-electron versus electron-phonon interactions in lattice models: Screening effects described by a density functional theory approach},\ }\href {https://doi.org/10.1103/PhysRevResearch.1.013017} {\bibfield  {journal} {\bibinfo  {journal} {Phys. Rev. Res.}\ }\textbf {\bibinfo {volume} {1}},\ \bibinfo {pages} {013017} (\bibinfo {year} {2019})}\BibitemShut {NoStop}%
\bibitem [{\citenamefont {Jain}\ \emph {et~al.}(2013)\citenamefont {Jain}, \citenamefont {Ong}, \citenamefont {Hautier}, \citenamefont {Chen}, \citenamefont {Richards}, \citenamefont {Dacek}, \citenamefont {Cholia}, \citenamefont {Gunter}, \citenamefont {Skinner}, \citenamefont {Ceder},\ and\ \citenamefont {Persson}}]{Jain2013}%
  \BibitemOpen
  \bibfield  {author} {\bibinfo {author} {\bibfnamefont {A.}~\bibnamefont {Jain}}, \bibinfo {author} {\bibfnamefont {S.~P.}\ \bibnamefont {Ong}}, \bibinfo {author} {\bibfnamefont {G.}~\bibnamefont {Hautier}}, \bibinfo {author} {\bibfnamefont {W.}~\bibnamefont {Chen}}, \bibinfo {author} {\bibfnamefont {W.~D.}\ \bibnamefont {Richards}}, \bibinfo {author} {\bibfnamefont {S.}~\bibnamefont {Dacek}}, \bibinfo {author} {\bibfnamefont {S.}~\bibnamefont {Cholia}}, \bibinfo {author} {\bibfnamefont {D.}~\bibnamefont {Gunter}}, \bibinfo {author} {\bibfnamefont {D.}~\bibnamefont {Skinner}}, \bibinfo {author} {\bibfnamefont {G.}~\bibnamefont {Ceder}},\ and\ \bibinfo {author} {\bibfnamefont {K.~A.}\ \bibnamefont {Persson}},\ }\bibfield  {title} {\bibinfo {title} {{Commentary: The materials project: A materials genome approach to accelerating materials innovation}},\ }\bibfield  {journal} {\bibinfo  {journal} {APL Materials}\ }\textbf {\bibinfo {volume} {1}},\ \href {https://doi.org/10.1063/1.4812323} {10.1063/1.4812323}
  (\bibinfo {year} {2013})\BibitemShut {NoStop}%
\bibitem [{\citenamefont {Shaltaf}\ \emph {et~al.}(2008)\citenamefont {Shaltaf}, \citenamefont {Durgun}, \citenamefont {Raty}, \citenamefont {Ghosez},\ and\ \citenamefont {Gonze}}]{PhysRevB.78.205203}%
  \BibitemOpen
  \bibfield  {author} {\bibinfo {author} {\bibfnamefont {R.}~\bibnamefont {Shaltaf}}, \bibinfo {author} {\bibfnamefont {E.}~\bibnamefont {Durgun}}, \bibinfo {author} {\bibfnamefont {J.-Y.}\ \bibnamefont {Raty}}, \bibinfo {author} {\bibfnamefont {P.}~\bibnamefont {Ghosez}},\ and\ \bibinfo {author} {\bibfnamefont {X.}~\bibnamefont {Gonze}},\ }\bibfield  {title} {\bibinfo {title} {{Dynamical, dielectric, and elastic properties of GeTe investigated with first-principles density functional theory}},\ }\href {https://doi.org/10.1103/PhysRevB.78.205203} {\bibfield  {journal} {\bibinfo  {journal} {Phys. Rev. B}\ }\textbf {\bibinfo {volume} {78}},\ \bibinfo {pages} {205203} (\bibinfo {year} {2008})}\BibitemShut {NoStop}%
\bibitem [{\citenamefont {Chan}\ \emph {et~al.}(2024)\citenamefont {Chan}, \citenamefont {Naik}, \citenamefont {Haber}, \citenamefont {Neaton}, \citenamefont {Louie}, \citenamefont {Qiu},\ and\ \citenamefont {da~Jornada}}]{doi:10.1021/acs.nanolett.4c01508}%
  \BibitemOpen
  \bibfield  {author} {\bibinfo {author} {\bibfnamefont {Y.-h.}\ \bibnamefont {Chan}}, \bibinfo {author} {\bibfnamefont {M.~H.}\ \bibnamefont {Naik}}, \bibinfo {author} {\bibfnamefont {J.~B.}\ \bibnamefont {Haber}}, \bibinfo {author} {\bibfnamefont {J.~B.}\ \bibnamefont {Neaton}}, \bibinfo {author} {\bibfnamefont {S.~G.}\ \bibnamefont {Louie}}, \bibinfo {author} {\bibfnamefont {D.~Y.}\ \bibnamefont {Qiu}},\ and\ \bibinfo {author} {\bibfnamefont {F.~H.}\ \bibnamefont {da~Jornada}},\ }\bibfield  {title} {\bibinfo {title} {{Exciton–Phonon Coupling Induces a New Pathway for Ultrafast Intralayer-to-Interlayer Exciton Transition and Interlayer Charge Transfer in $\text{WS}_2–\text{MoS}_2$ Heterostructure: A First-Principles Study}},\ }\href {https://doi.org/10.1021/acs.nanolett.4c01508} {\bibfield  {journal} {\bibinfo  {journal} {Nano Letters}\ }\textbf {\bibinfo {volume} {24}},\ \bibinfo {pages} {7972} (\bibinfo {year} {2024})},\ \bibinfo {note} {pMID: 38888269},\ \Eprint
  {https://arxiv.org/abs/https://doi.org/10.1021/acs.nanolett.4c01508} {https://doi.org/10.1021/acs.nanolett.4c01508} \BibitemShut {NoStop}%
\bibitem [{\citenamefont {Brown-Altvater}\ \emph {et~al.}(2020)\citenamefont {Brown-Altvater}, \citenamefont {Antonius}, \citenamefont {Rangel}, \citenamefont {Giantomassi}, \citenamefont {Draxl}, \citenamefont {Gonze}, \citenamefont {Louie},\ and\ \citenamefont {Neaton}}]{PhysRevB.101.165102}%
  \BibitemOpen
  \bibfield  {author} {\bibinfo {author} {\bibfnamefont {F.}~\bibnamefont {Brown-Altvater}}, \bibinfo {author} {\bibfnamefont {G.}~\bibnamefont {Antonius}}, \bibinfo {author} {\bibfnamefont {T.}~\bibnamefont {Rangel}}, \bibinfo {author} {\bibfnamefont {M.}~\bibnamefont {Giantomassi}}, \bibinfo {author} {\bibfnamefont {C.}~\bibnamefont {Draxl}}, \bibinfo {author} {\bibfnamefont {X.}~\bibnamefont {Gonze}}, \bibinfo {author} {\bibfnamefont {S.~G.}\ \bibnamefont {Louie}},\ and\ \bibinfo {author} {\bibfnamefont {J.~B.}\ \bibnamefont {Neaton}},\ }\bibfield  {title} {\bibinfo {title} {Band gap renormalization, carrier mobilities, and the electron-phonon self-energy in crystalline naphthalene},\ }\href {https://doi.org/10.1103/PhysRevB.101.165102} {\bibfield  {journal} {\bibinfo  {journal} {Phys. Rev. B}\ }\textbf {\bibinfo {volume} {101}},\ \bibinfo {pages} {165102} (\bibinfo {year} {2020})}\BibitemShut {NoStop}%
\bibitem [{\citenamefont {Giannozzi}\ \emph {et~al.}(2009)\citenamefont {Giannozzi}, \citenamefont {Baroni}, \citenamefont {Bonini}, \citenamefont {Calandra}, \citenamefont {Car}, \citenamefont {Cavazzoni}, \citenamefont {Ceresoli}, \citenamefont {Chiarotti}, \citenamefont {Cococcioni}, \citenamefont {Dabo}, \citenamefont {{Dal Corso}}, \citenamefont {Fabris}, \citenamefont {Fratesi}, \citenamefont {{de Gironcoli}}, \citenamefont {Gebauer}, \citenamefont {Gerstmann}, \citenamefont {Gougoussis}, \citenamefont {Kokalj}, \citenamefont {Lazzeri}, \citenamefont {Martin-Samos}, \citenamefont {Marzari}, \citenamefont {Mauri}, \citenamefont {Mazzarello}, \citenamefont {Paolini}, \citenamefont {Pasquarello}, \citenamefont {Paulatto}, \citenamefont {Sbraccia}, \citenamefont {Scandolo}, \citenamefont {Sclauzero}, \citenamefont {Seitsonen}, \citenamefont {Smogunov}, \citenamefont {Umari},\ and\ \citenamefont {Wentzcovitch}}]{QE}%
  \BibitemOpen
  \bibfield  {author} {\bibinfo {author} {\bibfnamefont {P.}~\bibnamefont {Giannozzi}}, \bibinfo {author} {\bibfnamefont {S.}~\bibnamefont {Baroni}}, \bibinfo {author} {\bibfnamefont {N.}~\bibnamefont {Bonini}}, \bibinfo {author} {\bibfnamefont {M.}~\bibnamefont {Calandra}}, \bibinfo {author} {\bibfnamefont {R.}~\bibnamefont {Car}}, \bibinfo {author} {\bibfnamefont {C.}~\bibnamefont {Cavazzoni}}, \bibinfo {author} {\bibfnamefont {D.}~\bibnamefont {Ceresoli}}, \bibinfo {author} {\bibfnamefont {G.~L.}\ \bibnamefont {Chiarotti}}, \bibinfo {author} {\bibfnamefont {M.}~\bibnamefont {Cococcioni}}, \bibinfo {author} {\bibfnamefont {I.}~\bibnamefont {Dabo}}, \bibinfo {author} {\bibfnamefont {A.}~\bibnamefont {{Dal Corso}}}, \bibinfo {author} {\bibfnamefont {S.}~\bibnamefont {Fabris}}, \bibinfo {author} {\bibfnamefont {G.}~\bibnamefont {Fratesi}}, \bibinfo {author} {\bibfnamefont {S.}~\bibnamefont {{de Gironcoli}}}, \bibinfo {author} {\bibfnamefont {R.}~\bibnamefont {Gebauer}}, \bibinfo {author} {\bibfnamefont
  {U.}~\bibnamefont {Gerstmann}}, \bibinfo {author} {\bibfnamefont {C.}~\bibnamefont {Gougoussis}}, \bibinfo {author} {\bibfnamefont {A.}~\bibnamefont {Kokalj}}, \bibinfo {author} {\bibfnamefont {M.}~\bibnamefont {Lazzeri}}, \bibinfo {author} {\bibfnamefont {L.}~\bibnamefont {Martin-Samos}}, \bibinfo {author} {\bibfnamefont {N.}~\bibnamefont {Marzari}}, \bibinfo {author} {\bibfnamefont {F.}~\bibnamefont {Mauri}}, \bibinfo {author} {\bibfnamefont {R.}~\bibnamefont {Mazzarello}}, \bibinfo {author} {\bibfnamefont {S.}~\bibnamefont {Paolini}}, \bibinfo {author} {\bibfnamefont {A.}~\bibnamefont {Pasquarello}}, \bibinfo {author} {\bibfnamefont {L.}~\bibnamefont {Paulatto}}, \bibinfo {author} {\bibfnamefont {C.}~\bibnamefont {Sbraccia}}, \bibinfo {author} {\bibfnamefont {S.}~\bibnamefont {Scandolo}}, \bibinfo {author} {\bibfnamefont {G.}~\bibnamefont {Sclauzero}}, \bibinfo {author} {\bibfnamefont {A.~P.}\ \bibnamefont {Seitsonen}}, \bibinfo {author} {\bibfnamefont {A.}~\bibnamefont {Smogunov}}, \bibinfo {author}
  {\bibfnamefont {P.}~\bibnamefont {Umari}},\ and\ \bibinfo {author} {\bibfnamefont {R.~M.}\ \bibnamefont {Wentzcovitch}},\ }\bibfield  {title} {\bibinfo {title} {{QUANTUM ESPRESSO: a modular and open-source software project for quantum simulations of materials}},\ }\href@noop {} {\bibfield  {journal} {\bibinfo  {journal} {Journal of Physics: Condensed Matter}\ }\textbf {\bibinfo {volume} {21}},\ \bibinfo {pages} {395502} (\bibinfo {year} {2009})}\BibitemShut {NoStop}%
\bibitem [{\citenamefont {Pizzi}\ \emph {et~al.}(2020)\citenamefont {Pizzi}, \citenamefont {Vitale}, \citenamefont {Arita}, \citenamefont {Bl{\"{u}}gel}, \citenamefont {Freimuth}, \citenamefont {G{\'{e}}ranton}, \citenamefont {Gibertini}, \citenamefont {Gresch}, \citenamefont {Johnson}, \citenamefont {Koretsune}, \citenamefont {Ibanez-Azpiroz}, \citenamefont {Lee}, \citenamefont {Lihm}, \citenamefont {Marchand}, \citenamefont {Marrazzo}, \citenamefont {Mokrousov}, \citenamefont {Mustafa}, \citenamefont {Nohara}, \citenamefont {Nomura}, \citenamefont {Paulatto}, \citenamefont {Ponc{\'{e}}}, \citenamefont {Ponweiser}, \citenamefont {Qiao}, \citenamefont {Th{\"{o}}le}, \citenamefont {Tsirkin}, \citenamefont {Wierzbowska}, \citenamefont {Marzari}, \citenamefont {Vanderbilt}, \citenamefont {Souza}, \citenamefont {Mostofi},\ and\ \citenamefont {Yates}}]{Pizzi2020}%
  \BibitemOpen
  \bibfield  {author} {\bibinfo {author} {\bibfnamefont {G.}~\bibnamefont {Pizzi}}, \bibinfo {author} {\bibfnamefont {V.}~\bibnamefont {Vitale}}, \bibinfo {author} {\bibfnamefont {R.}~\bibnamefont {Arita}}, \bibinfo {author} {\bibfnamefont {S.}~\bibnamefont {Bl{\"{u}}gel}}, \bibinfo {author} {\bibfnamefont {F.}~\bibnamefont {Freimuth}}, \bibinfo {author} {\bibfnamefont {G.}~\bibnamefont {G{\'{e}}ranton}}, \bibinfo {author} {\bibfnamefont {M.}~\bibnamefont {Gibertini}}, \bibinfo {author} {\bibfnamefont {D.}~\bibnamefont {Gresch}}, \bibinfo {author} {\bibfnamefont {C.}~\bibnamefont {Johnson}}, \bibinfo {author} {\bibfnamefont {T.}~\bibnamefont {Koretsune}}, \bibinfo {author} {\bibfnamefont {J.}~\bibnamefont {Ibanez-Azpiroz}}, \bibinfo {author} {\bibfnamefont {H.}~\bibnamefont {Lee}}, \bibinfo {author} {\bibfnamefont {J.~M.}\ \bibnamefont {Lihm}}, \bibinfo {author} {\bibfnamefont {D.}~\bibnamefont {Marchand}}, \bibinfo {author} {\bibfnamefont {A.}~\bibnamefont {Marrazzo}}, \bibinfo {author} {\bibfnamefont
  {Y.}~\bibnamefont {Mokrousov}}, \bibinfo {author} {\bibfnamefont {J.~I.}\ \bibnamefont {Mustafa}}, \bibinfo {author} {\bibfnamefont {Y.}~\bibnamefont {Nohara}}, \bibinfo {author} {\bibfnamefont {Y.}~\bibnamefont {Nomura}}, \bibinfo {author} {\bibfnamefont {L.}~\bibnamefont {Paulatto}}, \bibinfo {author} {\bibfnamefont {S.}~\bibnamefont {Ponc{\'{e}}}}, \bibinfo {author} {\bibfnamefont {T.}~\bibnamefont {Ponweiser}}, \bibinfo {author} {\bibfnamefont {J.}~\bibnamefont {Qiao}}, \bibinfo {author} {\bibfnamefont {F.}~\bibnamefont {Th{\"{o}}le}}, \bibinfo {author} {\bibfnamefont {S.~S.}\ \bibnamefont {Tsirkin}}, \bibinfo {author} {\bibfnamefont {M.}~\bibnamefont {Wierzbowska}}, \bibinfo {author} {\bibfnamefont {N.}~\bibnamefont {Marzari}}, \bibinfo {author} {\bibfnamefont {D.}~\bibnamefont {Vanderbilt}}, \bibinfo {author} {\bibfnamefont {I.}~\bibnamefont {Souza}}, \bibinfo {author} {\bibfnamefont {A.~A.}\ \bibnamefont {Mostofi}},\ and\ \bibinfo {author} {\bibfnamefont {J.~R.}\ \bibnamefont {Yates}},\ }\bibfield
  {title} {\bibinfo {title} {{Wannier90 as a community code: New features and applications}},\ }\bibfield  {journal} {\bibinfo  {journal} {Journal of Physics Condensed Matter}\ }\textbf {\bibinfo {volume} {32}},\ \href {https://doi.org/10.1088/1361-648X/ab51ff} {10.1088/1361-648X/ab51ff} (\bibinfo {year} {2020})\BibitemShut {NoStop}%
\bibitem [{\citenamefont {Lee}\ \emph {et~al.}(2023)\citenamefont {Lee}, \citenamefont {Ponc{\'e}}, \citenamefont {Bushick}, \citenamefont {Hajinazar}, \citenamefont {Lafuente-Bartolome}, \citenamefont {Leveillee}, \citenamefont {Lian}, \citenamefont {Lihm}, \citenamefont {Macheda}, \citenamefont {Mori}, \citenamefont {Paudyal}, \citenamefont {Sio}, \citenamefont {Tiwari}, \citenamefont {Zacharias}, \citenamefont {Zhang}, \citenamefont {Bonini}, \citenamefont {Kioupakis}, \citenamefont {Margine},\ and\ \citenamefont {Giustino}}]{Lee2023}%
  \BibitemOpen
  \bibfield  {author} {\bibinfo {author} {\bibfnamefont {H.}~\bibnamefont {Lee}}, \bibinfo {author} {\bibfnamefont {S.}~\bibnamefont {Ponc{\'e}}}, \bibinfo {author} {\bibfnamefont {K.}~\bibnamefont {Bushick}}, \bibinfo {author} {\bibfnamefont {S.}~\bibnamefont {Hajinazar}}, \bibinfo {author} {\bibfnamefont {J.}~\bibnamefont {Lafuente-Bartolome}}, \bibinfo {author} {\bibfnamefont {J.}~\bibnamefont {Leveillee}}, \bibinfo {author} {\bibfnamefont {C.}~\bibnamefont {Lian}}, \bibinfo {author} {\bibfnamefont {J.-M.}\ \bibnamefont {Lihm}}, \bibinfo {author} {\bibfnamefont {F.}~\bibnamefont {Macheda}}, \bibinfo {author} {\bibfnamefont {H.}~\bibnamefont {Mori}}, \bibinfo {author} {\bibfnamefont {H.}~\bibnamefont {Paudyal}}, \bibinfo {author} {\bibfnamefont {W.~H.}\ \bibnamefont {Sio}}, \bibinfo {author} {\bibfnamefont {S.}~\bibnamefont {Tiwari}}, \bibinfo {author} {\bibfnamefont {M.}~\bibnamefont {Zacharias}}, \bibinfo {author} {\bibfnamefont {X.}~\bibnamefont {Zhang}}, \bibinfo {author} {\bibfnamefont
  {N.}~\bibnamefont {Bonini}}, \bibinfo {author} {\bibfnamefont {E.}~\bibnamefont {Kioupakis}}, \bibinfo {author} {\bibfnamefont {E.~R.}\ \bibnamefont {Margine}},\ and\ \bibinfo {author} {\bibfnamefont {F.}~\bibnamefont {Giustino}},\ }\bibfield  {title} {\bibinfo {title} {{Electron--phonon physics from first principles using the EPW code}},\ }\href {https://doi.org/10.1038/s41524-023-01107-3} {\bibfield  {journal} {\bibinfo  {journal} {npj Computational Materials}\ }\textbf {\bibinfo {volume} {9}},\ \bibinfo {pages} {156} (\bibinfo {year} {2023})}\BibitemShut {NoStop}%
\bibitem [{\citenamefont {Nakamura}\ \emph {et~al.}(2021)\citenamefont {Nakamura}, \citenamefont {Yoshimoto}, \citenamefont {Nomura}, \citenamefont {Tadano}, \citenamefont {Kawamura}, \citenamefont {Kosugi}, \citenamefont {Yoshimi}, \citenamefont {Misawa},\ and\ \citenamefont {Motoyama}}]{Nakamura2021}%
  \BibitemOpen
  \bibfield  {author} {\bibinfo {author} {\bibfnamefont {K.}~\bibnamefont {Nakamura}}, \bibinfo {author} {\bibfnamefont {Y.}~\bibnamefont {Yoshimoto}}, \bibinfo {author} {\bibfnamefont {Y.}~\bibnamefont {Nomura}}, \bibinfo {author} {\bibfnamefont {T.}~\bibnamefont {Tadano}}, \bibinfo {author} {\bibfnamefont {M.}~\bibnamefont {Kawamura}}, \bibinfo {author} {\bibfnamefont {T.}~\bibnamefont {Kosugi}}, \bibinfo {author} {\bibfnamefont {K.}~\bibnamefont {Yoshimi}}, \bibinfo {author} {\bibfnamefont {T.}~\bibnamefont {Misawa}},\ and\ \bibinfo {author} {\bibfnamefont {Y.}~\bibnamefont {Motoyama}},\ }\bibfield  {title} {\bibinfo {title} {{RESPACK: An ab initio tool for derivation of effective low-energy model of material}},\ }\href {https://doi.org/10.1016/j.cpc.2020.107781} {\bibfield  {journal} {\bibinfo  {journal} {Computer Physics Communications}\ }\textbf {\bibinfo {volume} {261}},\ \bibinfo {pages} {107781} (\bibinfo {year} {2021})},\ \Eprint {https://arxiv.org/abs/2001.02351} {arXiv:2001.02351} \BibitemShut
  {NoStop}%
\bibitem [{\citenamefont {Monserrat}(2016)}]{PhysRevB.93.100301}%
  \BibitemOpen
  \bibfield  {author} {\bibinfo {author} {\bibfnamefont {B.}~\bibnamefont {Monserrat}},\ }\bibfield  {title} {\bibinfo {title} {Correlation effects on electron-phonon coupling in semiconductors: Many-body theory along thermal lines},\ }\href {https://doi.org/10.1103/PhysRevB.93.100301} {\bibfield  {journal} {\bibinfo  {journal} {Phys. Rev. B}\ }\textbf {\bibinfo {volume} {93}},\ \bibinfo {pages} {100301} (\bibinfo {year} {2016})}\BibitemShut {NoStop}%
\bibitem [{\citenamefont {Allen}\ and\ \citenamefont {Cohen}(1969)}]{PhysRev.177.704}%
  \BibitemOpen
  \bibfield  {author} {\bibinfo {author} {\bibfnamefont {P.~B.}\ \bibnamefont {Allen}}\ and\ \bibinfo {author} {\bibfnamefont {M.~L.}\ \bibnamefont {Cohen}},\ }\bibfield  {title} {\bibinfo {title} {Carrier-concentration-dependent superconductivity in {SnTe} and {GeTe}},\ }\href {https://doi.org/10.1103/PhysRev.177.704} {\bibfield  {journal} {\bibinfo  {journal} {Phys. Rev.}\ }\textbf {\bibinfo {volume} {177}},\ \bibinfo {pages} {704} (\bibinfo {year} {1969})}\BibitemShut {NoStop}%
\bibitem [{\citenamefont {Kriener}\ \emph {et~al.}(2020)\citenamefont {Kriener}, \citenamefont {Sakano}, \citenamefont {Kamitani}, \citenamefont {Bahramy}, \citenamefont {Yukawa}, \citenamefont {Horiba}, \citenamefont {Kumigashira}, \citenamefont {Ishizaka}, \citenamefont {Tokura},\ and\ \citenamefont {Taguchi}}]{PhysRevLett.124.047002}%
  \BibitemOpen
  \bibfield  {author} {\bibinfo {author} {\bibfnamefont {M.}~\bibnamefont {Kriener}}, \bibinfo {author} {\bibfnamefont {M.}~\bibnamefont {Sakano}}, \bibinfo {author} {\bibfnamefont {M.}~\bibnamefont {Kamitani}}, \bibinfo {author} {\bibfnamefont {M.~S.}\ \bibnamefont {Bahramy}}, \bibinfo {author} {\bibfnamefont {R.}~\bibnamefont {Yukawa}}, \bibinfo {author} {\bibfnamefont {K.}~\bibnamefont {Horiba}}, \bibinfo {author} {\bibfnamefont {H.}~\bibnamefont {Kumigashira}}, \bibinfo {author} {\bibfnamefont {K.}~\bibnamefont {Ishizaka}}, \bibinfo {author} {\bibfnamefont {Y.}~\bibnamefont {Tokura}},\ and\ \bibinfo {author} {\bibfnamefont {Y.}~\bibnamefont {Taguchi}},\ }\bibfield  {title} {\bibinfo {title} {{Evolution of Electronic States and Emergence of Superconductivity in the Polar Semiconductor GeTe by Doping Valence-Skipping Indium}},\ }\href {https://doi.org/10.1103/PhysRevLett.124.047002} {\bibfield  {journal} {\bibinfo  {journal} {Phys. Rev. Lett.}\ }\textbf {\bibinfo {volume} {124}},\ \bibinfo {pages} {047002}
  (\bibinfo {year} {2020})}\BibitemShut {NoStop}%
\bibitem [{\citenamefont {Momma}\ and\ \citenamefont {Izumi}(2011)}]{Momma:db5098}%
  \BibitemOpen
  \bibfield  {author} {\bibinfo {author} {\bibfnamefont {K.}~\bibnamefont {Momma}}\ and\ \bibinfo {author} {\bibfnamefont {F.}~\bibnamefont {Izumi}},\ }\bibfield  {title} {\bibinfo {title} {{{\it VESTA3} for three-dimensional visualization of crystal, volumetric and morphology data}},\ }\href {https://doi.org/10.1107/S0021889811038970} {\bibfield  {journal} {\bibinfo  {journal} {Journal of Applied Crystallography}\ }\textbf {\bibinfo {volume} {44}},\ \bibinfo {pages} {1272} (\bibinfo {year} {2011})}\BibitemShut {NoStop}%
\bibitem [{\citenamefont {Hamann}(2013)}]{Hamann2013}%
  \BibitemOpen
  \bibfield  {author} {\bibinfo {author} {\bibfnamefont {D.~R.}\ \bibnamefont {Hamann}},\ }\bibfield  {title} {\bibinfo {title} {{Optimized norm-conserving Vanderbilt pseudopotentials}},\ }\href {https://doi.org/10.1103/PhysRevB.88.085117} {\bibfield  {journal} {\bibinfo  {journal} {Physical Review B - Condensed Matter and Materials Physics}\ }\textbf {\bibinfo {volume} {88}},\ \bibinfo {pages} {1} (\bibinfo {year} {2013})},\ \Eprint {https://arxiv.org/abs/1306.4707} {1306.4707} \BibitemShut {NoStop}%
\bibitem [{\citenamefont {van Setten}\ \emph {et~al.}(2018)\citenamefont {van Setten}, \citenamefont {Giantomassi}, \citenamefont {Bousquet}, \citenamefont {Verstraete}, \citenamefont {Hamann}, \citenamefont {Gonze},\ and\ \citenamefont {Rignanese}}]{VanSetten2018}%
  \BibitemOpen
  \bibfield  {author} {\bibinfo {author} {\bibfnamefont {M.~J.}\ \bibnamefont {van Setten}}, \bibinfo {author} {\bibfnamefont {M.}~\bibnamefont {Giantomassi}}, \bibinfo {author} {\bibfnamefont {E.}~\bibnamefont {Bousquet}}, \bibinfo {author} {\bibfnamefont {M.~J.}\ \bibnamefont {Verstraete}}, \bibinfo {author} {\bibfnamefont {D.~R.}\ \bibnamefont {Hamann}}, \bibinfo {author} {\bibfnamefont {X.}~\bibnamefont {Gonze}},\ and\ \bibinfo {author} {\bibfnamefont {G.~M.}\ \bibnamefont {Rignanese}},\ }\bibfield  {title} {\bibinfo {title} {{The PSEUDODOJO: Training and grading a 85 element optimized norm-conserving pseudopotential table}},\ }\href {https://doi.org/10.1016/j.cpc.2018.01.012} {\bibfield  {journal} {\bibinfo  {journal} {Computer Physics Communications}\ }\textbf {\bibinfo {volume} {226}},\ \bibinfo {pages} {39} (\bibinfo {year} {2018})},\ \Eprint {https://arxiv.org/abs/1710.10138} {arXiv:1710.10138} \BibitemShut {NoStop}%
\bibitem [{\citenamefont {Perdew}\ \emph {et~al.}(1996)\citenamefont {Perdew}, \citenamefont {Burke},\ and\ \citenamefont {Ernzerhof}}]{pbe}%
  \BibitemOpen
  \bibfield  {author} {\bibinfo {author} {\bibfnamefont {J.~P.}\ \bibnamefont {Perdew}}, \bibinfo {author} {\bibfnamefont {K.}~\bibnamefont {Burke}},\ and\ \bibinfo {author} {\bibfnamefont {M.}~\bibnamefont {Ernzerhof}},\ }\bibfield  {title} {\bibinfo {title} {{Generalized Gradient Approximation Made Simple}},\ }\href@noop {} {\bibfield  {journal} {\bibinfo  {journal} {Physical Review Letters}\ }\textbf {\bibinfo {volume} {77}},\ \bibinfo {pages} {3865} (\bibinfo {year} {1996})}\BibitemShut {NoStop}%
\bibitem [{\citenamefont {Vogl}(1976)}]{PhysRevB.13.694}%
  \BibitemOpen
  \bibfield  {author} {\bibinfo {author} {\bibfnamefont {P.}~\bibnamefont {Vogl}},\ }\bibfield  {title} {\bibinfo {title} {Microscopic theory of electron-phonon interaction in insulators or semiconductors},\ }\href {https://doi.org/10.1103/PhysRevB.13.694} {\bibfield  {journal} {\bibinfo  {journal} {Phys. Rev. B}\ }\textbf {\bibinfo {volume} {13}},\ \bibinfo {pages} {694} (\bibinfo {year} {1976})}\BibitemShut {NoStop}%
\bibitem [{\citenamefont {Chernikov}\ \emph {et~al.}(2012)\citenamefont {Chernikov}, \citenamefont {Bornwasser}, \citenamefont {Koch}, \citenamefont {Chatterjee}, \citenamefont {B\"ottge}, \citenamefont {Feldtmann}, \citenamefont {Kira}, \citenamefont {Koch}, \citenamefont {Wassner}, \citenamefont {Lautenschl\"ager}, \citenamefont {Meyer},\ and\ \citenamefont {Eickhoff}}]{PhysRevB.85.035201}%
  \BibitemOpen
  \bibfield  {author} {\bibinfo {author} {\bibfnamefont {A.}~\bibnamefont {Chernikov}}, \bibinfo {author} {\bibfnamefont {V.}~\bibnamefont {Bornwasser}}, \bibinfo {author} {\bibfnamefont {M.}~\bibnamefont {Koch}}, \bibinfo {author} {\bibfnamefont {S.}~\bibnamefont {Chatterjee}}, \bibinfo {author} {\bibfnamefont {C.~N.}\ \bibnamefont {B\"ottge}}, \bibinfo {author} {\bibfnamefont {T.}~\bibnamefont {Feldtmann}}, \bibinfo {author} {\bibfnamefont {M.}~\bibnamefont {Kira}}, \bibinfo {author} {\bibfnamefont {S.~W.}\ \bibnamefont {Koch}}, \bibinfo {author} {\bibfnamefont {T.}~\bibnamefont {Wassner}}, \bibinfo {author} {\bibfnamefont {S.}~\bibnamefont {Lautenschl\"ager}}, \bibinfo {author} {\bibfnamefont {B.~K.}\ \bibnamefont {Meyer}},\ and\ \bibinfo {author} {\bibfnamefont {M.}~\bibnamefont {Eickhoff}},\ }\bibfield  {title} {\bibinfo {title} {Phonon-assisted luminescence of polar semiconductors: Fr\"ohlich coupling versus deformation-potential scattering},\ }\href {https://doi.org/10.1103/PhysRevB.85.035201}
  {\bibfield  {journal} {\bibinfo  {journal} {Phys. Rev. B}\ }\textbf {\bibinfo {volume} {85}},\ \bibinfo {pages} {035201} (\bibinfo {year} {2012})}\BibitemShut {NoStop}%
\bibitem [{\citenamefont {Cohen}\ and\ \citenamefont {Louie}(2016)}]{Cohen_Louie_2016}%
  \BibitemOpen
  \bibfield  {author} {\bibinfo {author} {\bibfnamefont {M.~L.}\ \bibnamefont {Cohen}}\ and\ \bibinfo {author} {\bibfnamefont {S.~G.}\ \bibnamefont {Louie}},\ }\href@noop {} {\emph {\bibinfo {title} {Fundamentals of Condensed Matter Physics}}}\ (\bibinfo  {publisher} {Cambridge University Press},\ \bibinfo {year} {2016})\BibitemShut {NoStop}%
\bibitem [{\citenamefont {Hele}\ \emph {et~al.}(2021)\citenamefont {Hele}, \citenamefont {Monserrat},\ and\ \citenamefont {Alvertis}}]{10.1063/5.0052247}%
  \BibitemOpen
  \bibfield  {author} {\bibinfo {author} {\bibfnamefont {T.~J.~H.}\ \bibnamefont {Hele}}, \bibinfo {author} {\bibfnamefont {B.}~\bibnamefont {Monserrat}},\ and\ \bibinfo {author} {\bibfnamefont {A.~M.}\ \bibnamefont {Alvertis}},\ }\bibfield  {title} {\bibinfo {title} {{Systematic improvement of molecular excited state calculations by inclusion of nuclear quantum motion: A mode-resolved picture and the effect of molecular size}},\ }\href {https://doi.org/10.1063/5.0052247} {\bibfield  {journal} {\bibinfo  {journal} {The Journal of Chemical Physics}\ }\textbf {\bibinfo {volume} {154}},\ \bibinfo {pages} {244109} (\bibinfo {year} {2021})}\BibitemShut {NoStop}%
\bibitem [{\citenamefont {Poncé}\ \emph {et~al.}(2025)\citenamefont {Poncé}, \citenamefont {Lihm},\ and\ \citenamefont {Park}}]{poncé2025verificationvalidationzeropointelectronphonon}%
  \BibitemOpen
  \bibfield  {author} {\bibinfo {author} {\bibfnamefont {S.}~\bibnamefont {Poncé}}, \bibinfo {author} {\bibfnamefont {J.-M.}\ \bibnamefont {Lihm}},\ and\ \bibinfo {author} {\bibfnamefont {C.-H.}\ \bibnamefont {Park}},\ }\href {https://arxiv.org/abs/2410.14319} {\bibinfo {title} {Verification and validation of zero-point electron-phonon renormalization of the bandgap, mass enhancement, and spectral functions}} (\bibinfo {year} {2025}),\ \Eprint {https://arxiv.org/abs/2410.14319} {arXiv:2410.14319 [cond-mat.mtrl-sci]} \BibitemShut {NoStop}%
\bibitem [{\citenamefont {Park}\ \emph {et~al.}(2020)\citenamefont {Park}, \citenamefont {Zhou}, \citenamefont {Jhalani}, \citenamefont {Dreyer},\ and\ \citenamefont {Bernardi}}]{PhysRevB.102.125203}%
  \BibitemOpen
  \bibfield  {author} {\bibinfo {author} {\bibfnamefont {J.}~\bibnamefont {Park}}, \bibinfo {author} {\bibfnamefont {J.-J.}\ \bibnamefont {Zhou}}, \bibinfo {author} {\bibfnamefont {V.~A.}\ \bibnamefont {Jhalani}}, \bibinfo {author} {\bibfnamefont {C.~E.}\ \bibnamefont {Dreyer}},\ and\ \bibinfo {author} {\bibfnamefont {M.}~\bibnamefont {Bernardi}},\ }\bibfield  {title} {\bibinfo {title} {Long-range quadrupole electron-phonon interaction from first principles},\ }\href {https://doi.org/10.1103/PhysRevB.102.125203} {\bibfield  {journal} {\bibinfo  {journal} {Phys. Rev. B}\ }\textbf {\bibinfo {volume} {102}},\ \bibinfo {pages} {125203} (\bibinfo {year} {2020})}\BibitemShut {NoStop}%
\bibitem [{\citenamefont {Chen}\ \emph {et~al.}(2021)\citenamefont {Chen}, \citenamefont {Wang}, \citenamefont {Rebec}, \citenamefont {Jia}, \citenamefont {Hashimoto}, \citenamefont {Lu}, \citenamefont {Moritz}, \citenamefont {Moore}, \citenamefont {Devereaux},\ and\ \citenamefont {Shen}}]{doi:10.1126/science.abf5174}%
  \BibitemOpen
  \bibfield  {author} {\bibinfo {author} {\bibfnamefont {Z.}~\bibnamefont {Chen}}, \bibinfo {author} {\bibfnamefont {Y.}~\bibnamefont {Wang}}, \bibinfo {author} {\bibfnamefont {S.~N.}\ \bibnamefont {Rebec}}, \bibinfo {author} {\bibfnamefont {T.}~\bibnamefont {Jia}}, \bibinfo {author} {\bibfnamefont {M.}~\bibnamefont {Hashimoto}}, \bibinfo {author} {\bibfnamefont {D.}~\bibnamefont {Lu}}, \bibinfo {author} {\bibfnamefont {B.}~\bibnamefont {Moritz}}, \bibinfo {author} {\bibfnamefont {R.~G.}\ \bibnamefont {Moore}}, \bibinfo {author} {\bibfnamefont {T.~P.}\ \bibnamefont {Devereaux}},\ and\ \bibinfo {author} {\bibfnamefont {Z.-X.}\ \bibnamefont {Shen}},\ }\bibfield  {title} {\bibinfo {title} {Anomalously strong near-neighbor attraction in doped 1d cuprate chains},\ }\href {https://doi.org/10.1126/science.abf5174} {\bibfield  {journal} {\bibinfo  {journal} {Science}\ }\textbf {\bibinfo {volume} {373}},\ \bibinfo {pages} {1235} (\bibinfo {year} {2021})}\BibitemShut {NoStop}%
\bibitem [{\citenamefont {Jiang}(2022)}]{PhysRevB.105.024510}%
  \BibitemOpen
  \bibfield  {author} {\bibinfo {author} {\bibfnamefont {M.}~\bibnamefont {Jiang}},\ }\bibfield  {title} {\bibinfo {title} {{Enhancing $d$-wave superconductivity with nearest-neighbor attraction in the extended Hubbard model}},\ }\href {https://doi.org/10.1103/PhysRevB.105.024510} {\bibfield  {journal} {\bibinfo  {journal} {Phys. Rev. B}\ }\textbf {\bibinfo {volume} {105}},\ \bibinfo {pages} {024510} (\bibinfo {year} {2022})}\BibitemShut {NoStop}%
\bibitem [{\citenamefont {Peng}\ \emph {et~al.}(2023)\citenamefont {Peng}, \citenamefont {Wang}, \citenamefont {Wen}, \citenamefont {Lee}, \citenamefont {Devereaux},\ and\ \citenamefont {Jiang}}]{PhysRevB.107.L201102}%
  \BibitemOpen
  \bibfield  {author} {\bibinfo {author} {\bibfnamefont {C.}~\bibnamefont {Peng}}, \bibinfo {author} {\bibfnamefont {Y.}~\bibnamefont {Wang}}, \bibinfo {author} {\bibfnamefont {J.}~\bibnamefont {Wen}}, \bibinfo {author} {\bibfnamefont {Y.~S.}\ \bibnamefont {Lee}}, \bibinfo {author} {\bibfnamefont {T.~P.}\ \bibnamefont {Devereaux}},\ and\ \bibinfo {author} {\bibfnamefont {H.-C.}\ \bibnamefont {Jiang}},\ }\bibfield  {title} {\bibinfo {title} {{Enhanced superconductivity by near-neighbor attraction in the doped extended Hubbard model}},\ }\href {https://doi.org/10.1103/PhysRevB.107.L201102} {\bibfield  {journal} {\bibinfo  {journal} {Phys. Rev. B}\ }\textbf {\bibinfo {volume} {107}},\ \bibinfo {pages} {L201102} (\bibinfo {year} {2023})}\BibitemShut {NoStop}%
\bibitem [{\citenamefont {Chen}\ \emph {et~al.}(2023)\citenamefont {Chen}, \citenamefont {Wang},\ and\ \citenamefont {Chen}}]{PhysRevB.108.064514}%
  \BibitemOpen
  \bibfield  {author} {\bibinfo {author} {\bibfnamefont {W.-C.}\ \bibnamefont {Chen}}, \bibinfo {author} {\bibfnamefont {Y.}~\bibnamefont {Wang}},\ and\ \bibinfo {author} {\bibfnamefont {C.-C.}\ \bibnamefont {Chen}},\ }\bibfield  {title} {\bibinfo {title} {{Superconducting phases of the square-lattice extended Hubbard model}},\ }\href {https://doi.org/10.1103/PhysRevB.108.064514} {\bibfield  {journal} {\bibinfo  {journal} {Phys. Rev. B}\ }\textbf {\bibinfo {volume} {108}},\ \bibinfo {pages} {064514} (\bibinfo {year} {2023})}\BibitemShut {NoStop}%
\bibitem [{\citenamefont {Han}\ \emph {et~al.}(2020)\citenamefont {Han}, \citenamefont {Kivelson},\ and\ \citenamefont {Yao}}]{PhysRevLett.125.167001}%
  \BibitemOpen
  \bibfield  {author} {\bibinfo {author} {\bibfnamefont {Z.}~\bibnamefont {Han}}, \bibinfo {author} {\bibfnamefont {S.~A.}\ \bibnamefont {Kivelson}},\ and\ \bibinfo {author} {\bibfnamefont {H.}~\bibnamefont {Yao}},\ }\bibfield  {title} {\bibinfo {title} {Strong coupling limit of the holstein-hubbard model},\ }\href {https://doi.org/10.1103/PhysRevLett.125.167001} {\bibfield  {journal} {\bibinfo  {journal} {Phys. Rev. Lett.}\ }\textbf {\bibinfo {volume} {125}},\ \bibinfo {pages} {167001} (\bibinfo {year} {2020})}\BibitemShut {NoStop}%
\bibitem [{\citenamefont {Onari}\ \emph {et~al.}(2004)\citenamefont {Onari}, \citenamefont {Arita}, \citenamefont {Kuroki},\ and\ \citenamefont {Aoki}}]{PhysRevB.70.094523}%
  \BibitemOpen
  \bibfield  {author} {\bibinfo {author} {\bibfnamefont {S.}~\bibnamefont {Onari}}, \bibinfo {author} {\bibfnamefont {R.}~\bibnamefont {Arita}}, \bibinfo {author} {\bibfnamefont {K.}~\bibnamefont {Kuroki}},\ and\ \bibinfo {author} {\bibfnamefont {H.}~\bibnamefont {Aoki}},\ }\bibfield  {title} {\bibinfo {title} {Phase diagram of the two-dimensional extended hubbard model: Phase transitions between different pairing symmetries when charge and spin fluctuations coexist},\ }\href {https://doi.org/10.1103/PhysRevB.70.094523} {\bibfield  {journal} {\bibinfo  {journal} {Phys. Rev. B}\ }\textbf {\bibinfo {volume} {70}},\ \bibinfo {pages} {094523} (\bibinfo {year} {2004})}\BibitemShut {NoStop}%
\bibitem [{\citenamefont {Ejima}\ and\ \citenamefont {Nishimoto}(2007)}]{PhysRevLett.99.216403}%
  \BibitemOpen
  \bibfield  {author} {\bibinfo {author} {\bibfnamefont {S.}~\bibnamefont {Ejima}}\ and\ \bibinfo {author} {\bibfnamefont {S.}~\bibnamefont {Nishimoto}},\ }\bibfield  {title} {\bibinfo {title} {Phase diagram of the one-dimensional half-filled extended hubbard model},\ }\href {https://doi.org/10.1103/PhysRevLett.99.216403} {\bibfield  {journal} {\bibinfo  {journal} {Phys. Rev. Lett.}\ }\textbf {\bibinfo {volume} {99}},\ \bibinfo {pages} {216403} (\bibinfo {year} {2007})}\BibitemShut {NoStop}%
\bibitem [{\citenamefont {Craciun}\ \emph {et~al.}(2009)\citenamefont {Craciun}, \citenamefont {Giovannetti}, \citenamefont {Rogge}, \citenamefont {Brocks}, \citenamefont {Morpurgo},\ and\ \citenamefont {van~den Brink}}]{PhysRevB.79.125116}%
  \BibitemOpen
  \bibfield  {author} {\bibinfo {author} {\bibfnamefont {M.~F.}\ \bibnamefont {Craciun}}, \bibinfo {author} {\bibfnamefont {G.}~\bibnamefont {Giovannetti}}, \bibinfo {author} {\bibfnamefont {S.}~\bibnamefont {Rogge}}, \bibinfo {author} {\bibfnamefont {G.}~\bibnamefont {Brocks}}, \bibinfo {author} {\bibfnamefont {A.~F.}\ \bibnamefont {Morpurgo}},\ and\ \bibinfo {author} {\bibfnamefont {J.}~\bibnamefont {van~den Brink}},\ }\bibfield  {title} {\bibinfo {title} {Evidence for the formation of a mott state in potassium-intercalated pentacene},\ }\href {https://doi.org/10.1103/PhysRevB.79.125116} {\bibfield  {journal} {\bibinfo  {journal} {Phys. Rev. B}\ }\textbf {\bibinfo {volume} {79}},\ \bibinfo {pages} {125116} (\bibinfo {year} {2009})}\BibitemShut {NoStop}%
\bibitem [{\citenamefont {Perepelitsky}\ \emph {et~al.}(2016)\citenamefont {Perepelitsky}, \citenamefont {Galatas}, \citenamefont {Mravlje}, \citenamefont {\ifmmode~\check{Z}\else \v{Z}\fi{}itko}, \citenamefont {Khatami}, \citenamefont {Shastry},\ and\ \citenamefont {Georges}}]{PhysRevB.94.235115}%
  \BibitemOpen
  \bibfield  {author} {\bibinfo {author} {\bibfnamefont {E.}~\bibnamefont {Perepelitsky}}, \bibinfo {author} {\bibfnamefont {A.}~\bibnamefont {Galatas}}, \bibinfo {author} {\bibfnamefont {J.}~\bibnamefont {Mravlje}}, \bibinfo {author} {\bibfnamefont {R.}~\bibnamefont {\ifmmode~\check{Z}\else \v{Z}\fi{}itko}}, \bibinfo {author} {\bibfnamefont {E.}~\bibnamefont {Khatami}}, \bibinfo {author} {\bibfnamefont {B.~S.}\ \bibnamefont {Shastry}},\ and\ \bibinfo {author} {\bibfnamefont {A.}~\bibnamefont {Georges}},\ }\bibfield  {title} {\bibinfo {title} {Transport and optical conductivity in the hubbard model: A high-temperature expansion perspective},\ }\href {https://doi.org/10.1103/PhysRevB.94.235115} {\bibfield  {journal} {\bibinfo  {journal} {Phys. Rev. B}\ }\textbf {\bibinfo {volume} {94}},\ \bibinfo {pages} {235115} (\bibinfo {year} {2016})}\BibitemShut {NoStop}%
\end{thebibliography}
\end{document}